\documentclass{aa}

\usepackage[varg]{txfonts}
\usepackage{amsmath}
\usepackage[caption=false]{subfig}
\usepackage[english,english]{babel}
\usepackage{amssymb,amsfonts,textcomp}
\usepackage{array}
\usepackage{supertabular}
\usepackage{hhline}
\usepackage{soul}
\usepackage[usenames,dvipsnames]{color}
\usepackage{hyperref}
\hypersetup{dvips, colorlinks=true, linkcolor=blue, citecolor=blue, filecolor=blue, urlcolor=blue}
\usepackage{graphicx}
\usepackage{orcidlink}
\bibpunct{(}{)}{;}{a}{}{,}

\begin{document}
\title{
Tracking on-the-fly massive black hole binary evolution and coalescence in galaxy simulations:  RAMCOAL
} 
\titlerunning{Coalescence of massive black holes in galaxies}
\authorrunning{Li et al.}

\author{Kunyang Li\inst{1}\orcidlink{0000-0002-0867-8946}
\and Marta Volonteri\inst{1}\orcidlink{0000-0002-3216-1322}
\and Yohan Dubois\inst{1}\orcidlink{0000-0003-0225-6387}
\and Ricarda Beckmann\inst{2}\orcidlink{0000-0002-2850-0192}
\and Maxime Trebitsch\inst{3}\orcidlink{0000-0002-6849-5375}
}
\institute{Institut d'Astrophysique de Paris, UMR 7095, CNRS, Sorbonne Universit\'e, 98 bis boulevard Arago, 75014 Paris, France 
\institute{Institute for Astronomy, University of Edinburgh, Royal Observatory, Edinburgh EH9 3HJ, UK}
\\
\email{likun@iap.fr}
\and 
Institute of Astronomy and Kavli Institute for Cosmology, University of Cambridge, Madingley Road, Cambridge CB3 0HA, United Kingdom
\and
LUX, Observatoire de Paris, Université PSL, Sorbonne Université, CNRS, 75014 Paris, France
}

\date{Accepted }
\abstract{
The detection of gravitational waves (GWs) from massive black hole binary (MBHB) coalescence motivates the development of a sub-grid model. We present RAMCOAL, integrated into the RAMSES code, which simulates the orbital evolution of MBHBs, accounting for stellar and gaseous dynamical friction (DF), stellar scattering, circumbinary disk interactions, and GW emission at scales below the simulation resolution. Unlike post-processing approaches, RAMCOAL tracks the real-time evolution of MBHBs within hydrodynamical simulations of galaxies using local quantities to model dynamics and accretion. This enables more accurate predictions of both GW signals and the properties of merging black holes. We validate RAMCOAL across isolated and merging galaxy setups at resolutions of 10, 50, and 100\,pc, with and without black hole accretion and feedback. In addition, we test the model in seven galaxy merger scenarios at 100 pc resolution. These tests demonstrate that RAMCOAL is largely resolution-independent and successfully captures the effects of DF from stars, dark matter, and gas, loss-cone scattering, viscous drag from circumbinary disks, and GW emission -- all within a realistic galactic environment, even at low resolutions.
With RAMCOAL, we can better estimate MBHB coalescence rates and the GW background, while providing insights into the electromagnetic counterparts of GW sources. This approach bridges the gap between electromagnetic observations and GW detection, offering a more comprehensive understanding of MBHB evolution in cosmological simulations.
}


\maketitle
\section{Introduction}
\label{sec:intro}
Massive black holes (MBHs), with masses ranging from $\sim 10^6$ to $10^{10} \, {\rm M}_{\odot}$, are found in the centers of most massive galaxies \citep{S1982, KR1995, M1998}. When two galaxies merge, their respective MBHs are brought together, eventually forming a bound system within the gravitational potential of the merger remnant. Once gravitationally bound, these MBH pairs can evolve into MBH binaries\footnote{We refer to a system of two MBHs as an MBH pair before they are gravitationally bound, and as an MBH binary (MBHB) after they become bound.} and continue decaying to smaller separations, ultimately becoming powerful sources of gravitational waves (GWs). The detection of GWs from stellar-mass black hole binaries by the Laser Interferometer Gravitational-Wave Observatory \citep[LIGO;][]{LIGO2016} marked the dawn of the GW astronomy era. Within the next $\sim 10-15$ years, the Laser Interferometer Space Antenna \citep[LISA;][]{LISA2017} is expected to detect GWs from merging MBHBs. Additionally, pulsar timing arrays (PTAs) have already detected GWs in the nHz range, likely originating from MBHBs \citep{Agazie_2023, EPTA2024, Reardon2023}.

LISA is expected to detect GWs from MBHB mergers with masses in the range of $10^4 - 10^7 \, {\rm M_\odot}$ out to high redshifts. With PTAs likely detecting stochastic GWs from the local population of more massive MBHBs ($>10^8 \, {\rm M_\odot}$), it has become increasingly important to interpret these findings and prepare for future LISA detections. Thus, a dedicated theoretical model for the evolution of MBHBs is critical for understanding their dynamical behavior and GW signatures.

The expected detection rate for PTA and LISA is governed not only by the frequency of galaxy mergers but also by the processes within the remnant galaxies that drive MBHs to small enough separations to form a binary. Understanding the evolution of MBHs in post-merger galaxies is therefore essential for predicting the GW signals that will be probed by future GW observatories.

\citet{BBR1980} established the framework for modeling the orbital decay of MBH pairs following a galactic merger. Depending on the separation and characteristics of the merging galaxies, MBH orbital decay can be driven by four primary physical mechanisms. Initially, when the MBHs are separated by $\sim 1$\,kpc, dynamical friction (DF) from gas and stars is expected to dominate their decay. DF arises as a massive object, such as an MBH, moves through a background medium. Gravitational deflection of gas \citep{O1999, KK2007} or collisionless particles \citep[e.g., stars or dark matter;][]{C1943, AM2012} forms an overdense wake that exerts a gravitational pull on the MBH, extracting its orbital energy.

The timescale for orbital decay due to DF is determined by the properties of the MBHs and their host galaxy. Key factors include the total mass, mass ratio, and initial orbits of the MBHs, as well as the distribution and kinematics of the surrounding gas and stars. Literature shows that, depending on the merger configuration, MBH pairing timescales can vary from $\sim 10^6$\,yr to longer than a Hubble time \citep{review2020}. Some merger remnants promote efficient MBH pairing and subsequent coalescence, while others may delay this process, potentially leading to the interaction of triple MBHs in future mergers \citep{Bonetti2018}. In some cases, MBHs may remain stranded, unable to form binaries at all \citep[e.g.,][]{Volonteri2003}.

Once two MBHs become gravitationally bound, stellar loss-cone scattering is expected to drive further orbital decay \citep[e.g.,][]{Q1996, QH1997, Y2002}. If the galaxy is sufficiently gas-rich, drag from a circumbinary disk may also influence the MBHB’s evolution at separations $\la 0.1$\,pc \citep{A2005}. At separations below $\sim 1000$ Schwarzschild radii, GW emission begins to dominate orbital decay, leading to coalescence \citep{KT1976, BBR1980}.

Previous studies have explored MBH pairing in stellar environments using N-body simulations \citep[e.g.,][]{QH1997, Y2002, B2006, KJM2011, K2013}, hydrodynamic simulations of MBHs interacting with gas \citep[e.g.,][]{E2005, D2007, C2009}, and semi-analytic models \citep[e.g.,][]{Volonteri2003,B2012, K2016, 2020MNRAS.495.4681I}. 
In cosmological simulations, many codes don't follow MBH dynamics on the fly and simply reposition MBHs \citep[see][for a review]{2022MNRAS.516..167B}. Other codes include gaseous DF  \citep{Dubois2013} and/or dynamical friction from stars and dark matter \citep{Tre2015,Hugo2019,Chen2021}. 

The dynamical evolution of MBH pairs at sub-kpc separations is still challenging to model in most cosmological simulations due to resolution limitations. Up to now, MBHB evolution and GW emission have typically been studied in post-processing, which disconnects the dynamical evolution from the evolving galactic environment, introducing significant uncertainties in the predicted GW signatures \citep[see][]{KL2018, Marta2020,LBB2022}. A notable exception is the KETJU code~\citep{Rantala2017}, which follows MBH binaries down to the GW emission stage in stellar environments \citep{Matias2023}.

To address these challenges, we developed RAMCOAL, a sub-grid module for MBHB dynamical evolution, integrated within the RAMSES code~\citep{Teyssier2002}. Unlike post-processing approaches, RAMCOAL evolves MBHB dynamics on-the-fly, incorporating physical processes such as DF from gas, stars, and dark matter, stellar hardening, viscous torques in circumbinary disks, and GW emission. RAMCOAL also includes MBH accretion and feedback.

This paper is organized as follows: In Sect.~\ref{sec:RAMCOAL}, we introduce the main features of RAMCOAL for evolving sub-grid MBH dynamics, accretion, and feedback. Section~\ref{sec:ramses} provides a brief overview of the RAMSES code in which RAMCOAL is implemented. In Sect.~\ref{sec:robust}, we present the robustness tests of RAMCOAL in isolated galaxy remnants with resolutions of $10$, $50$, and $100\,{\rm pc}$. In Sect.~\ref{sec:merger}, we extend the robustness tests to merging galaxy pairs and explore how galaxy properties influence MBHB coalescence time. Finally, we discuss the implications of our findings in Sect.~\ref{sec:discussion} and conclude in Sect.~\ref{sec:conclusion}.

\section{RAMCOAL: RAMSES sub-grid coalescence model of massive black hole binaries}
\label{sec:RAMCOAL}
RAMCOAL is a sub-grid model in RAMSES which simulates the MBHB evolution all the way to the final coalescence from the point where the separation between two MBHs become smaller than 4 times the simulation resolution ($4 \Delta x$). At this point RAMSES considers the two MBHs merged, something generally dubbed ``numerical merger'' \citep{Marta2020}. Before RAMCOAL is activated, MBHs evolve under the DF model in RAMSES \citep{Dubois2012,Hugo2019} which will be summarized briefly in the next section. RAMCOAL deals with the dynamical evolution and the separate mass growth of the two MBHs in the binary. It accounts also for how feedback affects the dynamical evolution of the binary. Feedback injected on larger scales is however not modified.   

The evolution of MBHs in RAMCOAL can be split into two stages, the first stage starts when the separation between two MBHs is $4 \Delta x$ and ends when two MBH become gravitationally bounded. Dynamical friction from stars and gas drives the evolution at the stage 1 of RAMCOAL. Stage 2 begins when two MBHs become gravitationally bounded (when the kinetic orbital energy of the binary is smaller than its potential energy) and the total gas and stellar mass enclosed by the binary orbit is smaller than twice the sum of the mass of the MBH pair, and ends when the MBHB coalescence happens. At the stage 2 of RAMCOAL, MBHB evolution is driven by loss-cone scattering, viscous drag from circumbinary disk, and GW emission. In this section, we introduce how these aforementioned physical mechanisms are modeled in RAMCOAL.

\begin{table*}
\caption{Stages of MBH dynamics in RAMSES and RAMCOAL.} \label{tab:stages}
\centering
\begin{tabular}{c|p{0.22\linewidth}|p{0.22\linewidth}|p{0.22\linewidth}}
\hline
\hline
Stage 0 &  Until MBHs are separated by  $4 \Delta x$  &  RAMSES & dynamical friction\\
\hline
Stage 1 &  After Stage 0 until MBHs are gravitationally bound & RAMCOAL & dynamical friction  \\
\hline
Stage 2 & After Stage 1 until coalescence &  RAMCOAL &  stellar hardening, migration in a circumbinary disc, GW emission \\
\hline
\end{tabular}
\end{table*}

\subsection{Numerical integration of the orbital parameters}
\label{subsec:orbEvol}

The orbital evolution of the MBHs due to DF is followed until the MBHB become gravitationally bounded. We solve the equation of motion for each MBH in the polar coordinates using an $8$th-order Runge-Kutta method. The MBHB evolution time step is sub-cycled with respect to the simulation time step. The time step in RAMCOAL is adaptive, with a maximum increase in semi-major axis or eccentricity of $20\%$ per time step, and capped at a maximum of $5\%$ of the simulation time step. We determined that the relative error in conservation of energy and angular momentum corresponding to this time step choice is smaller than $0.5\%$, which meets our error tolerance criterion. 

Over the extent of each simulation, the farthest and closest radial distance between the primary and secondary MBHs is recorded for every orbit in order to estimate the orbital semi-major axis, $a$, and eccentricity, $e$. As the galaxy potential is not proportional to $1/r$, the orbit
of the MBH is neither Keplerian nor closed. Thus, the computed $a$ and $e$ are only approximate values used to track the shape and size of the orbits.

\subsection{Dynamical friction force due to stars, dark matter, and gas}
\label{sub:sDF}
At the stage 1 of RAMCOAL, starting from $4 \Delta x$, the orbital evolution of MBHs is due to the DF from stars, dark matter (DM), and gas.
 We compute the DF force on both MBHs in a pair due to the collisionless component (stars and DM) following the approach laid out in equation~30 in \citet{AM2012}. 
 It accounts for the DF force contribution from particles moving slower than the MBH ($v_{\star}<v$, where $v$ is the velocity of the MBH and $v_\star$ is the velocity of the star or DM particle) and those moving faster than the MBH ($v_{\star}>v$), hence, leading to 
\begin{equation}
  \label{eq:bulgeforce}
\vec{F}_{\star} =\vec{F}^{(v_{\star}<v)}_{\star}+\vec{F}^{(v_{\star}>v)}_{\star}\, .
\end{equation}
If we define $\vec{\mathcal{F}}=-4\pi G^2 M^2 \bar{\rho}_{\rm star/ DM}(\vec{r}) (\vec{v}/{v^3})$,  where $\bar{\rho}_{\rm star}(\vec{r})$ and $\bar{\rho}_{\rm DM}(\vec{r})$ are the volume-weighted average stellar and DM density\footnote{The density is averaged within a resolution sphere with radius $4\Delta x$ centered at the position $\vec r$ of the MBH}, respectively, then, the stellar and DM DF on the MBH can be represented by:
\begin{equation}
\label{eq:bulgeforce1}
\vec{F}^{(v_{\star}<v)}_{\star} = \vec{\mathcal{F}} \int_{0}^{v} 4 \pi
f(v_{\star}) v_{\star}^2 \ln \left[\frac{p_{\mathrm{max}}}{G M} (v^2
  - v_{\star}^2)\right] dv_{\star}
\end{equation}
and
\begin{equation}
\vec{F}^{(v_{\star}>v)}_{\star} = \vec{\mathcal{F}} \int_{v}^{v_{\rm esc}} 4 \pi
  f(v_{\star}) v_{\star}^2 \left[\ln \left(\frac{v_{\star} + v}{v_{\star} - v}\right) -2\frac{v}{v_{\star}} \right]dv_{\star}\, ,
\label{eq:bulgeforce2}
\end{equation}
where $G$ is the gravitational constant, $M$ is the mass of the MBH, $f(v_{\star})$ is the velocity distribution of collisionless particles (assuming the star and DM particles around the MBHB have the same velocity distribution, that is $\sigma_{\star}$ and $f(v_{\star})$ are computed from a population of star and DM particles).
$v_{\rm esc}(r_{\rm com})=\sqrt{2 G(M_{\rm enc}(r_{\rm com})+\tilde M_{12})/r_{\rm com}}$ is the escape velocity, where $M_{\rm enc} (r_{\rm com})$ is the total star, DM and gas mass enclosed in the MBH orbit around the center of mass of the MBHB, $\tilde M_{12}=M_{1}M_{2}/(M_{1}+M_{2})$ is the reduced mass of the MBHB (here and in the following $M_1>M_2$), and $r_{\rm com}$ is the distance from each MBH to the center of mass of the binary. We set the maximum impact parameter to be $p_{\rm max}=4\Delta x$, and the minimum impact parameter to be $p_{\rm min}=\max(GM/(v^2- v_{\star}^2),R_{\rm sg})$, where $R_{\rm sg}$ is the radius of a self-gravitating $\alpha$-disk around the MBH \citep{Dotti2013}. This action is to avoid the artificial effect of zero stellar DF due to the unresolved stellar velocity distribution around the MBH.
According to equations~\ref{eq:bulgeforce1} and \ref{eq:bulgeforce2}, both the collisionless particles moving slower and faster than the MBH will be deflected into an overdensity wake trailing the MBH, pulling it backward.

In RAMCOAL neither the velocity distribution of collisionless particles nor the density profile is resolved, since by definition the subgrid model for MBHB DF is triggered once the distance between the two MBHs are below a resolution limit of $4\Delta x$. Therefore, in the sub-grid model, we set the velocity distribution of collisionless particles to be Maxwellian:
\begin{equation}
f{(v_\star)} = \frac{1}{(2\pi\sigma^2_\star)^{3/2}}\exp\left(-\frac{v^2_\star}{2\sigma^2_\star}\right)\, ,
\end{equation}
where 
\begin{equation}
\label{eq:maxw}
\sigma_\star=\sqrt{\frac{G(M_{\rm sp}+\tilde M_{12})}{R_{\rm sp}}}
\end{equation}
is the velocity dispersion of collisionless particles. $M_{\rm sp}=M_{\rm gas, sp}+M_{\rm star, sp}+M_{\rm DM, sp}$ is the total mass made of the sum of, respectively, the gas, star, and DM mass within the `resolution sphere' with a radius $R_{\rm sp}=4\Delta x$. The size of the resolution sphere is chosen to encompass the two MBHs:  the numerical merger happens when the two MBHs separated by $4\Delta x$, which is therefore considered as the radius of the sphere. 

The maximum density in a simulation depends on resolution, and at the typical resolutions of galaxy simulations it cannot reach the high densities typical of nuclear star clusters and molecular clouds. To lessen the impact of resolution, and therefore be able to apply this approach to simulations of different resolutions, we devise the following approach that mimics the evolution of a MBHB in such environments. 

We consider a MBHB is embedded in a nuclear star cluster and a  molecular cloud when the star formation criterion is triggered in the whole resolution sphere, i.e. when 
$\bar{\rho}_{i,\rm sp} > \rho_{\rm th}$, $\bar{\rho}_{i,\rm  sp}$ is the mass-weighted mean density in the resolution sphere ($i$ standing for either gas or star). $\rho_{\rm th}=n_{0}m_{\rm p}/X_{\rm H}$ is the mass density threshold for star formation, $m_{\rm p}$ is the proton mass, $X_{\rm H}=0.76$ is the hydrogen mass fraction, and $n_{0}$ is a free parameter representing the hydrogen number density threshold above which star formation (SF) is allowed. In this work, we set $n_{0}$ to be as $n_{0}=2^{\rm 3(\ell_{\rm max}-\ell_{\rm 0})}\times 1\,\rm H\, cm^{-3}$, where $\ell_{\rm max}$ is the maximum level of refinement\footnote{This scaling is constructed to obtain an identical mass resolution, proportional to $n_0/2^{3\ell_{\rm max}}$, across various spatial resolution runs.} and $\ell_0=11$ (corresponding to a cell size of $50\,\rm pc$ for these particular simulation setups).

To model this sub-grid environment, we assume the MBH finds itself in a region where star formation occurs with the typical characteristics of a molecular cloud and where stars follow an (identical) unresolved density profile. We assume the evolution of sub-grid star cluster is triggered upon the star formation criteria in RAMSES code but not in a single cell, in the whole resolution sphere of $4\Delta x$. This means that we require simultaneous and widespread star formation in all cells around the BH, a more stringent criterion than stochastic star formation in a single cell, and something that is likely to lead to a star cluster.
After the corresponding star formation criterion in RAMSES is met, the sub-grid gaseous or stellar density profile (represented by $i$ for gas or stars) evolves into a cored-isothermal profile
\begin{equation} \label{eq:subgrid_rho_gas}
  \hat \rho_i(\vec{r}) = 
\begin{cases}
  \rho_{i,{\rm c}} & {\rm if}\, r \leq r_{i,\rm c}\, ,\\
  {\rm max}(\rho_{i,{\rm c}} \left(\frac{r_{i,\rm c}}{r}\right)^{2},\bar{\rho}_{i,\rm sp} ) &  {\rm if}\,  r_{i,\rm c} < r \leq 4\Delta x \, ,\\
  \bar{\rho}_{i} & {\rm if}\, r > 4\Delta x\,  ,\\
\end{cases}
\end{equation}
where the core density is $\rho_{\rm gas,c}=N_{\rm scale}\rho_{\rm th}$, with $N_{\rm scale} = 250\times 2^{3(\ell_{\rm 0}-\ell_{\rm max})}$.

$r_{\rm c}$ is computed using the total gas mass within the resolution sphere. $N_{\rm scale}$ is a scaling factor making $\rho_{\rm gas,c}$ independent of resolution, with $\rho_{\rm gas,c} \sim 5\times 10^{\rm -22}\, {\rm g\,cm^{-3}}$ at all resolutions. Once the sub-grid density criteria is met, the core density gradually increases to the final value of $\rho_{\rm gas,c}$ within a timescale of $N_{\rm rise} T_{\rm sp}$, where $N_{\rm rise} = 2^{(\ell_{\rm max}-\ell_{\rm 0})}$ is a scaling factor, and $T_{\rm sp} = 2\pi \sqrt{R_{\rm sp}^3/(G M_{\rm sp})}$ is the orbital period at $R_{\rm sp}=4\Delta x$.
$M_{\rm sp}$ is the total mass in the resolution sphere.
 
$\bar{\rho}_{\rm gas, sp}$ and $\bar{\rho}_{\rm star,sp}$ are the mean gas and star density within the resolution sphere surrounding the center of mass of the MBHB, so their values change and evolve depending on the position of the center of mass of the MBHB within the galaxy. During the time when $\bar{\rho}_{\rm gas, sp} > \rho_{\rm th}$, the gas and stellar sub-grid density profile is updated every $N_{\rm rise} T_{\rm sp}$. Even if the gas sub-grid density profile (i.e.~equation~\ref{eq:subgrid_rho_gas}) is already triggered, it can also be de-triggered within an arbitrarily chosen value of $3$ fine time steps\footnote{We have explored the choice of de-trigger time, and found that, as long as it is a small non-zero number, the results are not affected. We choose $3$ fine time steps because it gives us a short time but not too short to make the de-trigger process non-physical.} when $\bar{\rho}_{\rm gas, sp} < \rho_{\rm th}$  (usually of the order of $\sim 9\times 10^4 \, \mathrm{yr}$ for a resolution of $50\,\rm pc$). After the first triggering of the sub-grid stellar density, it is is updated at each time step of the simulation. The sub-grid profile is determined by conserving mass in the resolution sphere, with a proviso for the connection with the outer region, as expressed in Eq.~\ref{eq:subgrid_rho_gas}. 

The de-triggering of the sub-grid gas density profile is either due to the feedback blowing away the gas reservoir or when the MBHB enters a relatively low gas density region in the galaxy. Either way, the gaseous environment within the resolution sphere is perturbed and a molecular cloud profile is no longer suitable for describing the sub-grid gas structure. Therefore we assume that the de-triggering of the sub-grid gas density profile is equal to $\bar{\rho}_{\rm gas, sp}$ everywhere in the resolution sphere for simplicity. Only the gas sub-grid density profile is de-triggered while that of the stellar component remains as the cored-power law once triggered. We note that the de-triggering process is much faster than when the sub-grid structure is triggered, since the triggering process is dominated by the gravitational collapsing of gas and star formation in the core area, while on the other hand, the de-triggering process only depends on how fast the feedback or ram pressure perturbs the gas reservoir in the resolution sphere. If the de-triggering condition is met during the triggering process, the de-triggering would take over the triggering process, and the gas sub-grid density profile would be de-triggered. The sub-grid gas structure can be re-triggered if the criteria is met again, and the stellar sub-grid structure will be updated to the new sub-grid profile from the old sub-grid profile as well in $N_{\rm rise} T_{\rm sp}$. 

We compute the gaseous DF force exerted onto a MBH following \citet{KK2007}. It takes into account the contribution to the DF force of the spiral density wake created by the MBH orbiting in the gas disk of the host galaxy.
\begin{equation}
  \label{eq:gdforce}
\vec{F}_{\rm gas} = -\frac{4\pi (GM)^2 \bar{\rho}_{\rm gas, sp}}{\Delta \bar v^2}(I_{\rm r} \vec{e}_{\rm r}+I_{\phi} \vec{e}_\phi)
\end{equation}
where the DF force is expressed in terms of radial ($\vec e_{\rm r}$) and azimuthal ($\vec e_\phi$) unit vectors (where $\vec e_\phi$ is pointing towards the same direction than that the MBH orbit velocity). $I_{\rm r}$ and $I_{\phi}$ are dimensionless functions
\begin{equation} \label{eq:ir}
  I_{\rm r} =
\begin{cases}
  {\cal M}^2 10^{3.51{\cal M} -4.22} &  {\rm if}\, {\cal M} < 1.1\, , \\
  0.5 \ln [9.33{\cal M}^2 ({\cal M}^2 - 0.95)] & {\rm if}\, 1.1 \leq
  {\cal M} < 4.4\, ,\\
  0.3{\cal M}^2 & {\rm if}\, {\cal M} \geq 4.4 \, ,
\end{cases}
\end{equation}
and
\begin{equation}
  \label{eq:iphi}
I_{\phi} =
\begin{cases}
0.7706 \ln \left(\frac{1+{\cal M}}{1.0004-0.9185{\cal M}}\right) & {\rm if}\,
{\cal M} < 1.0\, , \\
\ln \left [ 330 \left ( \frac{b_{\rm max}}{b_{\mathrm{min}}} \right ) \frac{({\cal M}-0.71)^{5.72}}{{\cal
    M}^{9.58}} \right ]
& {\rm if}\, 1.0 \leq {\cal M} < 4.4\, , \\
\ln \left(\frac{b_{\rm max}/b_{\mathrm{min}}}{0.11{\cal M}+1.65}\right) & {\rm if}\, {\cal M}\geq 4.4\, ,
\end{cases}
\end{equation}
of the Mach number ${\cal M}=\Delta \bar v/\bar{c}_{\rm s}$, where $\Delta \bar v = v_{\rm BH} - \bar{v}_{\rm gas}$ is the velocity of the MBH relative to the gas background, and $\bar{v}_{\rm gas}$ and $\bar{c}_{\rm s}$ are respectively the mass-weighted mean relative velocity and sound speed within the resolution sphere.
$b_{\rm max}/b_{\rm min}$ is the ratio of maximum to minimum impact parameter for the gas interacting with the MBH, which we set to $10$ \citep{BT1987}. The ratio $b_{\rm max}/b_{\rm min}$ provides a relative measure of the extent of the gaseous wake at any time in the calculation. The extent of the wake of is bound by the MBH orbit on one end and by the event horizon of the MBH on the other, but can be smaller than that if the wake is dynamic and its size fluctuates in time. We note that the azimuthal component of the DF force has a weak logarithmic dependence on this ratio, thus, assuming a constant ratio provides a satisfactory approximation \citep{BT1987}.
These expressions imply a radial component of the gaseous DF force that always points towards the center and an azimuthal component that points in the opposite direction from the velocity vector $\Delta \vec{v}$, both contributing to shrinking the binary separation.
Both $I_{\rm r}$ and $I_{\phi}$ peak sharply at ${\cal M} = 1$. Furthermore, since the strength of $\vec{F}_{\rm gas}$ is proportional to $I_{\rm r}/\mathcal{M}^2$ and $I_{\phi}/\mathcal{M}^2$ the gaseous DF force is small when the velocity difference between MBH and the gas is large (i.e.~when $\Delta v \ga 4 c_{\rm s}$).

\subsection{The radiation feedback effect}
\label{subsub:RF_effect}
When the MBHs undergo accretion, the radiation from this process can ionize the surrounding gas, and lead to the formation of an HII region along the orbit \citep{KK2007}. Meanwhile, the radiation pressure can cause a shell of dense ionized gas to form at the upstream of the MBH motion, generating a positive azimuthal DF force which speeds up the secondary MBH instead of slowing it down. The change in gaseous DF force under radiation feedback is: 

\begin{equation}
\label{eq:DF}
\vec{F}^{\prime}_{\rm gas, \phi} = -0.6\vec{F}_{\rm gas, \phi}\, ,\vec{F}^{\prime}_{\rm gas, r} = \vec{F}_{\rm gas, r}.
\end{equation}
However, the effect of radiation feedback on DF is only significant when the system satisfies the following conditions: $f_{\rm Edd} > 0.01$, ${\cal M}< 4$, and $(1+{\cal M}^2)\,M_{\rm BH}\,n_{\rm \infty}  <  10^9 \, {\rm M_\odot}\, {\rm cm}^{-3}$, where $f_{\rm Edd}$ is the ratio of accretion rate onto individual MBH over its Eddington accretion rate, $n_{\rm \infty}$ is the gas number density unaffected by the gravity of the MBH pair. When ${\cal M} \geq 4$, the dense shell upstream of the MBH becomes gravitationally unstable and collapses, restoring the DF to the value it would have in absence radiative feedback. Similarly, when $(1+{\cal M}^2)\,M_{\rm BH}\,n_{\rm \infty}>  10^9 \, {\rm M_\odot}\, {\rm cm}^{-3}$, the gravitational pressure of the gas becomes stronger than the opposing radiation pressure, and the DF is restored to the value and direction that it would have in the absence of radiative feedback \citep{PB2017, T2020}. 

\subsection{Loss-cone scattering}
\label{subsub:LC}
At stage 2, the MBHB shrinks under loss-cone (LC) scattering, circumbinary viscous drag, and GW emission. LC scattering dominates over DF in removing orbital energy \citep{BBR1980, AM2012}. Hardening of MBHB orbits by LC scattering can be approximately described by

\begin{equation}
\label{eq:LC1}
\left(\frac{{\rm d}f_{\rm orb}}{{\rm d}t}\right )_{\mathrm{LC}}=\frac{3G^{4/3}}{2(2\pi)^{2/3}}\frac{H \rho_{\rm inf}}{\sigma_{\rm \star}}M_{\rm bin}^{1/3}f_{\rm orb}^{1/3}
\end{equation}

and

\begin{equation}
\label{eq:LC2}
\left(\frac{{\rm d}e}{{\rm d}t}\right )_{\mathrm{LC}}=\frac{G^{4/3}}{(2\pi)^{2/3}}\frac{HK \rho_{\rm inf}}{\sigma_{\rm \star}}M_{\rm bin}^{1/3}f_{\rm orb}^{-2/3},
\end{equation}

where $f_{\rm orb}$ is the MBHB orbital frequency, $e$ is the eccentricity (see Sect.~\ref{subsec:orbEvol}), $\sigma_{\rm \star}$ is the stellar velocity dispersion in the resolution sphere (see equation~\ref{eq:maxw}), $\rho_{\rm inf}$ is the stellar density at $R_{\mathrm{inf}}$ (the influence density), where the mass enclosed in
the orbit is twice the binary mass $M_{\rm bin}=M_1+M_2$, and $H$ and $K$ are empirical numerical factors from three-body scattering experiments in tables 1 and 2 of \citet{Sesana2006}.

\subsection{Viscous drag in a circumbinary disk}
\label{subsub:VD}
We take into account the viscous drag due to the circumbinary disk when the semi-major axis of the MBHB is below $\sim 1$\,pc \citet{Haiman2009}, when the MBHB is in a gaseous circumbinary disk, viscous drag\footnote{Throughout the manuscript we refer to this evolution mechanism as `viscous drag', for simplicity. Note however that angular momentum transport at the inner edge of the circumbinary disk is mostly driven by gravitational torques from the binary and not the viscous torques.}  may significantly contribute or even 
dominate the evolution of the binary. 
\citet{Haiman2009} described how the orbit of a MBHB embedded in a
circumbinary \citet{SS1973} $\alpha$-disk evolves due to viscous drag, and how this evolution depends on the different physical conditions within the disk.

According to \citet{Haiman2009}, there are different regimes for a MBHB in a gap-opened, $\alpha$-disk depending on: (1) whether the radiation pressure or gas pressure balance the vertical gravity ($r^{\rm gas/rad}$); (2) whether the opacity is dominated by electron scattering or free-free absorption ($r^{\rm es/ff}$); (3) whether the binary is massive enough compared to the local disk mass (secondary-dominated or disk-dominated). The characteristic radii are defined as:
\begin{equation}
\label{eq:r_gas_rad}
r^{\rm gas/rad}=0.515\,M_7^{\rm 2/21} 10^{\rm 3} R_{\rm sch}
\end{equation}
and 
\begin{equation}
\label{eq:r_es_ff}
r^{\rm es/ff}=4.10\times 10^{\rm 3} R_{\rm sch}\, ,
\end{equation}
where $M_{\rm 7}$ is the binary mass in units of $10^{\rm 7} \, {\rm M_\odot}$ and $R_{\rm sch} = 2G M_{\rm bin}/c^{\rm 2}$ is the Schwarzschild radius corresponding to the binary mass. 

Based on the region and condition in the disk, it can be divided into three regions \citep{Shapiro1983}: (a) an inner region ($r<r^{\rm gas/rad}$) that is dominated by radiation pressure and electron scattering; (b) an intermediate region ($r^{\rm gas/rad}\leq r<r^{\rm es/ff}$) that is dominated by gas pressure and electron scattering; and (c) an outer region ($r\geq r^{\rm es/ff}$) that is dominated by gas pressure and free-free absorption. 

Within each region, there is the possibility that it is dominated by the secondary MBH ($r<r^{\rm \nu /s}$) or by the accretion disk ($r\geq r^{\rm \nu /s}$). The migration is in general slower when the system is in the `secondary-dominated' regime. The $r^{\rm \nu /s}$ radii in three regions of an $\alpha$-disk are defined in \citet{Haiman2009} as:
\begin{eqnarray}
\label{eq:r_v_s1}
r_{\rm in}^{\rm \nu/s}=3.61\,M_{\rm 7}^{\rm -2/7}\, q_{\rm s}^{\rm 2/7}\, 10^{\rm 3} R_{\rm sch}\; &{\rm if}& r\lesssim r^{\rm gas/rad}\, , \\
\label{eq:r_v_s2}r_{\rm mid}^{\rm \nu/s}=121\,M_{\rm 7}^{\rm -6/7}\,q_{\rm s}^{\rm 5/7}\, 10^{\rm 3} R_{\rm sch}\; &{\rm if}& r^{\rm gas/rad}\lesssim r\lesssim r^{\rm es/ff}\, , \\
\label{eq:r_v_s3}r_{\rm out}^{\rm \nu/s}=182\,M_{\rm 7}^{\rm -24/25}\,q_{\rm s}^{\rm 4/5}\, 10^{\rm 3} R_{\rm sch}\; &{\rm if}& r\gtrsim r^{\rm es/ff}\, .
\end{eqnarray}
Thus, there are six regimes and their orbital frequency evolution rate are listed below.

\noindent (1) Disk-dominated, inner region:
\begin{equation}
\label{eq:gas_drag1} \left ( \frac{{\rm d} f_{\rm orb}}{{\rm d} t} \right)_{\mathrm{VD}}=6.0 \times 10^{-8} M_{\rm 7}^{-2} r_{\rm 3}^{-5} \,\rm yr^{-2}
\end{equation}
if $r_{\rm in}^{\rm \nu/s}<r<r^{\rm gas/rad}$;

\noindent(2) $M_{\rm 2}$-dominated, inner region:
\begin{equation}
\label{eq:gas_drag2}
\left ( \frac{{\rm d} f_{\rm orb}}{{\rm d} t} \right)_{\mathrm{VD}}=2.8 \times 10^{-7} M_{\rm 7}^{-13/8} r_{\rm 3}^{-59/16} q_{\rm s}^{-3/8} \,\rm yr^{-2}
\end{equation}
if $r<r^{\rm gas/rad}$ and $r<r_{\rm in}^{\rm \nu/s}$;

\noindent(3) Disk-dominated, middle region:
\begin{equation}
\label{eq:gas_drag}
\left ( \frac{{\rm d} f_{\rm orb}}{{\rm d} t} \right)_{\mathrm{VD}}=2.9 \times 10^{-5} M_{\rm 7}^{-11/5} r_{\rm 3}^{-29/10} \,\rm yr^{-2}\, ,
\end{equation}
if $r^{\rm gas/rad}<r<r^{\rm es/ff}$ and $r>r_{\rm mid}^{\rm \nu/s}$;
where $f_{\rm orb}$ is the orbital frequency, $r_{\rm 3}$ is the orbital semi-major axis in units of $10^{\rm 3}R_{\rm sch}$, $q_{\rm s}=4q/(1+q)^{\rm 2}$ is the symmetric mass ratio, and $q=M_2/M_1<1$ is the mass ratio \citep{Haiman2009}. Note that this prescription implies that the MBHB orbit always shrinks under the influence of viscous drag, especially in the presence of stellar hardening suggested by some most recent simulations \citep{Cuadra2009, Roedig2012, Bortolas2021, Franchini2021,Amaro2023}

\noindent(4) $M_{\rm 2}$-dominated, middle region:
\begin{equation}
\label{eq:gas_drag4}
\left ( \frac{{\rm d} f_{\rm orb}}{{\rm d} t} \right)_{\mathrm{VD}}=2.3 \times 10^{-6} M_{\rm 7}^{-7/4} r_{\rm 3}^{-19/8} q_{\rm s}^{-3/8} \,\rm yr^{-2}\, ,
\end{equation}
if $r^{\rm gas/rad}<r<r^{\rm es/ff}$ and $r\leq r_{\rm mid}^{\rm \nu/s}$;

\noindent(5) Disk-dominated, outer region: 
\begin{equation}
\label{eq:gas_drag5}
\left ( \frac{{\rm d} f_{\rm orb}}{{\rm d} t} \right)_{\mathrm{VD}}=2.3 \times 10^{-5} M_{\rm 7}^{-11/5} r_{\rm 3}^{-11/4} \,\rm yr^{-2}\, ,
\end{equation}
if $r>r^{\rm es/ff}$ and $r>r_{\rm out}^{\rm \nu/s}$;

\noindent(6) $M_{\rm 2}$-dominated, outer region:
\begin{equation}
\label{eq:gas_drag6}
\left ( \frac{{\rm d} f_{\rm orb}}{{\rm d} t} \right)_{\mathrm{VD}}=1.6 \times 10^{-6} M_{\rm 7}^{-29/17} r_{\rm 3}^{-76/34} q_{\rm s}^{-3/8} \,\rm yr^{-2}\, ,
\end{equation}
if $r>r^{\rm es/ff}$ and $r\leq r_{\rm out}^{\rm \nu/s}$.

The eccentricity evolution due to viscous drag can be complex and cannot be trivially reduced to a prescription for a single dominant regime. \citet{Roedig2011} shows that if the incoming eccentricity
of the MBHB on a prograde orbit is $>0.04$ then there is a limiting
eccentricity in the range $[0.6,0.8]$ that the binary reaches during its
interaction with the circumbinary disk. Thus, if one of our model
MBHBs has a prograde orbit with an eccentricity larger than $0.04$
while viscous drag dominates the evolution, we then randomly assign the eccentricity 
between $0.6$ and $0.8$ after one viscous timescale (measured at
the separation where viscous drag begins to dominate the evolution). If however
the eccentricity of the orbit is less than $0.04$ when viscous drag takes over
the orbital decay, the eccentricity remains fixed until GW emission
takes over the orbital evolution.  

For MBHBs in retrograde orbits, there are three possibilities for the
eccentricity evolution that depend on the value of the eccentricity
when the MBH reaches this stage \citep{Roedig2014}. If the MBHB is in a near circular orbit (i.e.~$e<0.04$), then its eccentricity will not change due to viscous drag. However, if $0.04 \leq e < 0.8$, the eccentricity then increases as $\approx 0.09e-0.0034$ per orbit. Finally, if $e \geq 0.8$, a disk-binary interaction causes the binary to leave the disk plane, tilt, and converge to a prograde orbit with limiting eccentricity in the range of $[0.6,0.8]$. The timescale for this transition corresponds $\sim 10$ viscous timescales to reach the final steady state due to the reversal of the orbital direction from retrograde to prograde
\citep{Roedig2014}. More recent works find qualitatively similar results: that circular binaries remain circular and that eccentric binaries tend to evolve toward a threshold eccentricity, which exact value depends on the thermodynamic properties of the disk and was found to be close to 0.4 by \citet{Dora2021} and \citet{Zrake2021}.

\subsection{The radiation feedback effect in the circumbinary Disk}
\label{subsub:RF_s2}
To model the effect of AGN feedback on the dynamical evolution of MBHB embedded in a sub-grid circumbinary disk \citep{volonteri&delvalle2018}, we set a free parameter $\eta_{\rm 2} = 10^{\rm -2}$. If the total accretion rate of the MBHB (see Sect.~\ref{sec:accretion}) is larger than $\eta_{\rm 2} \dot{M}_{\rm Edd}$, we exclude the contribution from the circumbinary disk in shrinking the orbit and evolving the orbital eccentricity. 

\subsection{Gravitational waves}
\label{sub:GW}
The last stage of orbital decay is dominated by GW emission, described following \citet{Peters1964} 

\begin{eqnarray}
  \label{eq:GW1}
\left(\frac{{\rm d}f_{\rm orb}}{{\rm d}t}\right )_{\mathrm{GW}}=\frac{96\, (2\pi)^{\rm 8/3}}{5c^{\rm 5}}(GM_{\rm chirp})^{\rm 5/3}f_{\rm orb}^{\rm 11/3}\,{\cal F}(e),
\end{eqnarray}

and

\begin{eqnarray}
  \label{eq:GW2}
\left(\frac{{\rm d}e}{{\rm d}t}\right )_{\mathrm{GW}}=\frac{(2\pi)^{\rm 8/3}}{15c^{\rm 5}}(GM_{\rm chirp})^{\rm 5/3}f_{\rm orb}^{\rm 8/3}\,{\cal G}(e),
\end{eqnarray}

where $M_{\rm chirp}=(M_{\rm 1}M_{\rm 2})^{\rm 3/5} /(M_{\rm 1}+M_{\rm 2})^{\rm 1/5}$
is the source frame chirp mass and the factors $\mathcal{F}$ and $\mathcal{G}$ are

\begin{eqnarray}
  \label{eq:fe}
{\cal F}(e)=\frac{1+73/24e^{\rm 2}+37/96e^{\rm 4}}{(1-e^{\rm 2})^{\rm
    7/2}}
\end{eqnarray}

and

\begin{eqnarray}
  \label{eq:ge}
{\cal G}(e)=\frac{304e+121e^{\rm 3}}{(1-e^{\rm 2})^{\rm 5/2}}.
\end{eqnarray}

\subsection{Accretion in RAMCOAL}
\label{sec:accretion}
In RAMSES, once the two MBHs reach a separation smaller than $4\Delta x$, a `numerical' merger happens: the two MBHs become one bigger MBH positioned at the center of mass of the MBHB and from there on treated as one MBH in the simulation represented by a sink particle. The accretion rate of this numerically-merged MBH is:

\begin{equation}
\label{eq:BHL}
\dot{M}_{\rm BHL, bin}=\frac{4\pi G^2 M_{\rm bin}^2 \bar{\rho}_{\rm gas, sp}}{(\bar{c}_{\rm s}^2 + \bar v_{\rm rel, CoM}^2)^{\rm 3/2}},
\end{equation}

where $v_{\rm rel, CoM}$ is the relative speed between the gas and the center of mass of the MBH pair, which is also the position of the numerically merged sink by definition.
In RAMCOAL, the evolution of the MBHB is split into two stages according to whether or not the DF dominates the decay. The stage 1 is from when the separation of two MBHs reaches $4\Delta x$ to the point when the MBHs become gravitationally bounded. In stage 1, where DF dominates the orbital decay, we assume that both MBHs accrete at the Bondi-Hoyle-Lyttleton rate, and they add up to the accretion rate of the numerically-merged MBHB (i.e.~equation~\ref{eq:BHL}):
\begin{equation}
\label{eq:BHL_ramcoal}
\dot{M}_{j,\rm  stage 1}=f_{\rm norm} \frac{4\pi G^2 M_{j}^2  \hat \rho_{\rm gas}(r_{j})}{(\bar{c}_{\rm s}^2 + \bar v_{{\rm rel}, j}^2)^{\rm 3/2}}\, , j=1, 2
\end{equation}
where $f_{\rm norm}$ is a normalisation factor so that $\dot{M}_{1, \rm stage 1}+\dot{M}_{2, \rm stage 1} = \dot{M}_{\rm BHL, bin}$, and $\hat \rho_{\rm gas}(r_{j})$ is defined in Eq.~\ref{eq:subgrid_rho_gas}.

The stage 2 is from when the MBHB becomes gravitationally bounded to the final coalescence of the MBHB defined at the innermost stable orbit ($\sim 6 R_{\rm sch}$). At stage 2, the stellar scattering, viscous drag from the circumbinary disk, and the GW emission dominates over the DF in shrinking the binary. At stage 2 we adopt the preferential accretion model from \citet{Duffell2020} which is fitted from simulations of a broad range of mass ratios:

\begin{equation}
\frac{\dot{M}_{\rm 2, stage 2}}{\dot{M}_{\rm 1, stage 2}}=\frac{1}{0.1 + 0.9q} \, ,
\end{equation}
where we still impose $\dot{M}_{1, \rm stage 2}+\dot{M}_{2, \rm stage 2} = \dot{M}_{\rm BHL, bin}$. It is immediate to see that accretion during stage 2 will lead to an increase of the mass ratio $q$.

The feedback effect on accretion at stage 1 is taken care of by continuously triggering the gas sub-grid density profile when the mean gas density is high and de-triggering when it is low, hence mimicking the feedback regulation of gas reservoir around the MBHs. At stage 2, the dynamical contribution and accretion from the circumbinary disk is set to be zero \citep{volonteri&delvalle2018} when the total accretion rate of the MBHB is larger than $\eta_{2} \dot{M}_{\rm Edd}$. In this work, we use $\eta_{2}=0.01$. We tested that the mass and dynamical evolution of MBHs are not very sensitive to the choice of $\eta_{2}$, and defer the detailed study of its effect to future works. 

\section{The RAMSES code}
\label{sec:ramses}
RAMSES is an adaptive mesh refinement code \citep{Teyssier2002} which solves the Euler equations using the second order MUSCL-Hancock scheme. The gas is set to be composed of monoatomic particles with adiabatic index $\gamma = 5/3$. Collisionless particles including DM, stellar and BH particles are evolved using a particle-mesh solver for gravity with a cloud-in-cell interpolation. The size of which is that of the local cell for BHs and stars ($\Delta x$). The cloud-in-cell interpolation for DM particles is larger than that of BHs and stars due to their larger particle masses in order to smooth their contribution to numerical shot noise. This usage of cloud-in-cell interpolation solves the problem of BHs scattering off heavy DM particles as the DM distribution is smoothed in space.

When temperature is higher than $10^4\,\rm K$, gas is cooled by the hydrogen, helium, and metals following cooling curves from \citet{Sutherland1993}. When the gas temperature is below $10^4\,\rm K$ and above the minimum temperature of $10\,\rm K$, we implement the fitting functions from \citet{Rosen1995}. Heating of the gas comes from a uniform UV background at redshift 0 following \citet{haardt&madau96}. 

Star formation is triggered in regions where the gas number density is larger than $n_{0}$, in a Poisson random process \citep{rasera&teyssier06, dubois&teyssier08winds} following the Schmidt relation with a star formation efficiency $\varepsilon_\star$ depending on the local gravo-turbulent properties of the gas (see \citealp{Dubois2021} for details). Since the star formation model requires drawing a sequence of pseudo-random numbers, we used different values of the initial random seed number to estimate the numerical variance of the results. 

The implementation of feedback from supernovae is as described in \citet{Kimm&cen2014}. Immediately after star particles becomes older than $5\,\rm Myr$, it is assumed supernovae release a specific energy of $\eta_{\rm SN} \times 10^{\rm 50}\, {\rm erg}\, {\rm M_\odot}^{-1}$ for a given star particle mass, where $\eta_{\rm SN} = 0.2$. The amount of energy and momentum deposited into gas depends on the local density and metallicity, in order to capture either the Sedov or the snow-plough expansion phase of the explosion. Additionally we impose that the normalisation of the star formation efficiency increases linearly from $0$ to its canonical value within a $100\,{\rm Myr}$ timescale, to prevent the galaxy from exploding due to the initial peak of star formation.

The accretion of MBHs in RAMSES is as described in \citet{Dubois2012}, where Eddington-limited BHL accretion rate is used. The AGN feedback is bi-model, including a kinetic mode, which is active at a low accretion rate ($f_{\rm Edd} \leq 0.01$) in the form of a bipolar jet, where all the energy and momentum is ejected at $10^4\, \rm km\,s^{-1}$  into a cylindrical region of a height that is twice the radius which size is chosen to be equal to $\Delta x$ in this work. The other mode of AGN feedback is active at high accretion rate ($f_{\rm Edd}>0.01$), this thermal feedback is in the form of disk winds and radiation, releasing $15\%$ of the bolometric luminosity (with a radiation efficiency $\varepsilon_{\rm r}=0.1$) as thermal energy into the surrounding gas reservoir within a sphere of radius $r_{\rm thm}$ which we also set to be $\Delta x$ in this work. This accretion and feedback occur on the sink particle, and mass is then distributed between the two MBHs as described in Eq.~\ref{eq:BHL_ramcoal}.

Before the separation between two MBHs reaches the resolution limit, the gas DF is exerted on the MBH as $F_{\rm DF}= f_{\rm gas} 4 \pi \alpha \bar{\rho}_{\rm gas} (G M_{\rm BH}/\bar {c}_{\rm s})^2$, where $\bar{\rho}_{\rm gas}$ is the mean gas density within a sphere of radius $4 \, \Delta x$ and $f_{\rm gas}$ is a function of the mach number that follows the prediction of \citet{O1999}. $f_{\rm gas}$ is in a range between 0 and 2 for an assumed Coulomb logarithm of 3  \citep{chaponetal13}. In these simulations, the gas DF is not boosted (contrary to e.g.~\citealp{Dubois2014a}) as suggested by detailed turbulent box simulations~\citep{Lescaudron23}. Also the gas DF in RAMCOAL takes into account the contribution of the spiral density wake due to the orbital motion of MBHs in the binary.

Above the simulation resolution limit, the implementation of DF from collisionless particles (i.e.~stars and DM) is described in Sect.~2 of \citet{Hugo2019}. Not only the contribution to DF from slow moving particles but also that from fast moving particles is taken into account \citep{AM2012, DA2017}. All of the quantities needed to estimate collisionless particles DF are measured in a sphere $S$ with a radius $4 \Delta x$ centered on the MBH, where $\Delta x$ is the minimum grid size and also the resolution limit, consistent with the implementation of gas DF. The implementation of DF from collisionless particles for the shrinking of the pair (RAMCOAL part) is treated the same way as for the dynamical friction applying to the center of mass of the pair (RAMSES part). The difference is, in RAMSES, the DF due to each particle is resolved, and velocity distributions are estimated explicitly on the grid:

\begin{equation}
\label{eq:fv}
4\pi v^2 f(v)= \frac{3}{256\pi \Delta x^3}\sum_{i \in S } m_{i} \delta (v_{i}-v)\, ,
\end{equation}

where $m_{i}$ is mass of particle $i$ within the sphere $S$, $v_{i}$ is the particle velocity, and $\delta$ is the Dirac function. The DF due to star and DM particles is calculated through the same scheme but separately due to the difference in particle mass and velocity distribution of star and DM particles.

For the collisionless particle DF above the resolution limit, the Coulomb logarithm is $\ln(4\Delta x/r_{\rm def})$ where $r_{\rm def}$ is the minimum impact parameter for an orbital deflection of $90$ degrees to happen, which can be approximated by the influence radius $r_{\rm inf} = GM_{\rm BH}/v_{\rm BH} ^2$ \citep{Hugo2019}. The sub-grid DF is set to zero when $4\Delta x \leq r_{\rm def}$ as the force is accounted for directly \citep{Beckmann2018}. 

To summarize, the implementation of gas DF in RAMCOAL and RAMSES is generally the same. The mean gas density used in the DF calculation is determined in the same way in both RAMCOAL and RAMSES. However, the gas DF in RAMCOAL is more advanced, as it accounts for the spiral shape of the gaseous wake. As for the collisionless particle DF, the implementation is identical in both RAMCOAL and RAMSES. The difference lies in the treatment: RAMSES resolves the mass, relative velocity, and DF due to each individual stellar and dark matter particle, while RAMCOAL uses the mean density and assumes a Maxwellian velocity distribution for particles to calculate DF due to stars and dark matter.

\section{Tests of the RAMCOAL implementation in an isolated galaxy with two massive black holes}
\label{sec:robust}
In order to test the resolution convergence  in RAMCOAL, we run RAMCOAL on a pair of MBHs in an isolated galaxy embedded in its DM halo with different spatial resolutions.

\subsection{Initial conditions and refinement strategy}
\label{sec:IC}
The galaxy is a composite system of DM, stars, a central MBH, and gas, which is initialised with the MAKEDISK initial conditions generator~\citep{Springel05} adapted for RAMSES simulations~\citep[see][]{Rosdahl17}. The DM halo follows a spherical Navarro-Frenk-White \citep{NFW1997} density profile with a total virial mass $M_{\rm vir}=1.1\times 10^{\rm 11} \, {\rm M_\odot}$, a concentration parameter of $10$, and a virial radius of $95$\,kpc at redshift zero. The number of DM particles is $10^{\rm 6}$ leading to a DM particle mass of $m_{\rm DM} = 1.1 \times 10^{5} \, {\rm M_\odot}$.

The baryonic component of the galaxy contains a stellar bulge and a gas and stellar disk. The bulge has a total mass of $3.3\times 10^{\rm 9}\, {\rm M_\odot}$, and follows a spherical Hernquist density profile \citep{Hernquist1990} with a scale radius equals to $293$\,pc ($10\%$ of the disk scale radius).
The disk is described by an exponential surface density profile in cylindrical radius with a scale radius of $2.93$\,kpc, which is sharply cut at a cylindrical radius that is $5$ times larger. The profile is proportional to ${\rm sech}^2(z/(2H_z))$ in the vertical $z$-direction, where the scale height $H_z=293$\,pc, and the cutoff height is $5$ times larger. The total gas mass is $1.1\times 10^{\rm 9}\, {\rm M_\odot}$ which is $14.3\%$ of the total disk mass $7.78\times 10^{\rm 9}\, {\rm M_\odot}$, with a disk stellar mass of $6.67\times 10^{\rm 9}\, {\rm M_\odot}$. 
 The total number of particles in the bulge and in the disk are respectively  $5\times 10^{\rm 5}$, and $10^{\rm 6}$ leading to a star particle mass of $m_{\rm \star,d} = 6.7 \times 10^{3} \, {\rm M_\odot}$.

The galaxy is set at the center of a box with size of $100 \,\rm kpc$. 
The coarse grid is made of $128^{\rm 3}$ cells.
The grid is continuously refined or de-refined according to a pseudo-Lagrangian refinement criterion, with mass resolution of $8\times 10^4\,{\rm M_\odot}$, down to a minimum cell size of $\Delta x\simeq 10$, $50$, or $100\,{\rm pc}$ (with corresponding levels of refinement of $\ell_{\rm max}=13$, $11$, and $10$ respectively).
In addition to the criterion on mass resolution, we also refined in the dense interstellar medium ($n>5 \rm \,H\,cm^{-3}$) on the Jeans length $\lambda_{\rm J}$ when the cell size is larger than $\lambda_{\rm J}/4$.
The galaxy is relaxed for $100$\,Myr without MBHs at first. A pair of MBHs with the primary mass of $M_{\rm 1} = 3.33\times 10^{\rm 6}\, {\rm M_\odot}$ and the secondary mass of $M_{\rm 2} = 1.67\times 10^{\rm 6} \, {\rm M_\odot}$ are put into the relaxed galaxy. The primary is put at the center of the galaxy, and the secondary is in the disk mid-plane and $500$\,pc away from the primary with an initial tangential velocity of $10\,\rm km\,s^{-1}$ in corotation with the galaxy disk (the circular velocity in the galaxy at $500\,{\rm pc}$ is $\sim 130\,\rm km\,s^{-1}$).  

\subsection{Semi-analytic model}

To show the robustness of the predictions of MBHB evolution with RAMCOAL, we also compare to estimates from the semi-analytical model (SAM) of \citet{LBB20a,LBB2022}. We briefly summarize the SAM here.

In the SAM, we assume a galaxy merger produces a single remnant, with a DM halo, a stellar bulge and a gas disk\footnote{We omit the stellar disk since its impact on the orbital evolution from DF is negligible \citep{LBB20a}.}, which includes the MBH pair. The MBHs are set up in the same way as in the RAMCOAL simulation. The galaxy remnant density profile used in the SAM intends to be as close to that of the simulated galaxy as possible for the purpose of comparison. The DM halo and the non-rotating stellar bulge both follow a cored power-law density profile \citep[e.g.,][]{BT1987}:
\begin{equation} 
\rho_{i}(r) = 
  \rho_{i,0} \left (\frac{\max[r,10{\rm pc}]}{R_{0}}\right )^{-\alpha_{i}}\, ,
\label{eq:SAM_bulge}
\end{equation}
where $i={\rm b}$ or $\rm DM$ stand for the bulge and DM component respectively.

The gas surface density profile is fitted by the six-parameter core-Sersic profile \citep{Graham2003,Trujillo2004}

\begin{equation}
\label{eq:SAM_disk}
\Sigma_{\rm SAM}(r) = \Sigma_{\rm 0} \left [1+\left ( \frac{r_{\rm br}}{r}\right )^{\rm \alpha}\right ]^{\frac{\gamma}{\alpha}} {\rm exp} \left[ -b\left(\frac{r^{\rm \alpha}+r_{\rm br}^{\rm \alpha}}{r_{\rm e}^{\rm \alpha}}\right)^{\frac{1}{\alpha n}}\right]\, ,
\end{equation}

where $\Sigma_{\rm 0}$ is the normalisation surface density, $r_{\rm b}$ is the break disk radius at which separates the inner power law with a logarithm slope  of $\gamma$ and the outer S\'ersic profile with the index $n$ and the effective radius $r_{\rm e}$, while $\alpha$ controls how sharp the break at $r_{\rm br}$ is. 

\begin{figure}[t]
    \includegraphics[width=0.5\textwidth]{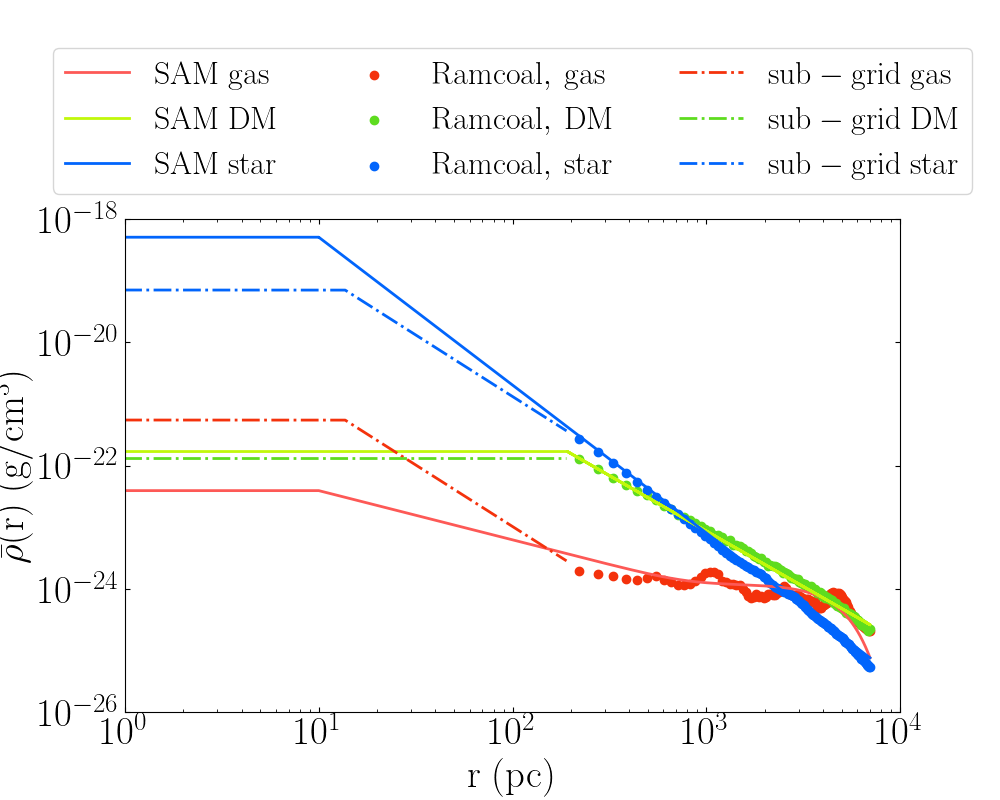}
\caption{The mean density profile of the isolated galaxy in RAMSES, the sub-grid density profile in RAMCOAL, and the fitted density profile used in the semi-analytic model. }
\label{fig:SAM_IC}
\end{figure}

To ensure the SAM galactic density profile can represent the RAMSES galaxy when the MBH pair starts evolving in the simulation, we perform a shrinking-sphere average to get the mean density profile of DM, star, and a shrinking-disk average to get the mean density profile of gas in the RAMSES galaxy at the output when two MBHs are added. We fit equations~\ref{eq:SAM_bulge} and \ref{eq:SAM_disk} to the density profiles thus obtained to calibrate the SAM density profile. Figure~\ref{fig:SAM_IC} illustrates the mean density profile of each component of the RAMSES galaxy and the corresponding fitted SAM density profile. The fitted parameters are: $R_{\rm 0} = 293\,{\rm pc}$, $\rho_{\rm b, 0} = 1.5\times 10^{\rm -22}\,{\rm g\,cm^{-3}}$, $\rho_{\rm DM, 0} = 7.9\times 10^{\rm -23}\,{\rm g\,cm^{-3}}$, $\alpha_{\rm b} = 2.4$, and $\alpha_{\rm DM} = 1.8$. A $10\,{\rm pc}$ core is included for the SAM stellar density profile, and a $200$\,pc core is included for the DM density profile to avoid infinite density at the center. The different core sizes are chosen to match the sub-grid model in RAMCOAL. In RAMCOAL, we do not assume any sub-grid density profile for DM; instead, we use the mean DM density within the resolution sphere ($R_{\rm res} = 200\,{\rm pc}$) for the calculation of DF. To ensure compatibility, the DM density profile in SAM is given a $200$\,pc core.
The following parameter values are used in equation~\ref{eq:SAM_disk} to fit the mean gas density (averaged using shrinking disks) in RAMSES outside the resolution sphere: $\Sigma_{\rm 0} = 2.14\times 10^{\rm -3} \, {\rm g\,cm^{-2}}$, $r_{\rm br}=800\, {\rm pc}$, $\alpha =4$, $\gamma =0.8$, $n=3$, and $r_{\rm e} = 5\, {\rm kpc}$. The scale height of the disk is $293\, {\rm pc}$.

The sub-grid density profiles of gas and stars used in the SAM and in RAMCOAL (equation~\ref{eq:subgrid_rho_gas}) within the resolution sphere are plotted in Fig.~\ref{fig:SAM_IC} for the simulation with resolution $50\,{\rm pc}$. As shown in the plot, the stellar density dominates over that of DM and gas within the resolution sphere of radius $4\Delta x=200 \,{\rm pc}$. The SAM stellar density is slightly higher than the sub-grid stellar density within the resolution sphere, due to the minor difference in core radius of sub-grid and SAM density profile. The core radius of the sub-grid gas density profile is calculated using the total gas mass within the resolution sphere when the star-formation criteria is satisfied, which is $13.7$\,pc in this case. Note the decay time from the SAM can change accordingly with the choice of core radius of the density profile.

In the SAM, the orbital evolution of the secondary MBH due to gas, star, and DM DF is tracked until the influence
radius of the MBHB, at which point the orbital decay
is dominated by loss-cone scattering, viscous drag from a
circumbinary disk, and GW emission. The calculation ends when the separation is smaller than the innermost stable circular orbits of the two MBHs. 

The calculation of the orbital decay due to DF in the SAM is
described in detail by \citet{LBB20a, LBB20b}. The DF force exerted by star and DM is calculated using equations~(5)-(7) in \citet{LBB20a}, following the work of \citet{AM2012}. The velocity distribution of collisionless particles is assumed to be Maxwellian (see \citealp{LBB20b}, equation~2) with the same velocity dispersion as in RAMCOAL. Since gaseous
DF depends on the Mach number of the moving body \citep[e.g.,][]{KK2007}, the sound speed of the gas must be defined. The temperature of the gaseous disk is uniformly taken to be $10^4\, {\rm K}$ which is above the minimum temperature required by the Toomre stability criterion \citep{T1964}. The gaseous DF force on the secondary MBH is then computed using equations~10--12 of \citet{LBB20a}, which results in a gaseous DF force that is strongest when the velocity difference between the secondary MBH and gas disk is close to the sound speed \citep{O1999,KK2007}.

The hardening of MBHB orbits by LC scattering uses the same  equations as in RAMCOAL (i.e.~equations~\ref{eq:LC1} and \ref{eq:LC2}). The viscous drag from a
circumbinary disk and GW emission is also calculated using the same method as in RAMCOAL (i.e.~equations~\ref{eq:r_gas_rad}--\ref{eq:gas_drag6} and equations~\ref{eq:GW1}--\ref{eq:ge}, respectively). 

\subsection{Resolution tests}
\label{sec:resolution}
\begin{figure*}[t]
  \centering
  \begin{tabular}{@{}cc@{}}
    \includegraphics[width=0.9\textwidth]{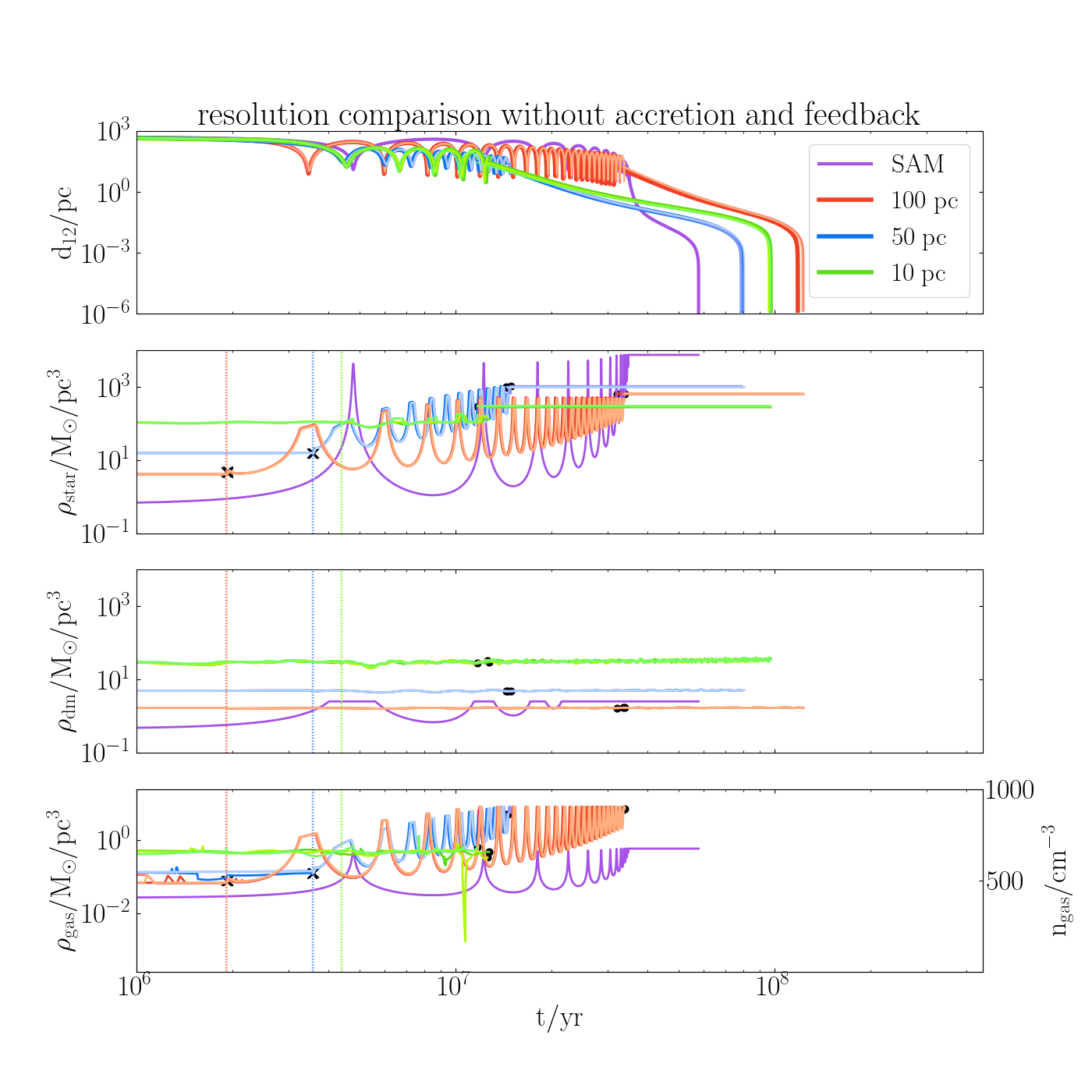}
  \end{tabular}
\caption{The MBHB orbital decay results of the single isolated disc setup simulated using RAMCOAL at $10$, $50$, and $100$\,pc resolution together with the semi-analytic counterpart without accretion and feedback (with colors as indicated in the top panel). Three runs with different star-formation random seeds in the galaxy are performed at each resolution to gauge the effect of stochasticity in the galactic environment on the MBHB orbital decay. The top panel shows the MBH separation time evolution. The second, third, and fourth panels show the stellar, DM, and gas density, respectively, used to calculate the DF on the secondary MBH. The vertical dotted lines in each panel indicate when the MBHB evolution is handed over to RAMCOAL from RAMSES at the boundary of the resolution sphere. The black crosses indicate the triggering of sub-grid density profiles. The black dots indicate the transition from stage 1 to stage 2 in RAMCOAL.}
\label{fig:iso_noa}
\end{figure*}

\begin{figure*}[t]
  \centering
  \begin{tabular}{@{}cc@{}}
    \includegraphics[width=0.9\textwidth]{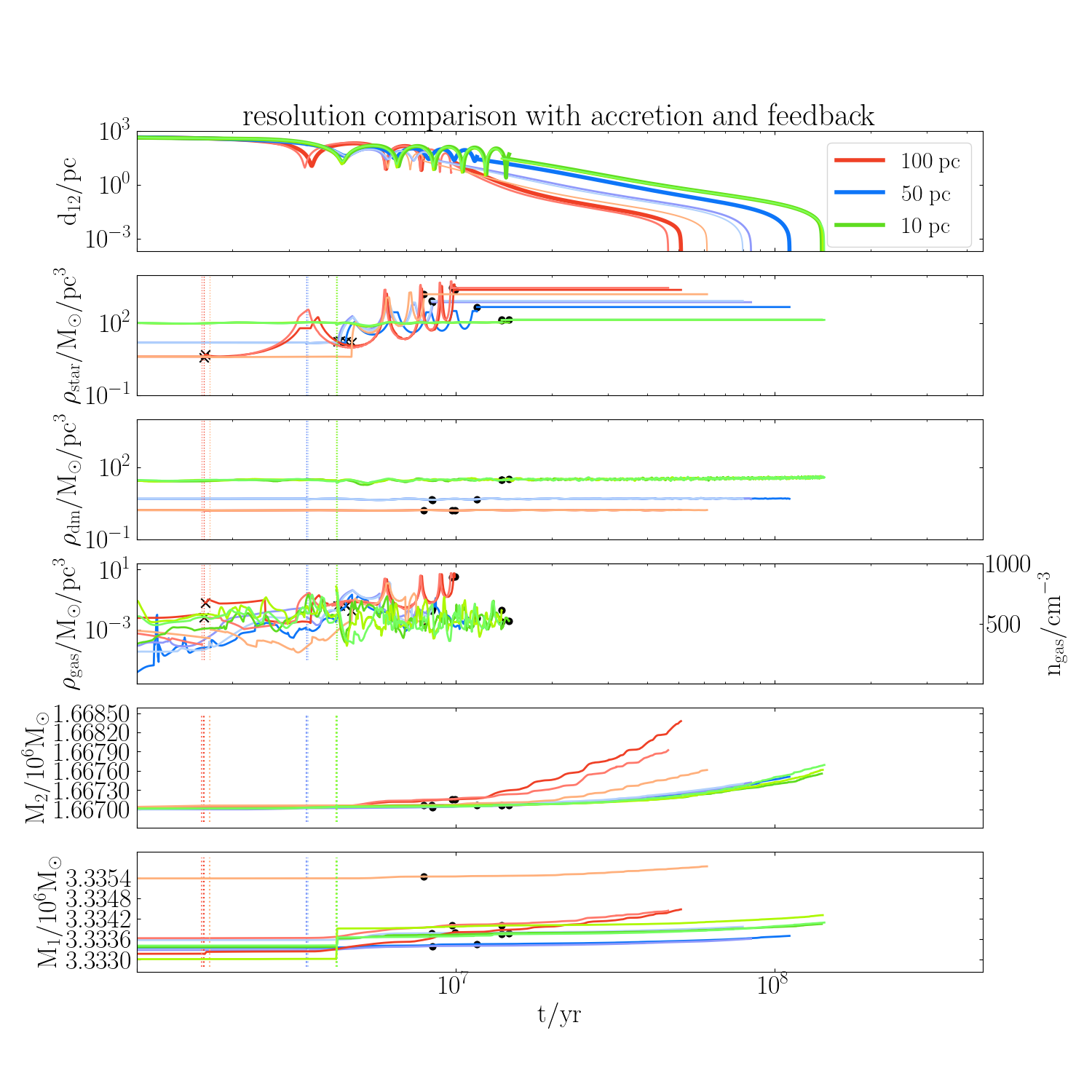}
  \end{tabular}
\caption{The same as Fig.~\ref{fig:iso_noa}, but for the counterpart results with accretion and feedback. The top panel shows the MBH separation time evolution. The second, third, and fourth panels show the stellar, DM, and gas density, respectively, used to calculate the DF on the secondary MBH. The black crosses indicate the triggering of sub-grid density profiles. The fifth and sixth panels show the mass growth of the primary and secondary MBHs, respectively. The black dots indicate the transition from stage 1 to stage 2 in RAMCOAL. }
\label{fig:iso_acc}
\end{figure*}

Figure~\ref{fig:iso_noa} illustrates the MBHB orbital decay simulated using RAMCOAL with $10$, $50$, and $100$\,pc resolution together with the SAM counterpart serving as the lower boundary in the coalescence time. For each resolution, three runs with different star-formation random seed in the galaxy are performed to demonstrate the effect of randomness in the galactic environment realization on the MBHB coalescence. 

The top panel shows the time evolution of the separation between two MBHs from the initial $500$\,pc separation to the final coalescence. The coalescence time of all three resolution runs are very close (within a factor of $2$), and there is no trend in coalescence time with resolution: since it decreases from $100$\,pc to $50$\,pc resolution, and it increases from $50$\,pc to $10$\,pc resolution. The star-formation random seed (see Sect.~\ref{sec:ramses}), which we varied to sample the numerical uncertainty of the result, does not lead to a big difference due to the short simulation time, within which nearly no star formation happened. Figure~\ref{fig:iso_noa} also shows the star, DM, and gas density used in the calculation of DF on the secondary MBH in the coalescence process. 

The vertical lines indicate the entry of the MBHB into the $4\Delta x$ resolution sphere, marking the start of the new sub-grid model for MBHB dynamical evolution. We define this evolutionary process in two stages: Stage 1 begins when the MBHB enters the resolution sphere and continues until the two MBHs become gravitationally bound. Stage 2 covers the period from when they become gravitationally bound until their final coalescence. 

The oscillatory pattern in density at stage 1 is due to the orbit of the secondary MBH in the sub-grid density profile, the triggering of which is indicated by the black dots in stellar and gaseous density panels. There is no oscillations in the stellar and gaseous density at $10$\,pc resolution, because the sub-grid density profile criteria is not satisfied at this resolution. Since $10$\,pc resolution is already high enough to resolve the inner sub-structure, there is no need for compensating with sub-grid density profiles. This however means that the density used in the calculation is the mean density within the resolution sphere, rather than a profile with a core and a slope, therefore oscillations within the resolution sphere do not reflect in an oscillating density. Following the oscillatory structure in density at stage 1, the plateau in density represents the influence stellar density used in calculation of loss-cone scattering at stage 2.
The stellar density dominates over the other two components in the entire coalescence process, and the slightly assumed higher central density in the SAM (cf. Fig.~\ref{fig:SAM_IC} and related text) explains the shorter coalescence time. As mentioned before, there is no sub-grid density profile for DM in RAMCOAL, which is represented by almost flat lines in the third panel.

With MBH accretion and feedback activated in both the large-scale RAMSES and sub-grid RAMCOAL simulations, the results are presented in Fig.~\ref{fig:iso_acc}. There is a greater variation in coalescence times across different resolutions. This increased variation is primarily due to differences in MBH mass growth and sub-grid stellar density, both of which are influenced by AGN feedback efficiency, which in turn depends on resolution.

As shown in the fourth and fifth panels of Fig.~\ref{fig:iso_acc}, the sub-grid gas density profile in the 100\,pc resolution runs starts to form (indicated by the density rising and the appearance of a zig-zag pattern). However, this formation is not triggered in the higher-resolution runs, where the zig-zag patterns reflect the average gas density within the resolution sphere, which moves with the center of mass of the MBHB in the galaxy. In the higher-resolution runs, feedback is more effective at expelling gas, leading to significantly lower sub-grid gas and stellar densities. This, in turn, results in less mass growth and lower mass ratios in the 50\,pc and 10\,pc resolution runs compared to the 100\,pc runs. The 100 pc runs exhibit shorter coalescence times due to the less efficient AGN feedback. As shown in \citet{Negri2017}, feedback efficiency varies with resolution, with lower resolutions requiring higher efficiency. This is because, at lower resolutions, the number of gas elements affected by the energy injected from BH feedback is fixed, but the mass of gas to be heated and swept up is larger, making the heating process less effective. In this plot, all resolution runs use the same feedback efficiency, resulting in lower resolution runs being less affected by AGN feedback compared to higher resolution runs.
We will revisit the issue of resolution-dependent feedback efficiency in the next section.

\section{Massive black holes coalescence in merging galaxies}
\label{sec:merger}
Through the resolution test in the previous section, we have proven the robustness of RAMCOAL. In this section, we simulate the MBHB coalescence in a more realistic environment: galaxy mergers.  
\subsection{Numerical setup}
To set up the initial conditions for the galaxy merger simulations, we first generate two individual disk galaxies and their host halos using the method described in Sect.~\ref{sec:IC}. Each galaxy (and their host halo), containing a central MBH at its center, is then placed into a co-planar orbit with an initial separation of $1.1$ times the sum of the cut-off radii of the two galaxies. The primary more massive galaxy is positioned at the center of the simulation box with zero initial velocity, while the secondary galaxy is given a prograde tangential velocity equal to two-thirds of the circular velocity at its position.

A list of the merger parameters is provided in Table~\ref{tab:params}. In Sects.~\ref{sub:100} and \ref{sub:all}, simulations are performed using 1:2 and 1:10 mass ratios ($q=0.5$ and $q=0.1$), where the mass ratio scales the halo, the galaxy, and the MBH masses.
For instance, for the so-called ``M1q0.5fg0.1'' merger simulation, there is a mass ratio $q=0.5$, $M_{\rm 1} = 3.33\times 10^{\rm 6}\, {\rm M_\odot}$, $M_{\rm 2} = 1.67\times 10^{\rm 6} \, {\rm M_\odot}$, $M_{\rm vir,1} = 1.1\times 10^{11}\, {\rm M_\odot}$, $M_{\rm vir,2} = 5.5 \times 10^{\rm 10} \, {\rm M_\odot}$ (with galaxy masses changed accordingly, as well as halo and galaxy sizes), and a galaxy gas-to-baryon mass fraction of $f_{\rm g}=0.1$, as outlined in Table~\ref{tab:params}. The initial tangential velocity of the secondary galaxy is $120\; {\rm km/s}$ in simulation M1q0.5fg0.1.
In Sect.~\ref{subsec:8pairs}, galaxy and MBH mass ratios, and gas fractions are varied to test the effect of changing these parameters on coalescence times.

\begin{figure*}[t]
           \includegraphics[width=0.33\textwidth , trim={0 0cm  0 0cm},clip]{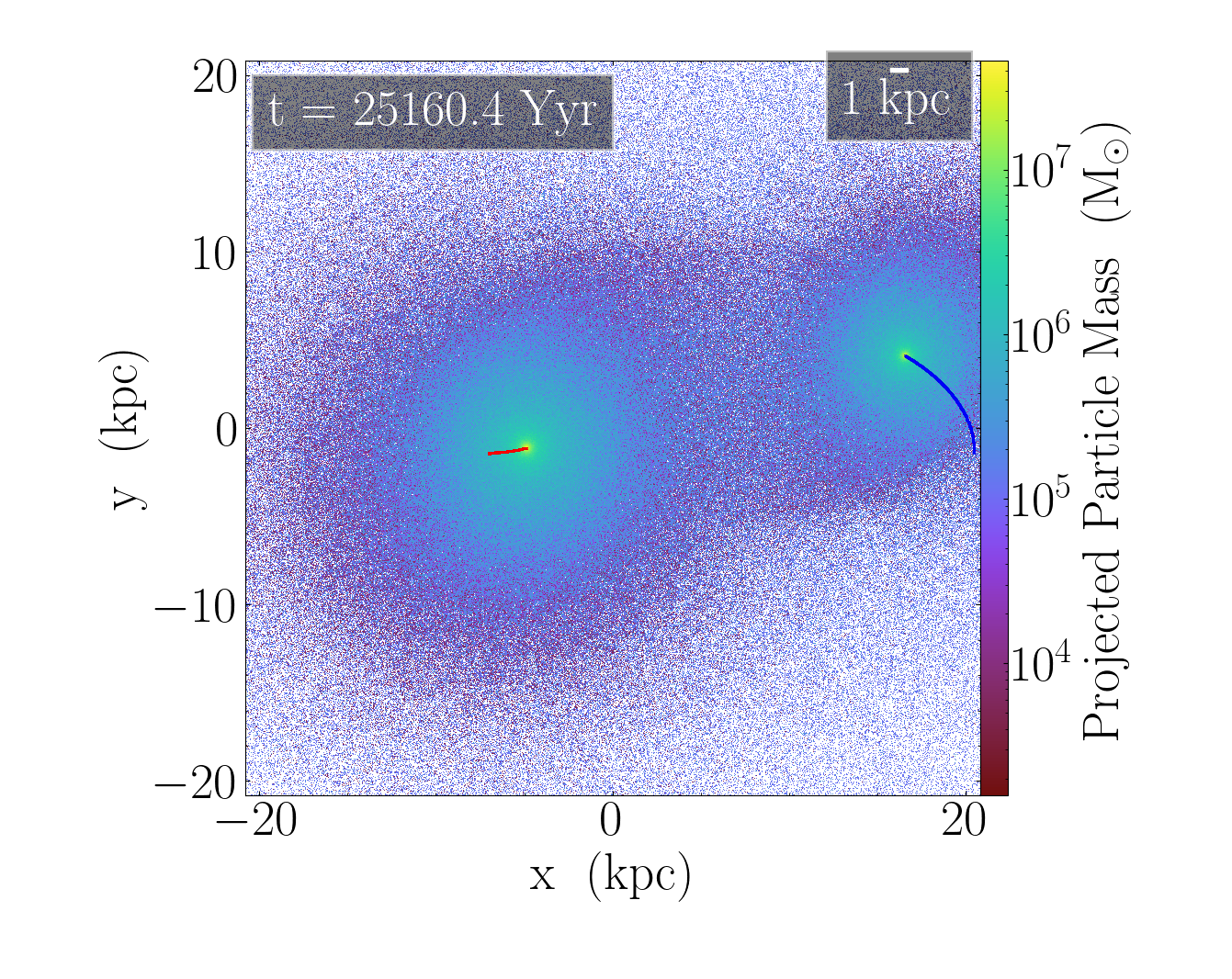}
           \includegraphics[width=0.33\textwidth , trim={0 0cm  0 0cm},clip]{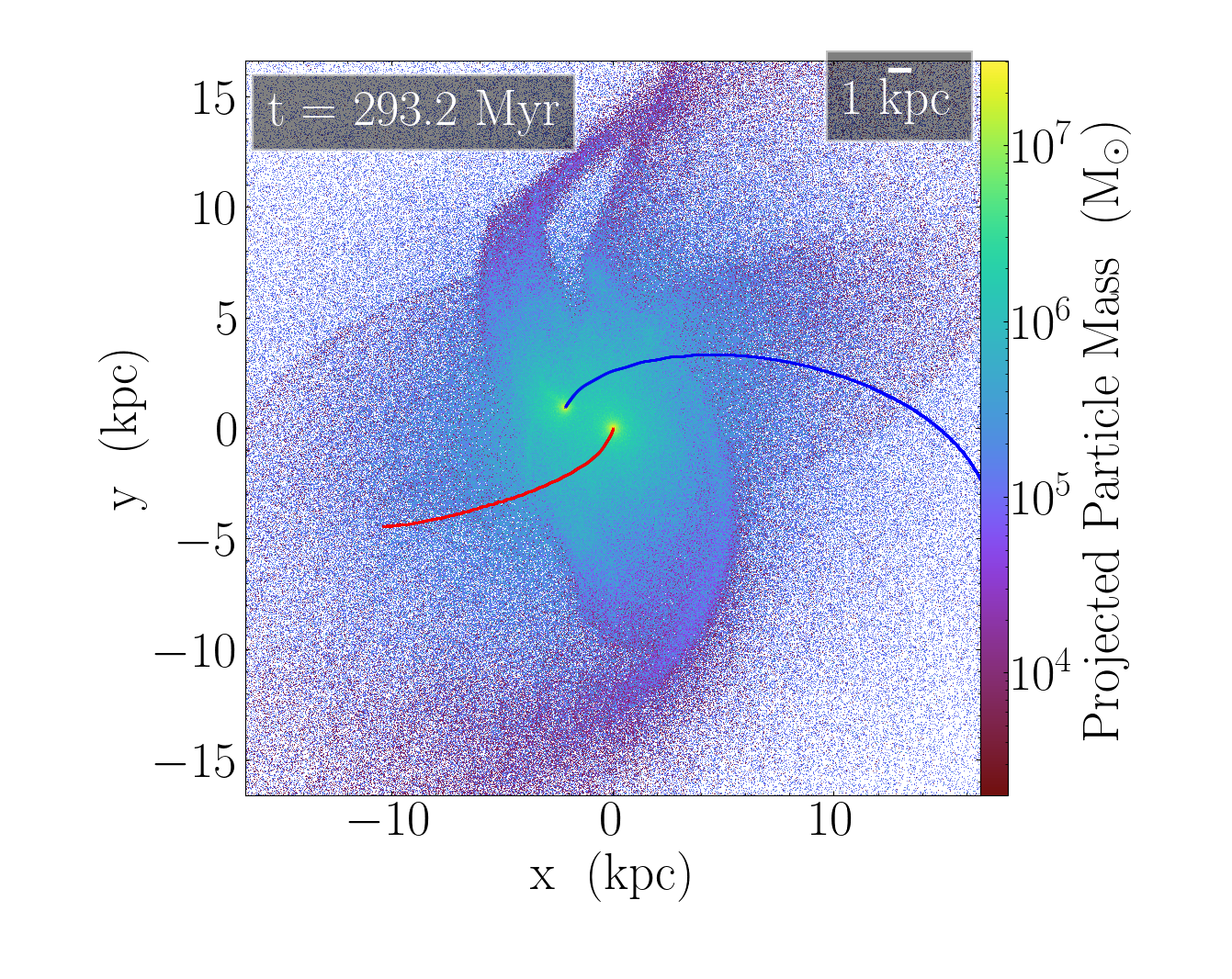}
           \includegraphics[width=0.33\textwidth , trim={0 0cm  0 0cm},clip]{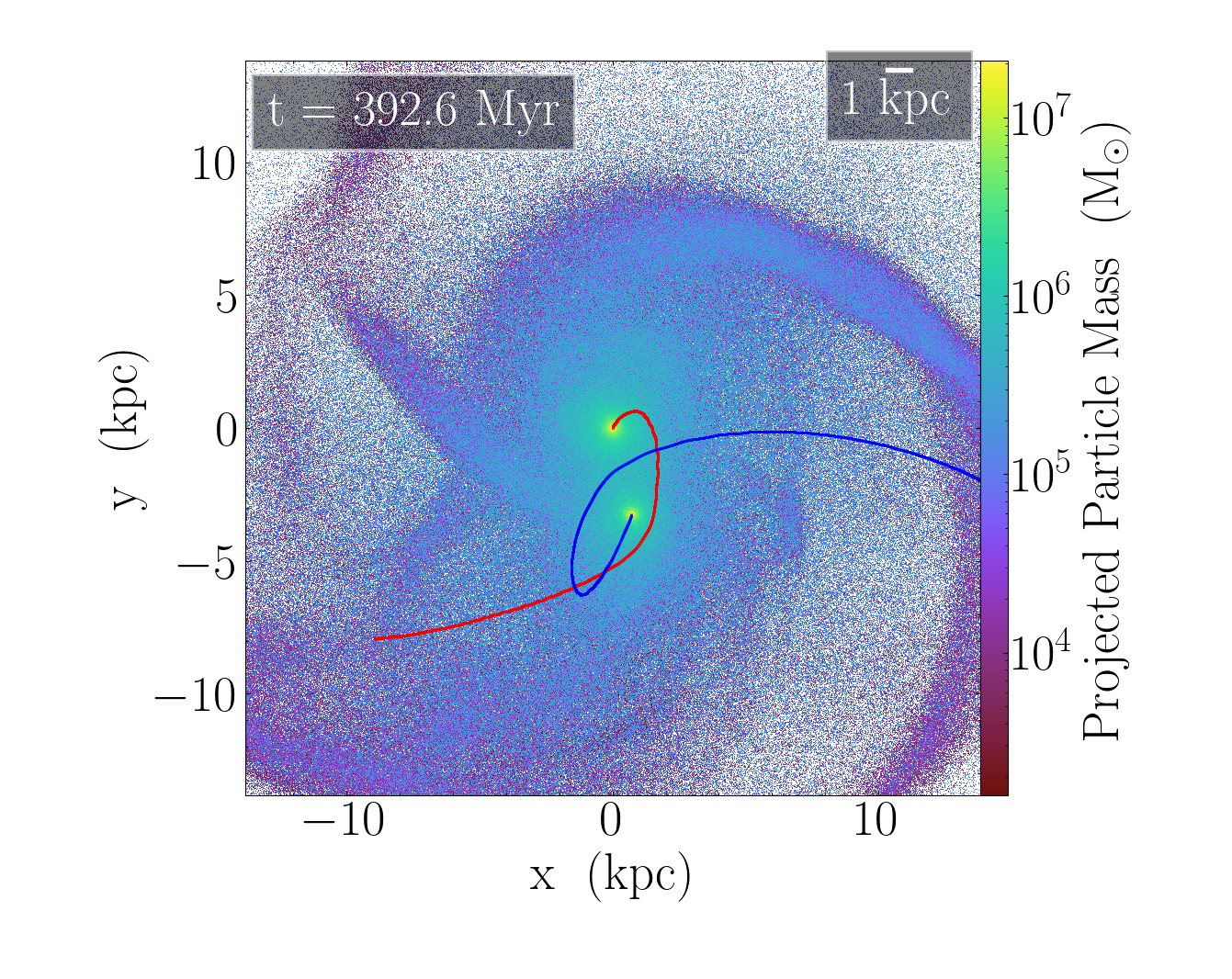}\vspace{-0.9cm}
           \includegraphics[width=0.33\textwidth , trim={0 0cm  0 0cm},clip]{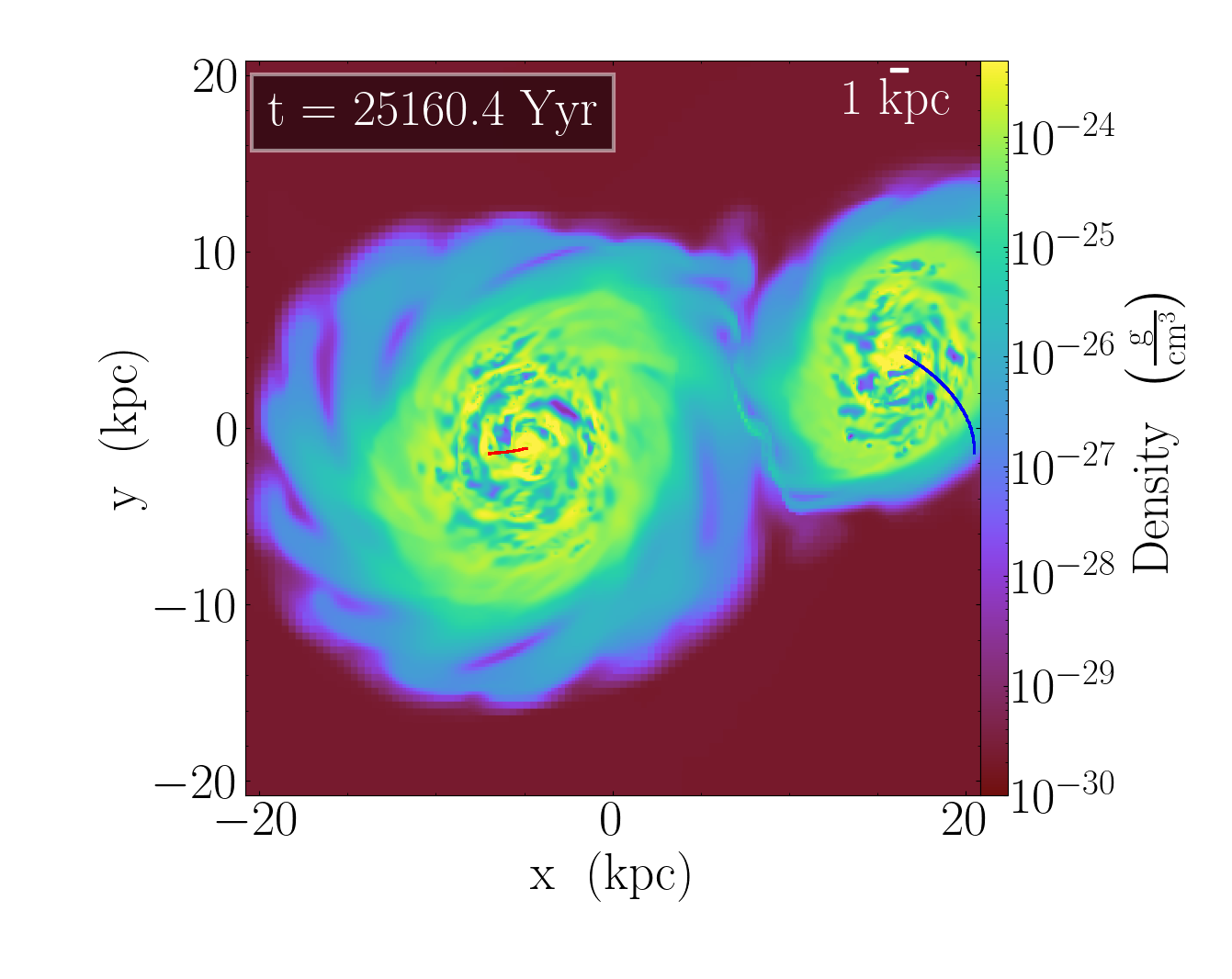}
           \includegraphics[width=0.33\textwidth , trim={0 0cm  0 0cm},clip]{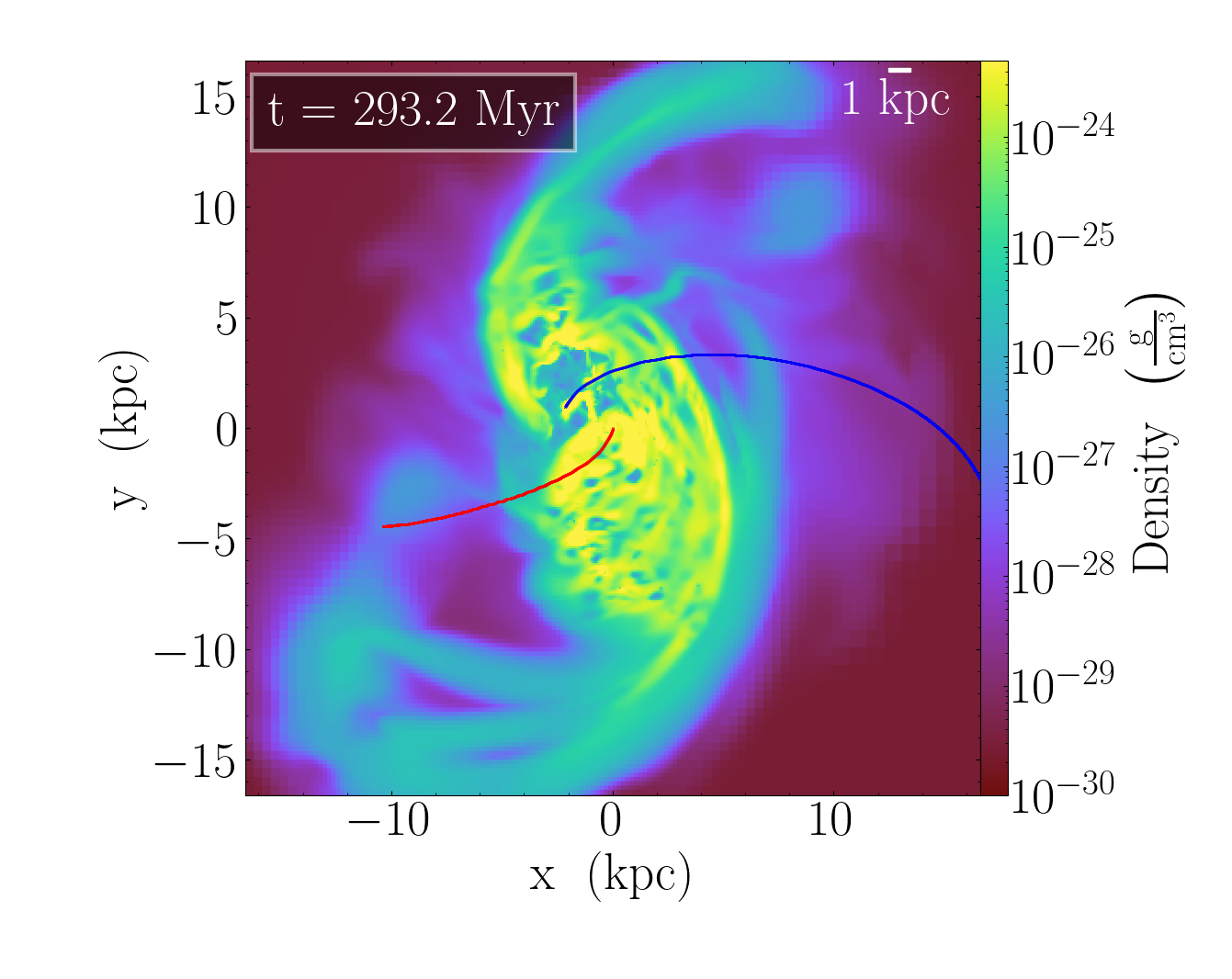}
           \includegraphics[width=0.33\textwidth , trim={0 0cm  0 0cm},clip]{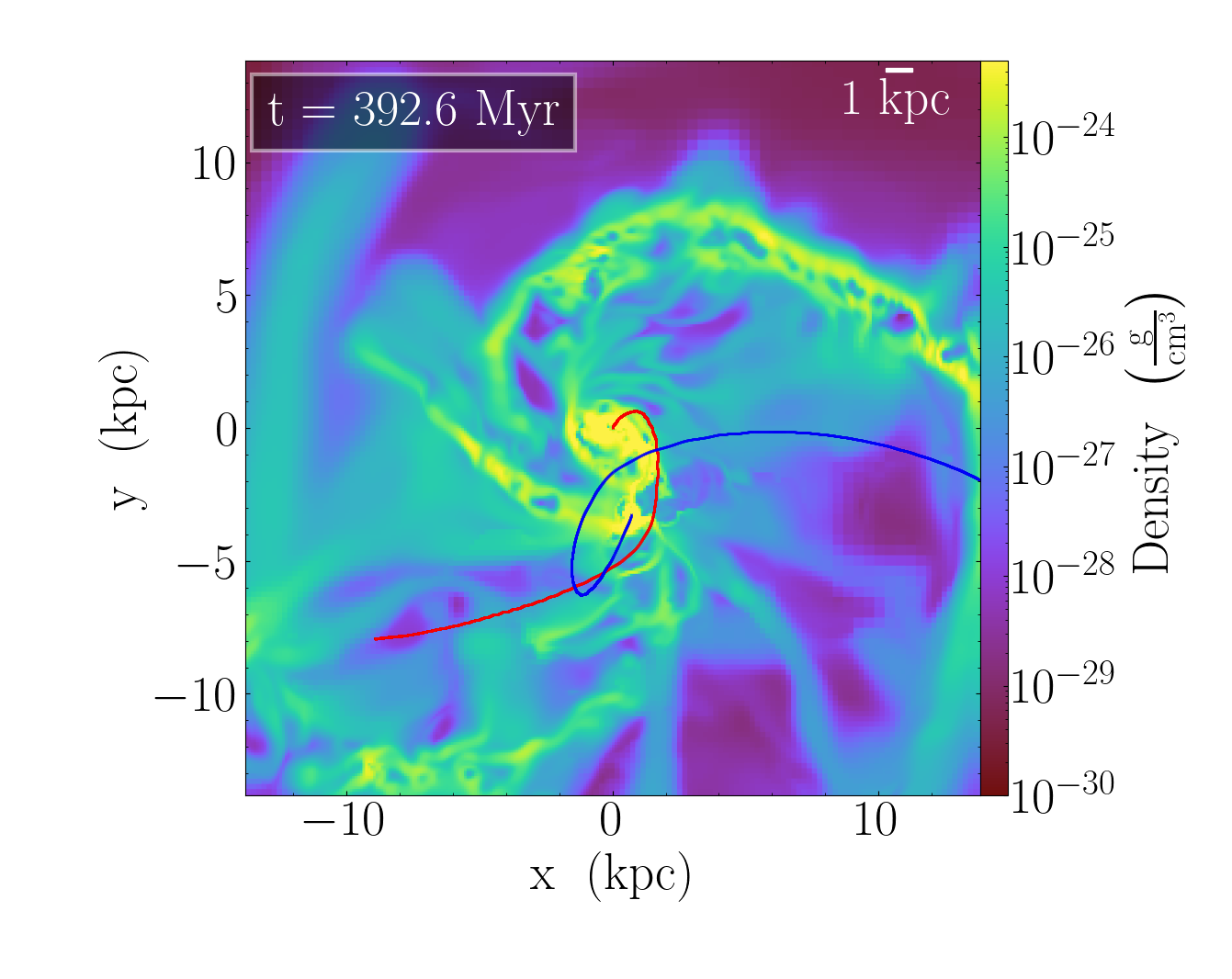}
           \includegraphics[width=0.33\textwidth , trim={0 0cm  0 0cm},clip]{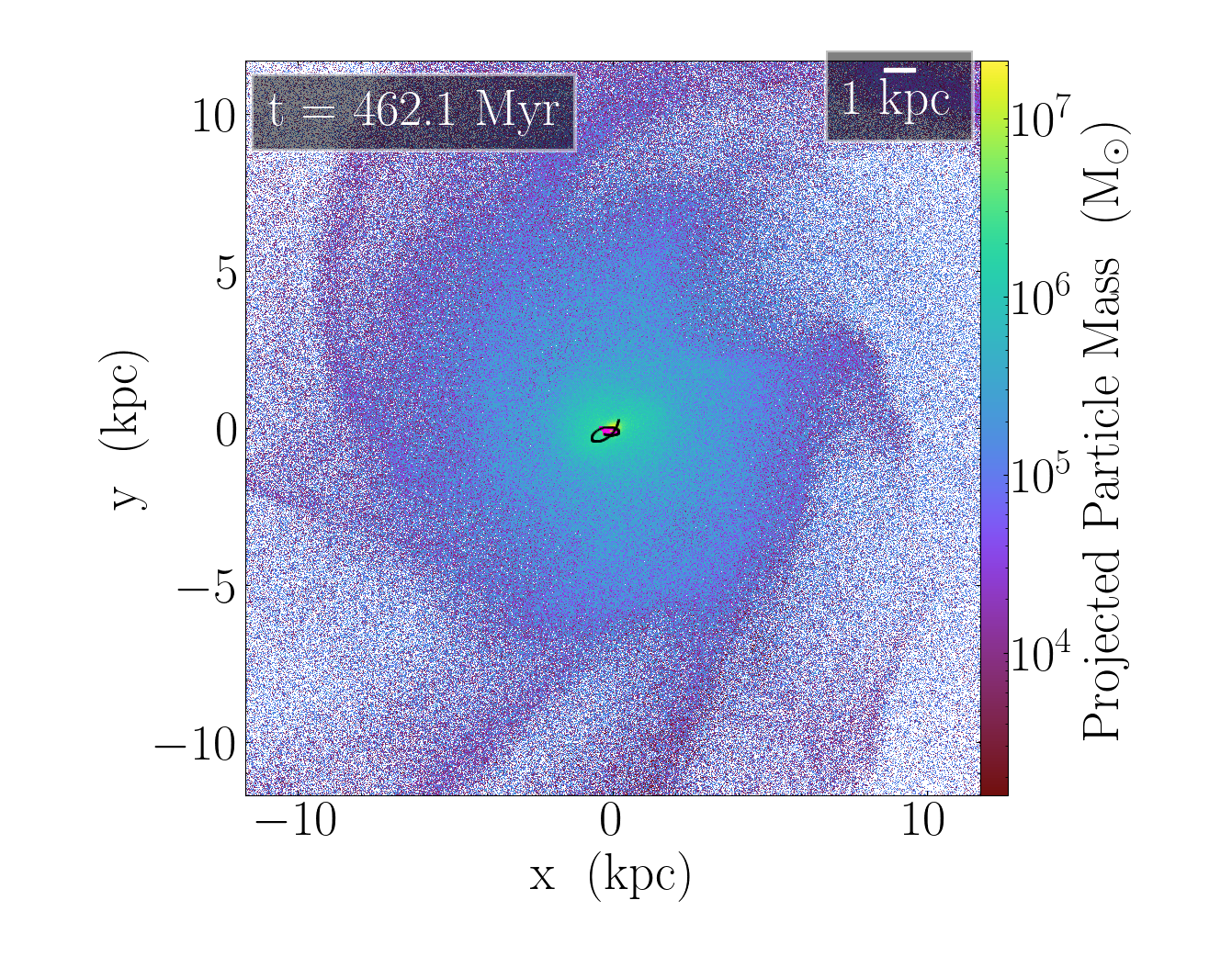}
           \includegraphics[width=0.33\textwidth , trim={0 0cm  0 0cm},clip]{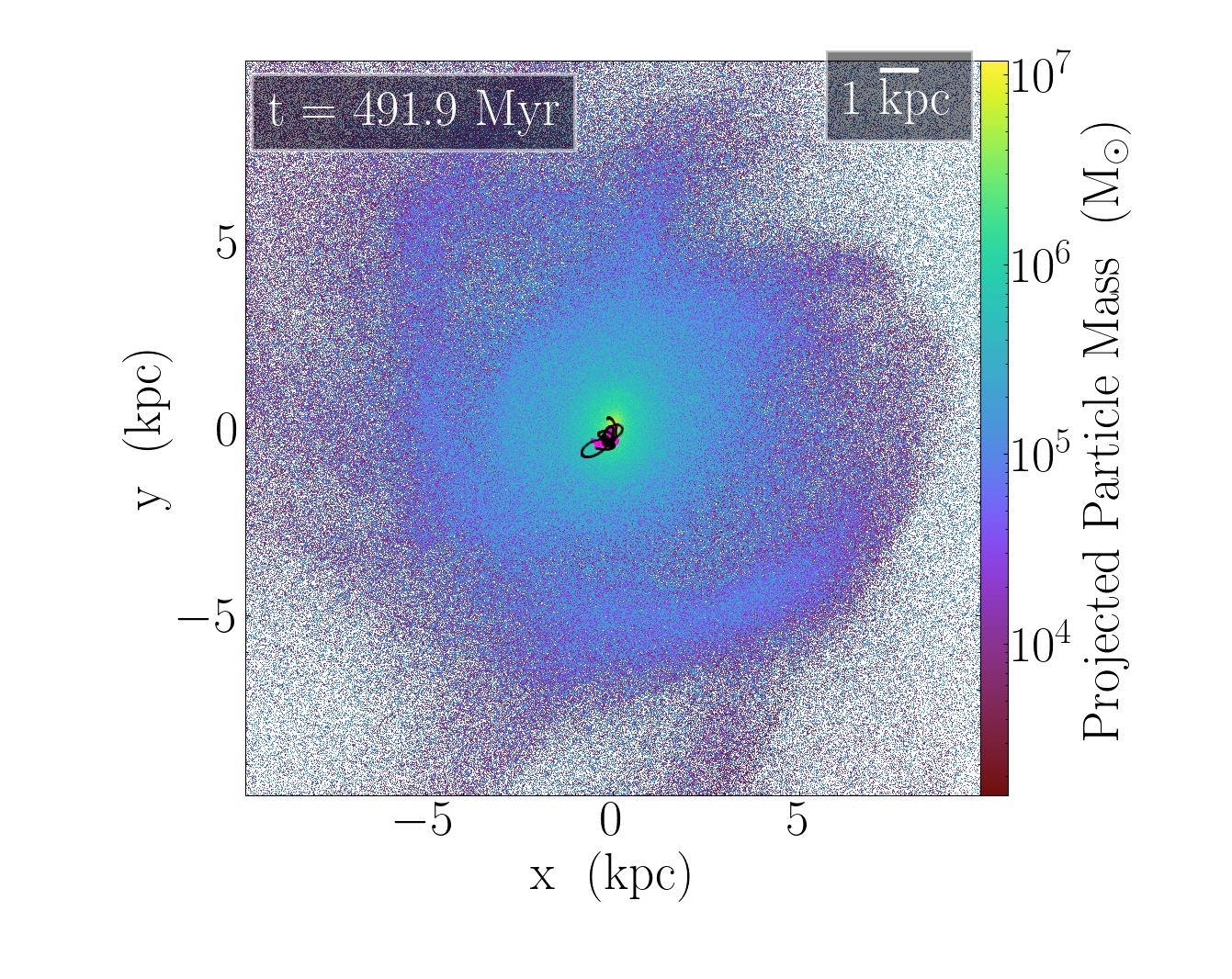}
           \includegraphics[width=0.33\textwidth , trim={0 0cm  0 0cm},clip]{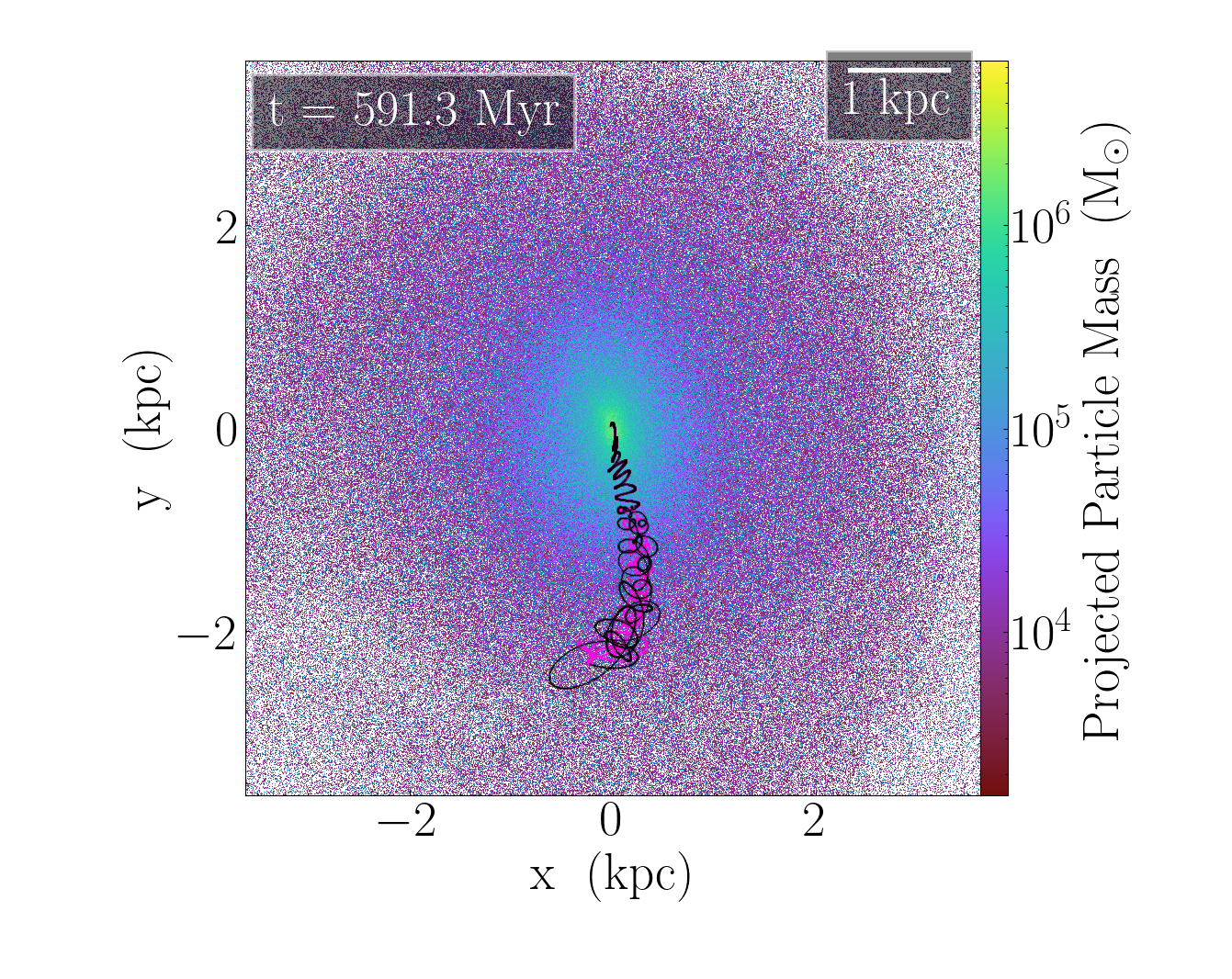}\vspace{-0.9cm}
           \includegraphics[width=0.33\textwidth , trim={0 0cm  0 0cm},clip]{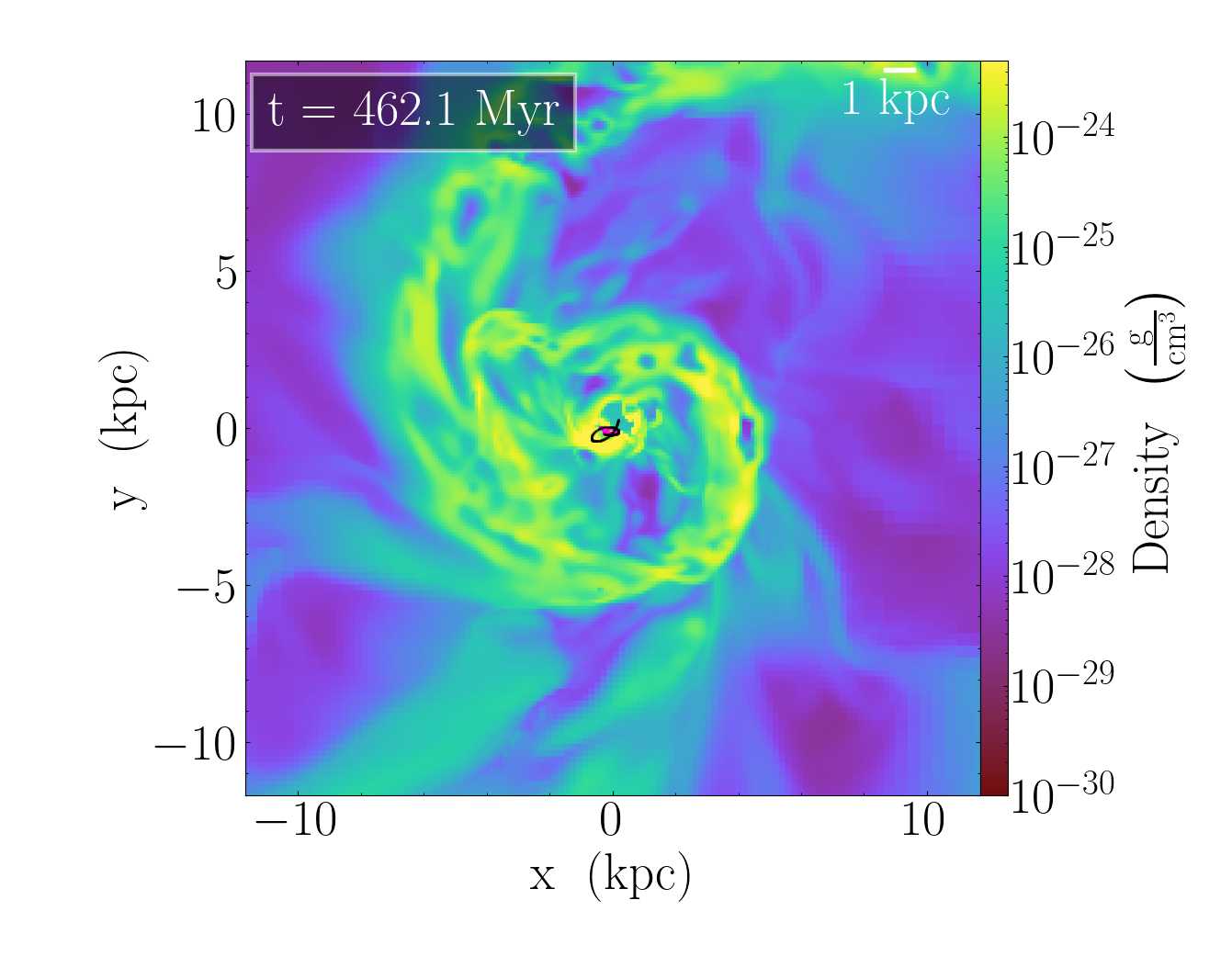}
           \includegraphics[width=0.33\textwidth , trim={0 0cm  0 0cm},clip]{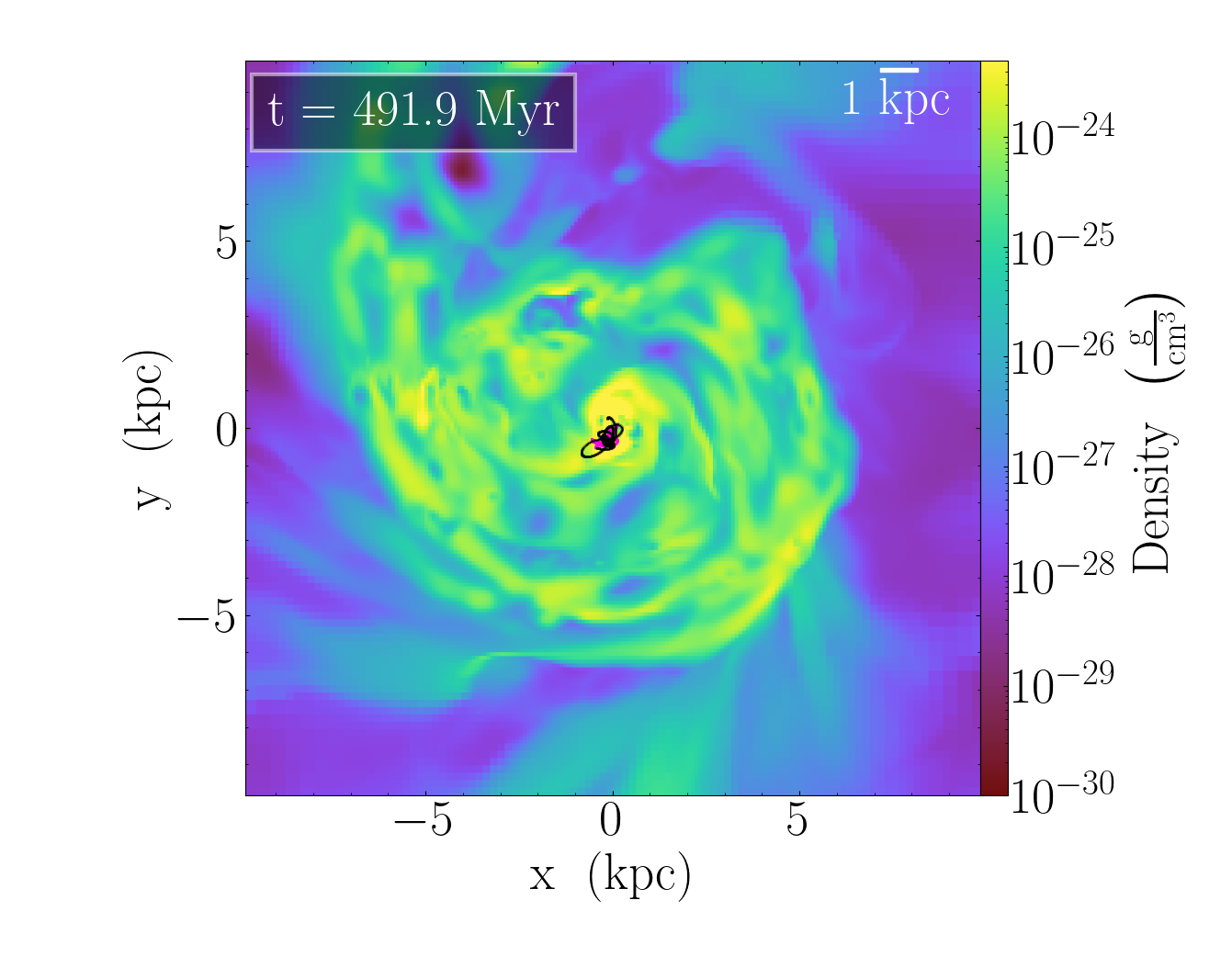}
           \includegraphics[width=0.33\textwidth , trim={0 0cm  0 0cm},clip]{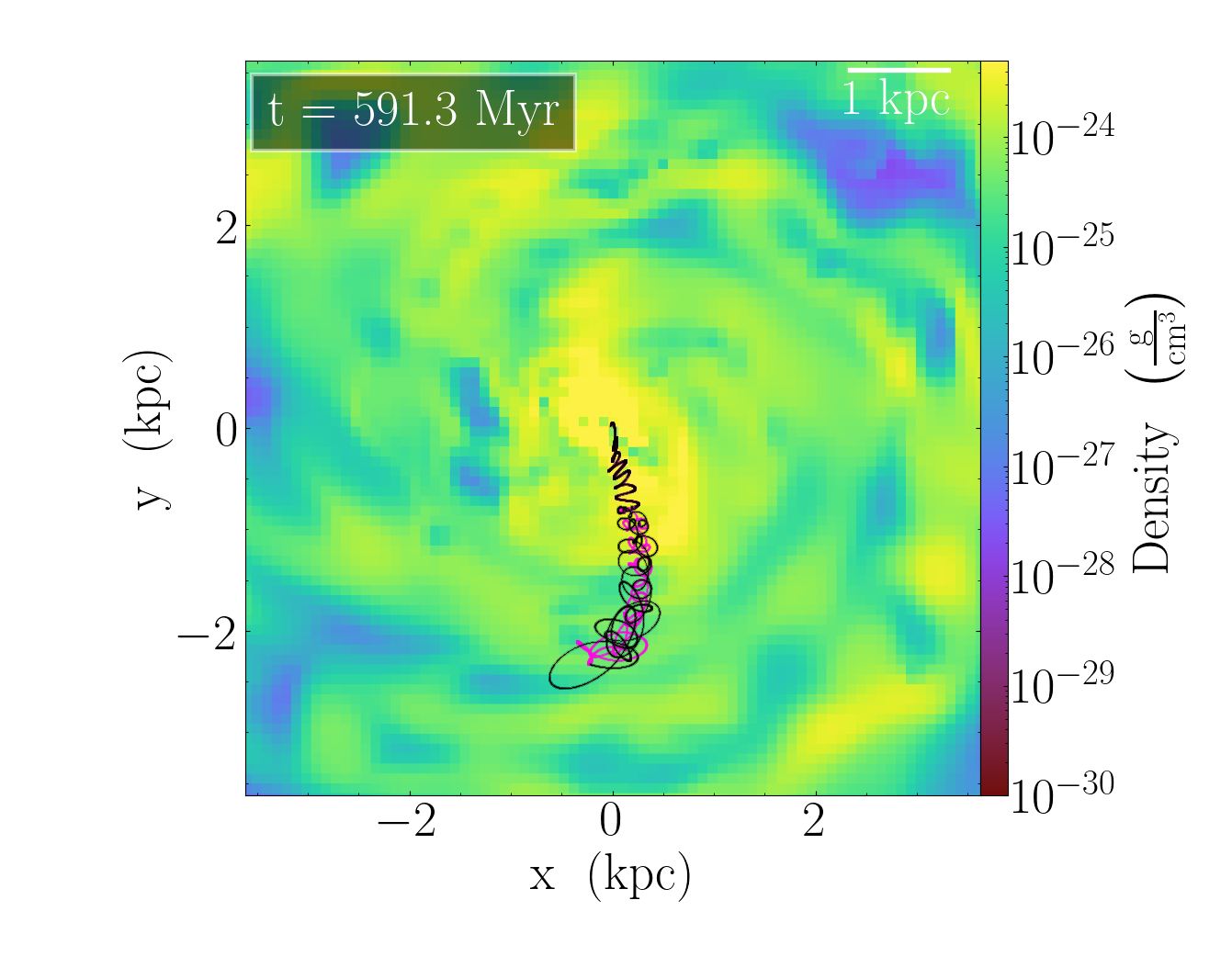}
\caption{The galaxy merger and coalescing trajectory of both MBHs on top of the stellar (first and third rows) and gaseous (second and fourth rows) density in simulation M1q0.5fg0.1 (see Table~\ref{tab:params}) at $100$\,pc resolution with accretion and feedback. In the sixth panels in the time sequences, the center of mass of the binary is not shown coincidentally to the center of mass of the galaxy except at the final time step. This is to more easily appreciate the orbital evolution of the binary. }
\label{fig:movie}
\end{figure*}

\subsection{Robustness tests with 100\,pc resolution}
\label{sub:100}
To explore the effect of initial star formation locations in the galaxy, as well as the impact of MBH accretion and feedback, we run a suite at $100$\,pc resolution using the M1q0.5fg0.1 setup (Table~\ref{tab:params}), with 10 different star formation random seed numbers (referred to as "seed1-10" in the following text), both with and without accretion and feedback. In Fig.~\ref{fig:movie}, we present the coalescing trajectory of MBHs superimposed on the evolution of gas and star density from one example of a $100$\,pc simulation with accretion and feedback.

\begin{figure*}[t]
    \centering
    \includegraphics[width=0.48\textwidth]{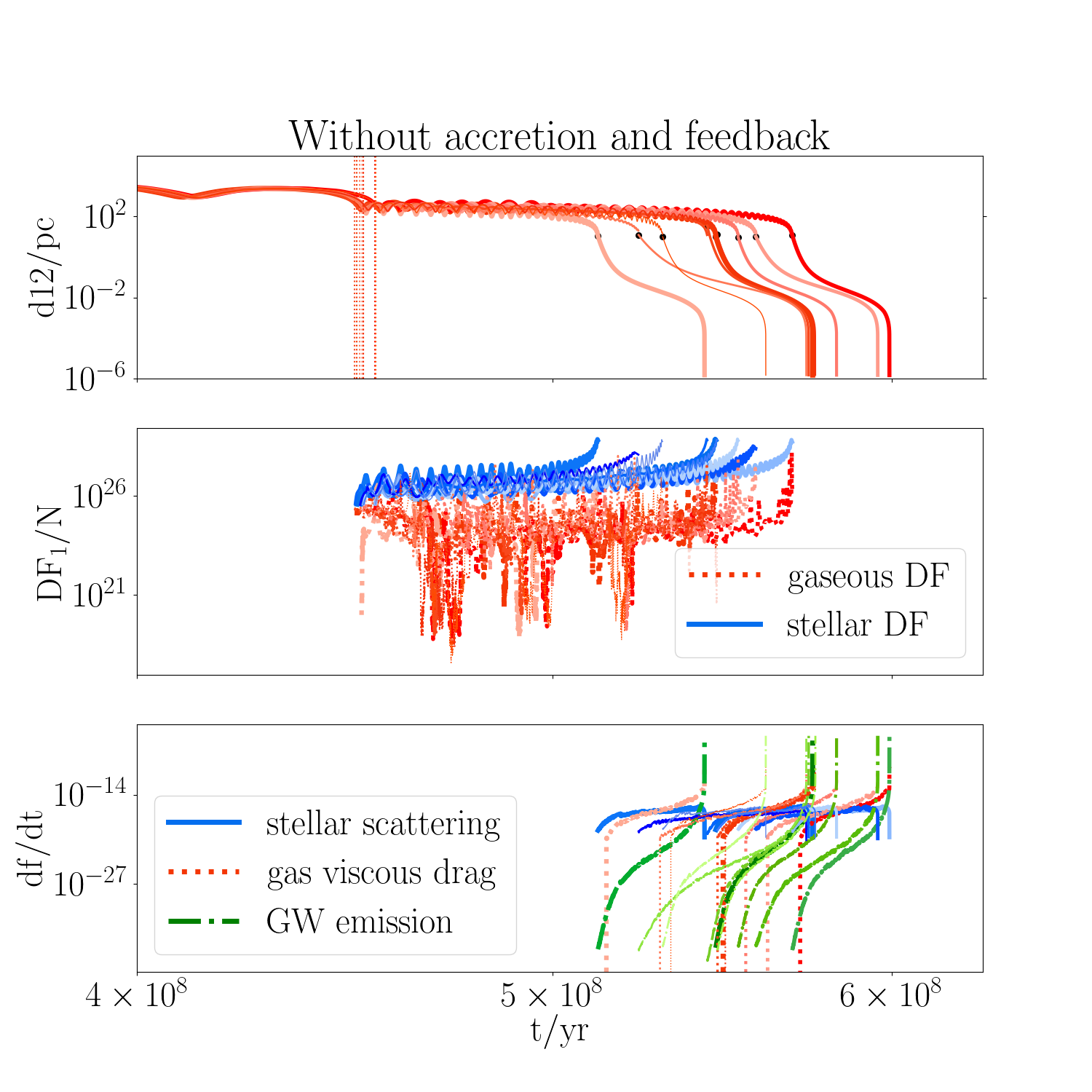}
    \includegraphics[width=0.48\textwidth]{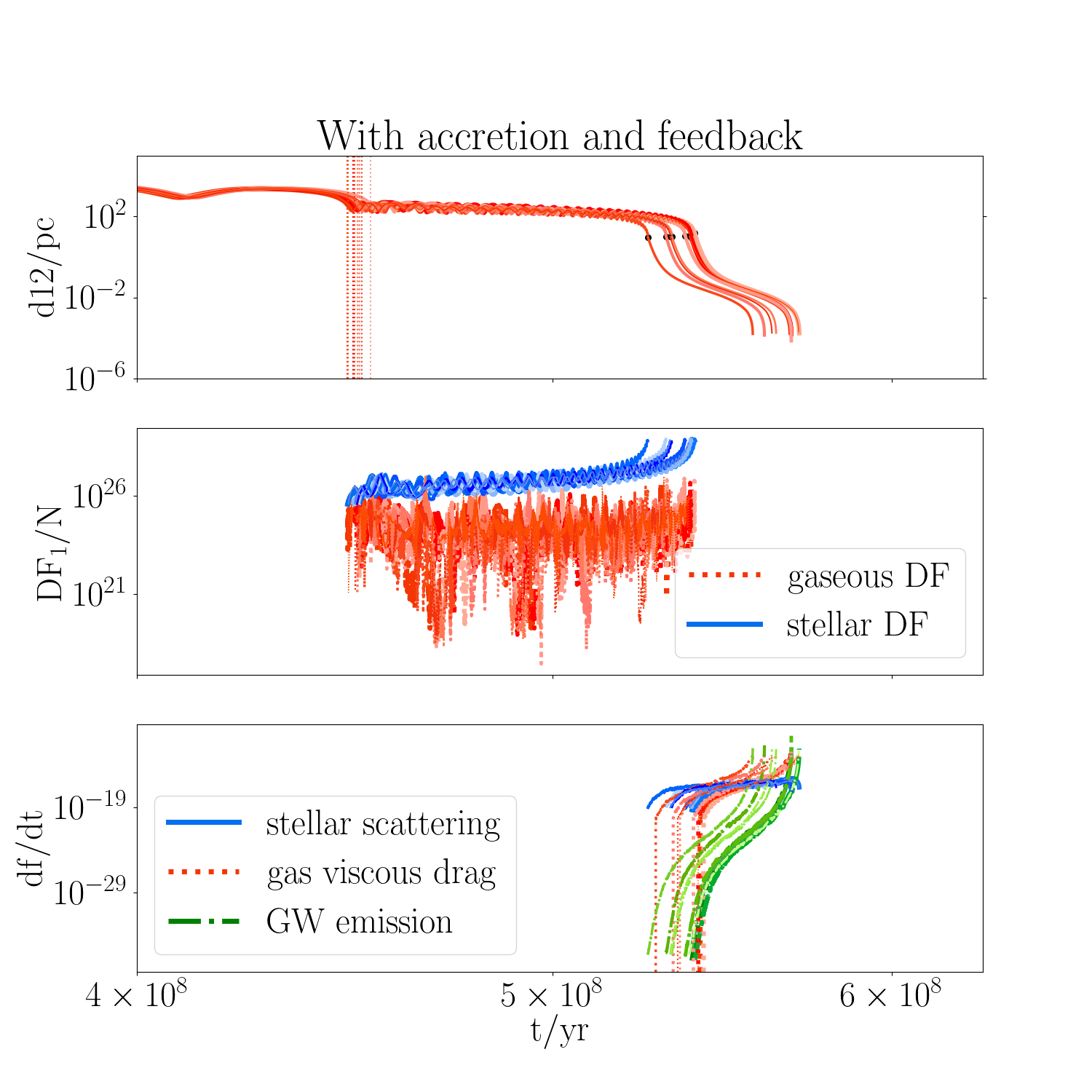}
    
\caption{The evolution of the separation between the two MBHs for the two merging disc galaxies setup without (top left panel) and with (top right panel) MBH accretion and feedback at $100$\,pc resolution. Ten runs with different star-formation random seeds in the galaxy are performed at each resolution to demonstrate the effect of stochasticity in the galactic environment on the MBHB orbital decay. The vertical dotted lines indicate when the MBHB evolution is handed over to RAMCOAL from RAMSES at the boundary of the resolution sphere. The black dots indicate the transition from stage 1 to stage 2 in RAMCOAL. Middle panels: the time evolution of dynamical friction from gas and collisionless particles: stars and dark matter. Bottom panels: the acceleration ${\rm d}f/{\rm d}t$ due to three physical mechanisms dominating stage 2 evolution: the stellar scattering, viscous torque in circumbinary disc, and GW emission for all $10$ runs of simulation M1q0.5fg0.1 at $100$\,pc resolution without (left) and with (right) accretion and feedback.}
\label{fig:10seed_100pc_d}
\end{figure*}

\begin{figure*}[t!]
    \centering
    \includegraphics[width=0.48\textwidth]{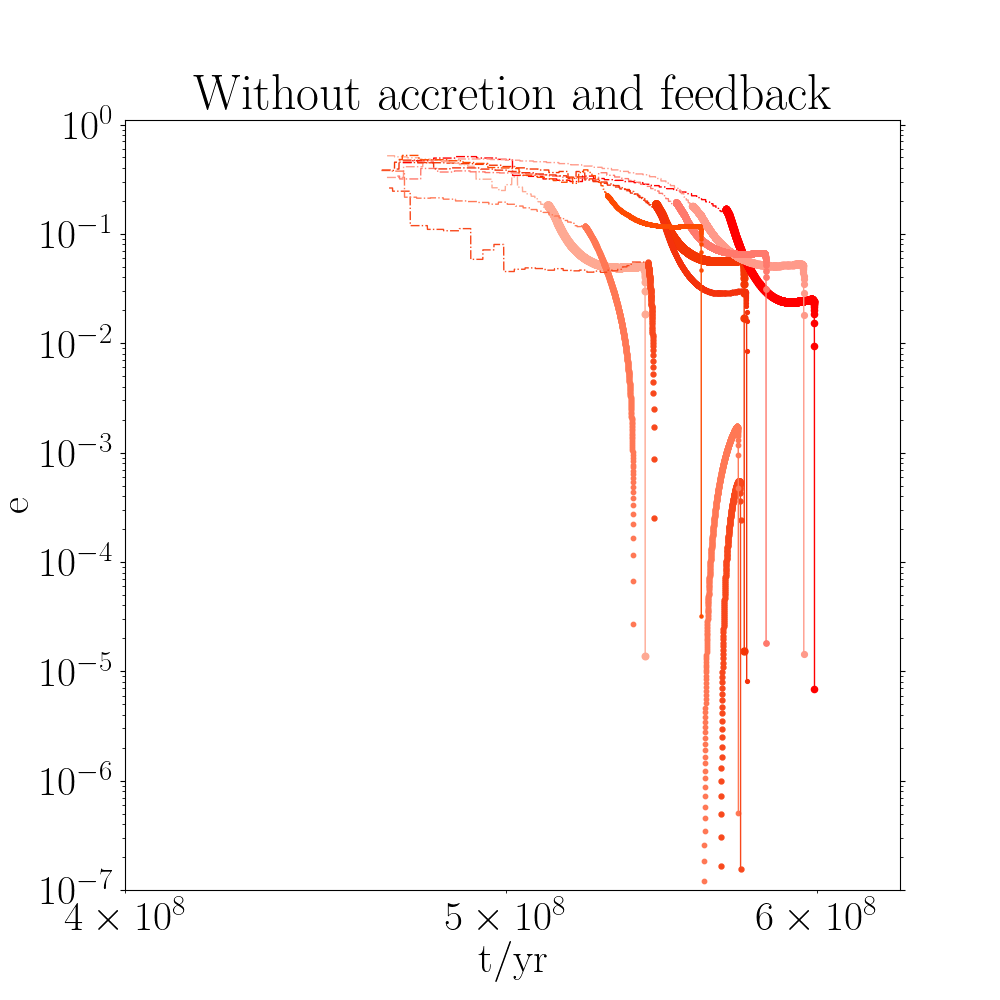}
    \includegraphics[width=0.48\textwidth]{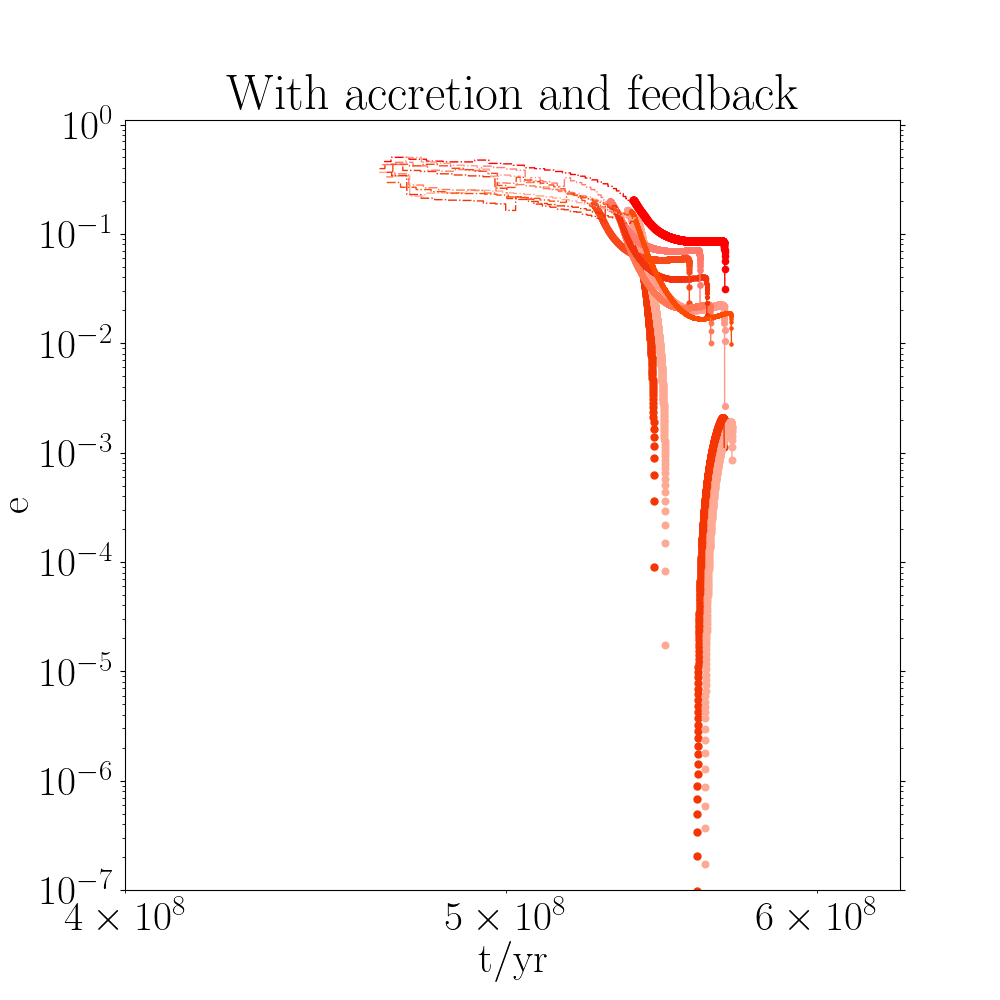}
\caption{The evolution of the orbital eccentricity between the two MBHs for the two merging disc galaxies setup without (left panel) and with (right panel) MBH accretion and feedback at $100$\,pc resolution. The dashed lines illustrate the evolution in RAMSES above the resolution limit. The change from dashed lines to dots indicates when the MBHB evolution is handed over to RAMCOAL at the boundary of the resolution sphere. }
\label{fig:10seed_100pc_e}
\end{figure*}

\begin{figure*}[t]
  \centering
    \includegraphics[width=0.48\textwidth]{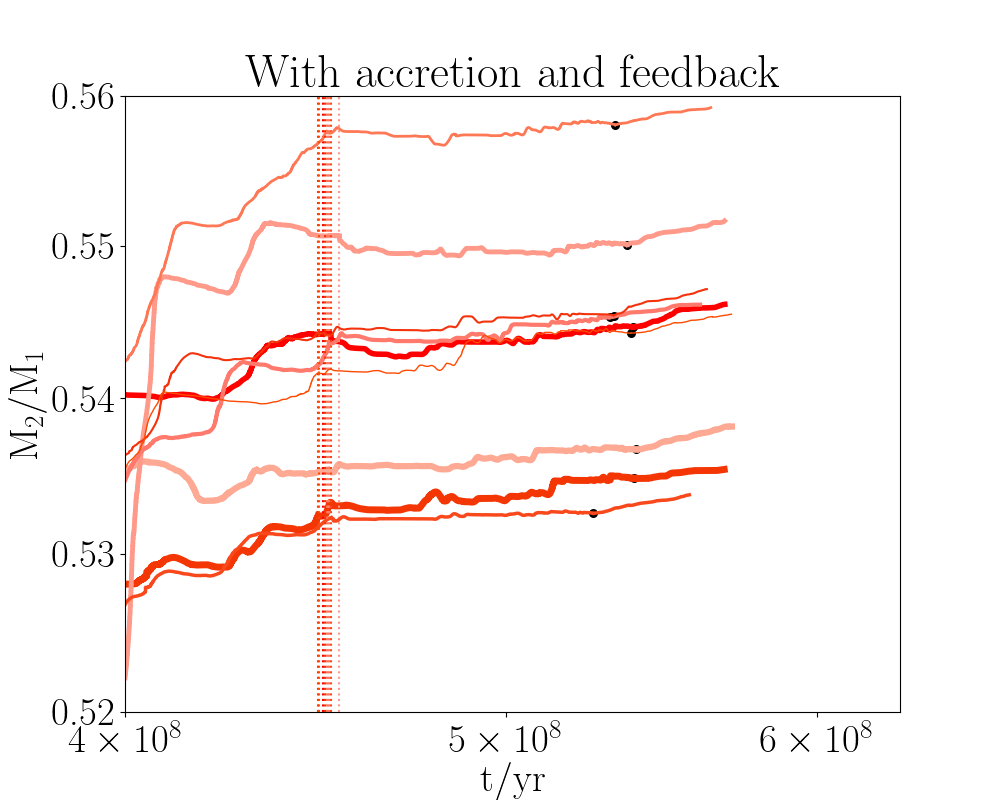}
    \includegraphics[width=0.48\textwidth]{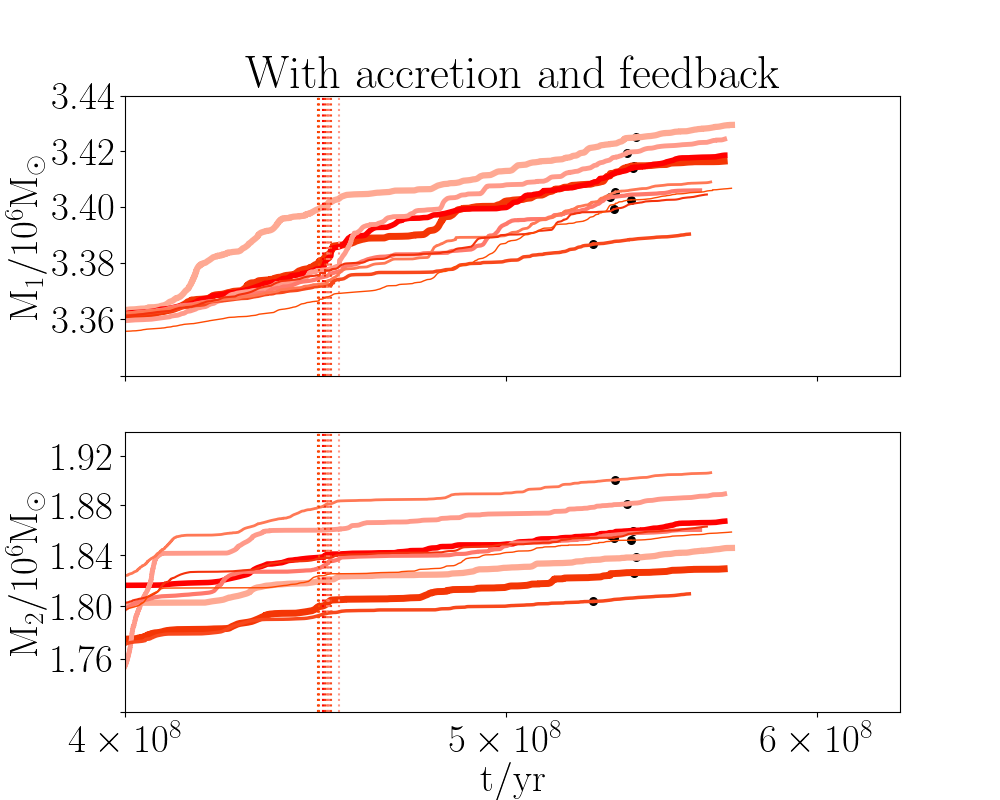}
\caption{The evolution of the MBH mass ratio ($M_{\rm 2}/M_{\rm 1}$) (left panel) and of their individual mass (right panels) for the two merging disc galaxies setup with MBH accretion and feedback at $100$\,pc resolution. The MBH masses are plotted in units of $10^{6} \, \rm M_{\odot}$.}
\label{fig:10seed_100pc_q}
\end{figure*}

The evolution of the separation between the two MBHs, with and without accretion and feedback, is shown in the top panels of Fig.~\ref{fig:10seed_100pc_d}. The coalescence time without MBH accretion and feedback is $5.75\times 10^{8}$\,yr with a fractional standard deviation of $0.01$ due to random variations in the gaseous and stellar structures. In contrast, the coalescence time with MBH accretion and feedback is $5.66\times 10^{8}$\,yr with a fractional standard deviation of $0.003$. The simulations with accretion and feedback stop at larger separations in the top right panel of Fig.~\ref{fig:10seed_100pc_d} than in the top left panel. This is because the innermost stable circular orbit (ISCO) radius in simulations with accretion is larger and hence the binary evolution stops at a larger separation compared to their accretion-free counterpart simulations in the left panel, since the ISCO radius is proportional to the binary mass which increases in the process through accretion.

On average, the coalescence time is shorter in the case of MBH accretion and feedback because accretion increases the mass of both MBHs as well as the mass ratio between them. The increase in MBH mass enhances the effectiveness of DF, accelerating the coalescence process. In the simulations with AGN feedback, it blows away the gas reservoir and periodically disrupts the sub-grid gas density profile, which can reduce the magnitude of gas DF and, in some cases, even reverse its direction, causing the MBHs to accelerate rather than decelerate. However, based on our results, this radiation feedback effect is subdominant because stellar density dominates over gas density, and radiation feedback does not affect stellar DF. Overall, in the competition between MBH accretion and feedback, accretion shortens the coalescence time more than feedback slows it down.

The fractional standard deviation of coalescence time in the case of MBH accretion and feedback is nearly one-third of that without MBH accretion and feedback. Without MBH accretion and feedback, encounters with gas clumps along the MBH trajectory are more likely, which can enhance gas DF and accelerate orbital decay. This phenomenon occurs in the run with the shortest coalescence time in the MBH no-accretion, no-feedback case (as shown in the top left panel of Fig.~\ref{fig:10seed_100pc_d}). The encounter with a gas clump in this run is also indicated by a sudden drop in eccentricity, as seen in the left panel of Fig.~\ref{fig:10seed_100pc_e}. The higher gas density in the clump increases the gaseous DF, speeding up orbital decay and circularizing the orbit.

In the absence of AGN feedback, the gas reservoir near the MBHs is not regulated, making it more susceptible to the randomness in galaxy realization. This leads to greater variation in both coalescence time and eccentricity (at stage 1) without accretion and feedback. In contrast, the coalescence time in the case of accretion and feedback shows a smaller sensitivity to variations in galaxy realization. This is because AGN feedback regulates the gas reservoir around both MBHs, reducing the presence of individual dense gas clumps. Furthermore, the sub-grid gaseous density profile is lower and is primarily determined by the MBH accretion and feedback parameters, thereby reducing the impact of random variations in galaxy realization.

For a comparison of the forces from each galactic component at different stages of the MBH orbital evolution. The evolution of the dynamical friction magnitude (stage 1) and acceleration (${\rm d}f/{\rm d}t$) in stage 2 is shown in the middle and bottom panels of Fig.~\ref{fig:10seed_100pc_d}. The top panels show the time evolution of dynamical friction from gas and star without (left) and with (right) accretion and feedback. The dynamical friction from stars is larger than that from gas most of the time due to a higher stellar density in the galaxy remnant. However, without the accretion and feedback (left panel), the gas dynamical friction goes through spikes of magnitude as high as the star dynamical friction when encounter high density gas clumps. On the other hand, with the effect of feedback shown in the right panel, the peaks in gas dynamical friction are smeared out, since the AGN feedback regularizes the gas reservoir around the MBH.

The bottom panels show the time evolution of frequency change rate due to three physical mechanisms in stage 2 evolution: the stellar scattering, viscous torque in circumbinary disc, and GW emission. The stellar scattering dominates the evolution at the beginning of stage 2 when the MBHs just become bounded. The viscous drag from circumbinary discs gradually increases and starts to dominate at some point, followed by the GW emission and the coalescence. This is because the acceleration due to viscous torque in circumbinary disc is proportional to $a_{\rm orb}^{-5}$ (see equation~\ref{eq:gas_drag1}), while that due to stellar scattering is proportional to $f_{\rm orb}^{1/3}$, which for circular orbits corresponds to $a_{\rm orb}^{-1/2}$. As the orbit shrinks, the separation decreases and the frequency increases, but due to the power difference, the acceleration due to viscous torque dominates over stellar scattering for a short period of time ($\sim 20\%$ of stage 2 evolution time) before GW emission takes over. We tested the difference in coalescence time due the change in the circumbinary disc survival threshold value ($\eta_{\rm 2}$) ranging from $0.01$ to $1$, and shown that the change in the coalescence time is sensitive to the value of $\eta_{\rm 2}$ only in the last $10\,\rm Myr$ before the coalescence when viscous torque dominates the evolution. By turning off the viscous torque completely, the coalescence time is increased by $15\%$. The results in this work are not sensitive to the value of $\eta_{\rm 2}$ because the general Eddington rate in our simulations is below $0.01$, in this regime the viscous torque is always efficient. 

This discussion shows that both stellar scattering and gas viscous drag contribute significantly to the MBHB hardening process. To this point, most simulations only include either stellar scattering or gas drag in circumbinary disc due to simulation method and/or high computational cost. The results from our simulation strengthens the importance of simulating both stellar hardening and migration in circumbinary discs in the evolution of MBHBs.

The eccentricity evolution with and without MBH accretion and feedback is shown in Fig.~\ref{fig:10seed_100pc_e}. The sharp decrease in eccentricity in the last few time steps before coalescence is caused by GW emission, which efficiently circularizes the orbit. It should be noted that the eccentricity value at stage 1 is not precise. Because the MBH pair is not yet gravitational bounded due to the large quantity of intervening stellar mass, which causes the Keplerian definition of eccentricity to oscillate and not instantaneously match exactly the actual MBH binary orbit. However, the overall trend is consistent with the expected orbital geometry. In some runs, the eccentricity reaches a turnover point before the drop caused by GW emission. This temporary increase in eccentricity is due to viscous drag in the circumbinary disk. As mentioned in Sect.~\ref{subsub:VD}, viscous drag in the circumbinary disk tends to increase the MBHB eccentricity to a converging value between $[0.6,0.8]$. Once GW emission dominates over viscous drag, the eccentricity decreases again due to circularization. The eccentricity at coalescence is $1.2\times 10^{-5}$ with a fractional standard deviation of $0.27$ without MBH accretion and feedback. With MBH accretion and feedback, the coalescence eccentricity is $0.013$, with a fractional standard deviation of $0.28$. Note that the ISCO radius in simulations with accretion is larger and hence the binary evolution stops at a larger separation compared to their accretion-free counterpart simulations, since the ISCO radius is proportional to the binary mass which increases in the process through accretion.

In the case without MBH accretion and feedback, the eccentricity evolves to a smaller value earlier in the process, as shown by the spread in eccentricity in the left panel of Fig.~\ref{fig:10seed_100pc_e} at around $5\times 10^{8}$\,yr. This is related to gaseous DF, which helps to circularize the orbit. As demonstrated in~\cite{LBB20a} (see their figure 3), in the absence of radiation feedback, gaseous DF reduces orbital eccentricity and causes it to converge to a range of $[0.3, 0.5]$, depending on the relative speed of the gas disk with respect to the MBH. On the other hand, in the case of MBH accretion and feedback, the gas around the MBHs is blown away, reducing the effect of gaseous DF.  Consequently, only stellar DF contributes to circularizing the orbit in the presence of MBH accretion and feedback, leading to a slower decrease in eccentricity compared to the case without MBH accretion and feedback, where both stellar and gaseous DF act to circularize the orbit at stage 1.

Lastly, the evolution of the mass ratio and MBH mass in the case of MBH accretion and feedback is shown in Fig.~\ref{fig:10seed_100pc_q}. The initial mass ratio is $0.5$, and the mass ratio at coalescence is $0.54$, with a fractional standard deviation of $0.005$. Most of the mass ratio growth occurs before the MBHs enter the resolution sphere (stage 0). 
This is in broad agreement with \citet{C2015} and \citet{Gabor2016}, who have shown that during the second passage of two merging galaxies, merger-induced tidal torques cause gas to lose angular momentum and flow inward, leading to enhanced MBH accretion and bursts of star formation. The mass ratio growth is less significant once both MBHs enter the resolution sphere for two reasons. First, the sub-grid gas density profile follows a core-power law, meaning the gas density around the primary MBH is higher than around the secondary MBH. Second, even if the sub-grid density profile is not triggered, the secondary MBH tends to move faster relative to the background gas, leading to a lower accretion rate onto the secondary MBH. These factors make accretion less prominent in stages 1 and 2.

\begin{figure*}[t!]
    \centering
    \includegraphics[width=0.48\textwidth]{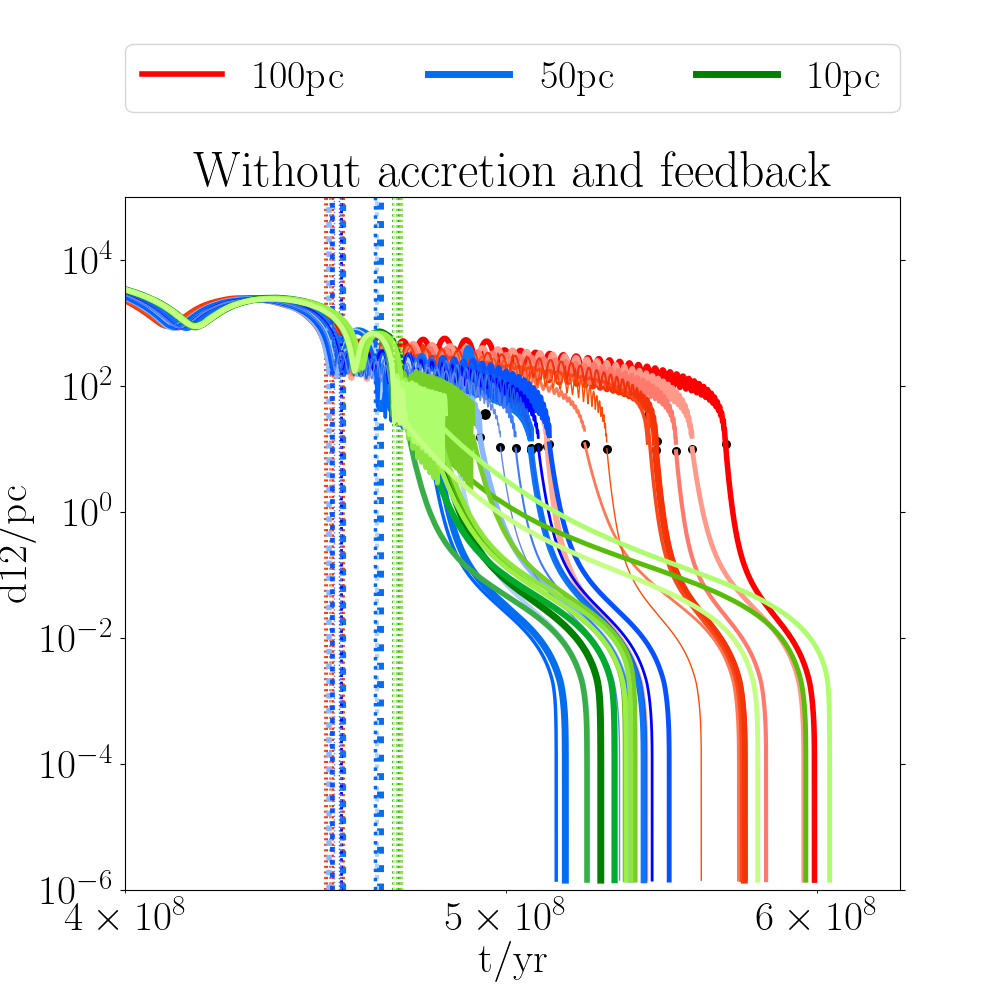}
    \includegraphics[width=0.48\textwidth]{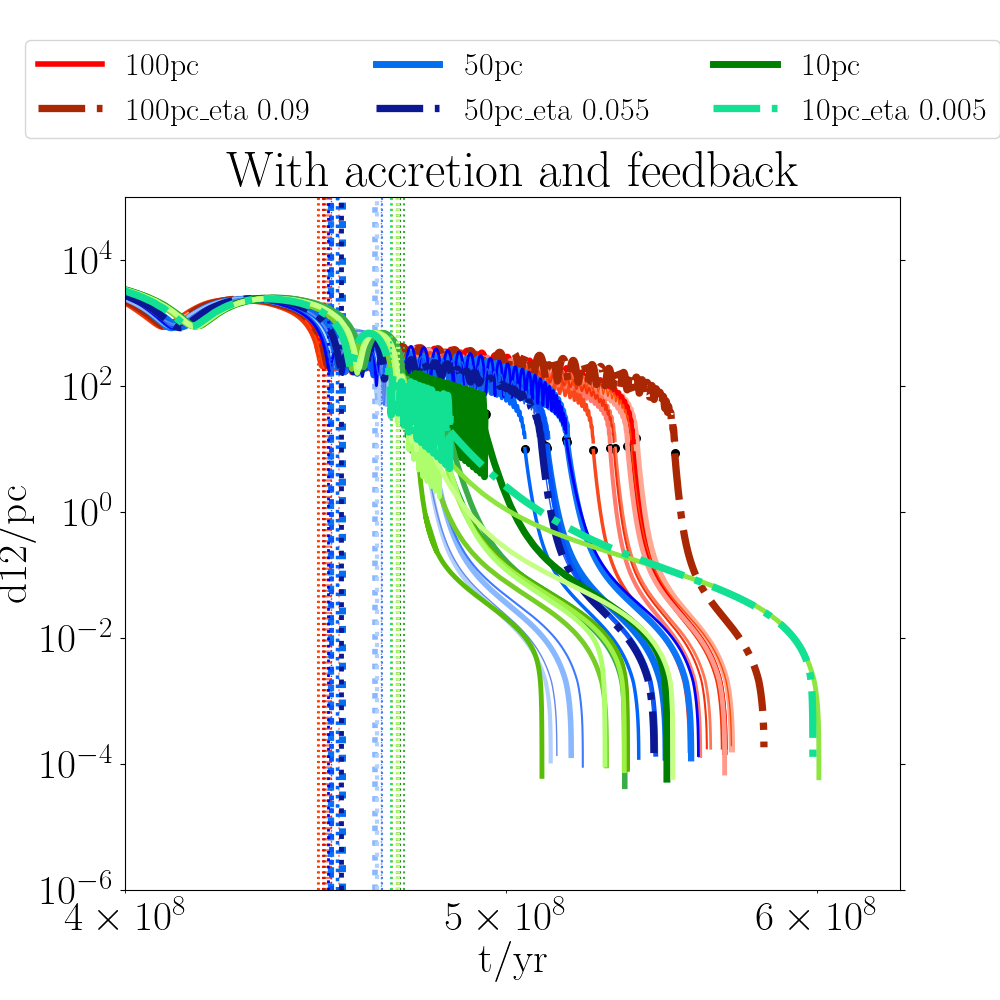}
\caption{The separation evolution for the two merging disc galaxies setup M1q0.5fg0.1 without (left panel) and with (right panel) MBH accretion and feedback at $100$, $50$, and $10$\,pc resolutions. $10$ runs are shown for each resolution with different initial star formation locations for the purpose of assessing the influence of small stochastic effects. The vertical dotted lines indicate when the MBHB evolution is handed over to RAMCOAL from RAMSES at the boundary of the resolution sphere. One additional run with resolution-dependent AGN feedback efficiency at each resolution is also plotted to illustrate the effect of feedback efficiency on the results. }
\label{fig:100_50_d12}
\end{figure*}

\begin{figure*}[t!]
    \centering
    \includegraphics[width=0.48\textwidth]{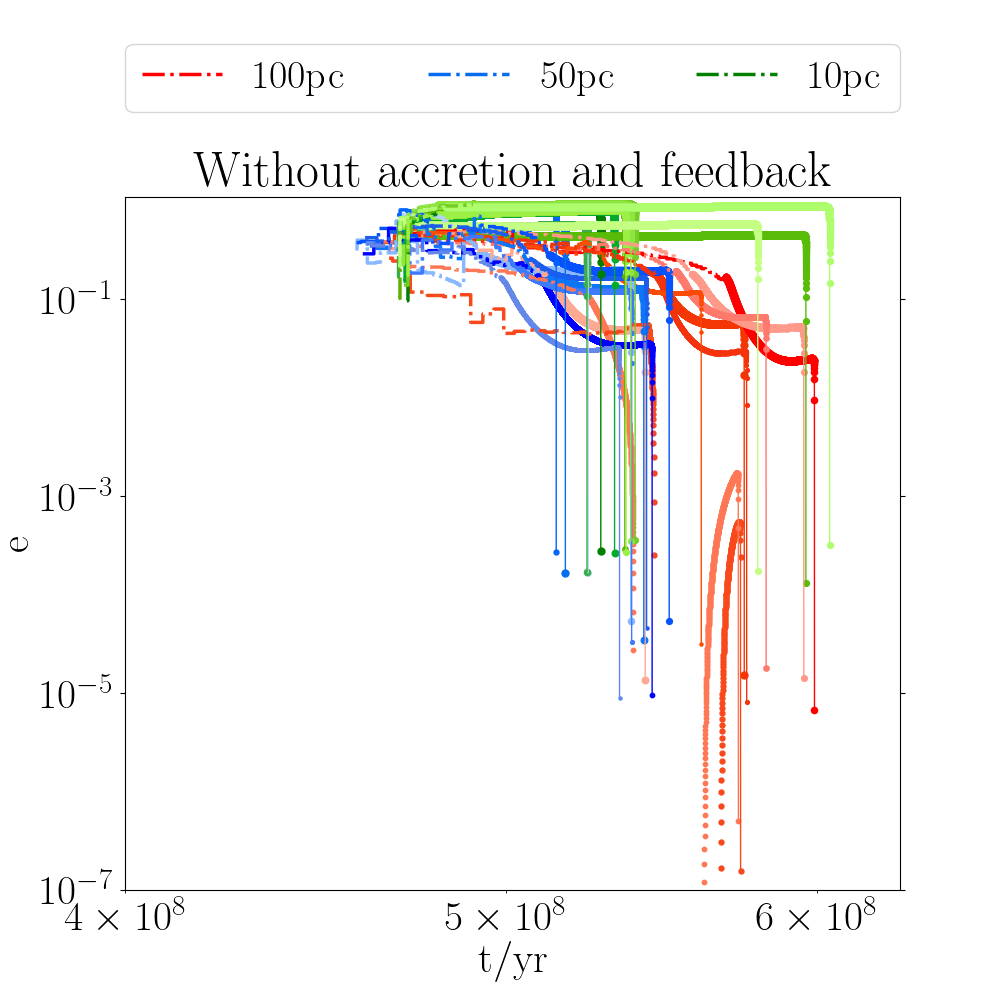}
    \includegraphics[width=0.48\textwidth]{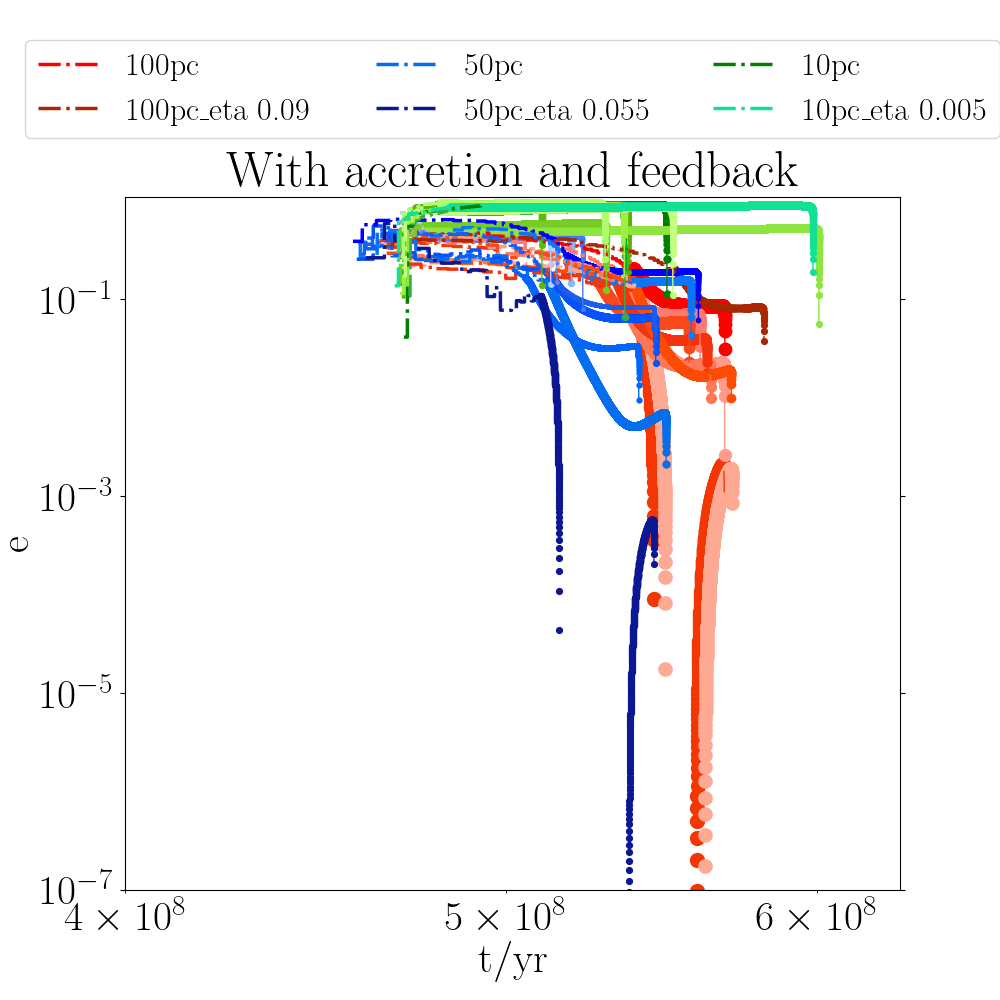}
\caption{The eccentricity evolution in merging galaxy simulation M1q0.5fg0.1 without (left) and with (right) MBH accretion and feedback at $100$, $50$, and $10$\,pc resolutions, $10$ runs for each resolution with different star formation random seeds for the purpose of gauging stochastic variations. The dashed lines illustrate the evolution in general RAMSES above the resolution limit. The change from dashed lines to dots indicates when the MBHB evolution is handed over to RAMCOAL from RAMSES at the boundary of the resolution sphere. One additional run with resolution-dependent AGN feedback efficiency at each resolution is also plotted to illustrate the effect of feedback efficiency on the results.}
\label{fig:100_50_e}
\end{figure*}

\begin{figure*}[t]
  \centering
    \includegraphics[width=0.48\textwidth]{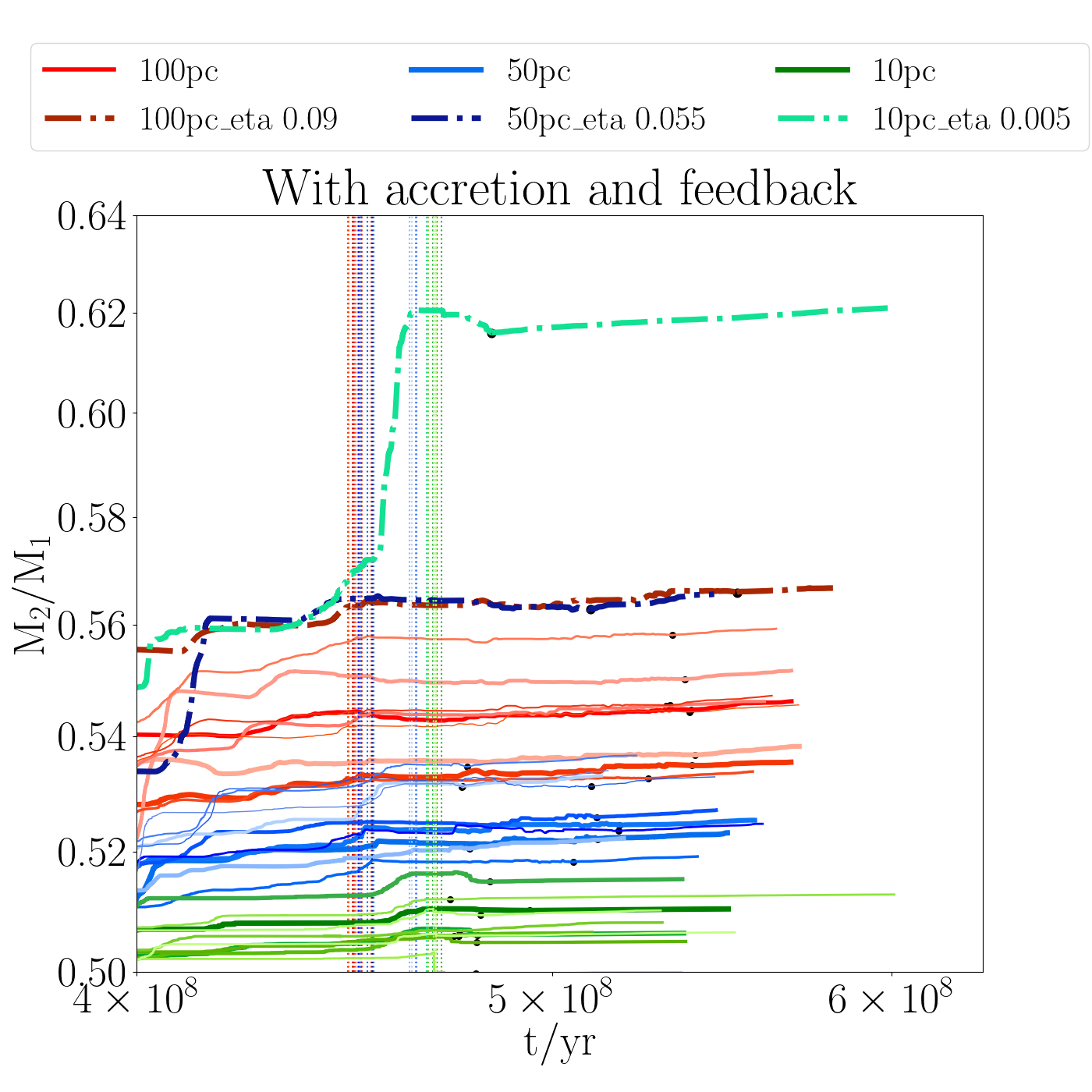}
    \includegraphics[width=0.48\textwidth]{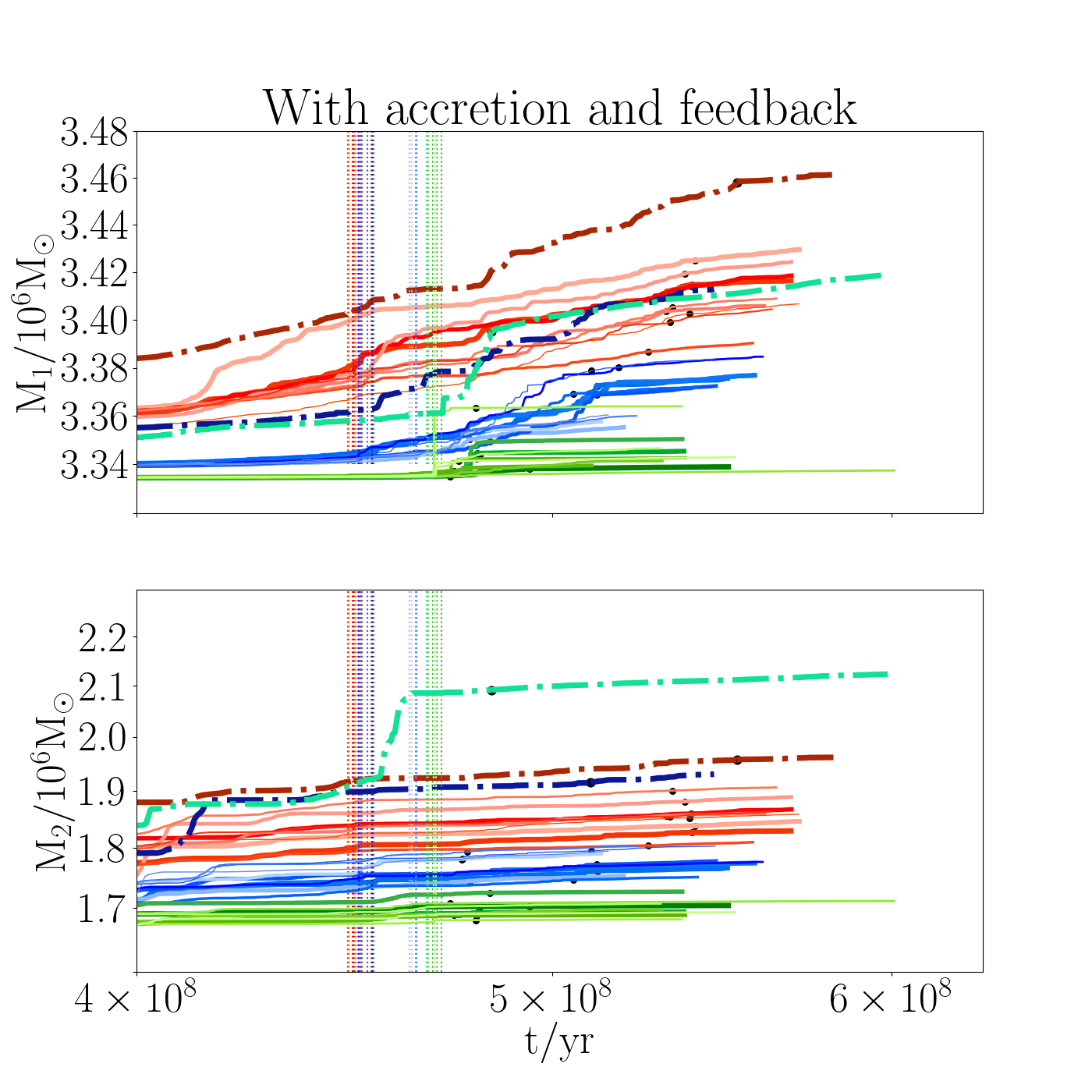}
\caption{The mass ratio and individual MBH mass evolution in the merging galaxy simulation M1q0.5fg0.1 with MBH accretion and feedback at $100$, $50$, and $10$\,pc resolutions, with $10$ runs for each resolution with different star formation random seeds. One additional run with resolution-dependent AGN feedback efficiency at each resolution is also plotted to illustrate the effect of feedback efficiency. The vertical dotted lines indicate when the MBHB evolution is handed over to RAMCOAL from RAMSES at the boundary of the resolution sphere.}
\label{fig:100_50_q}
\end{figure*}

\subsection{Resolution tests}
\label{sub:all}
\begin{table}
    \centering
    \caption{Results of MBHB coalescence in the merging galaxy M1q0.5fg0.1 with or without accretion and feedback at various resolutions.}
    \begin{tabular}{cccccc}
    \hline
    $\Delta x$ & acc. and & $\bar{t}_{\rm coa}$ & $\delta t_{\rm coa}$ & $\bar{q}$ & $\delta q$\\
    (pc) & feedback & ({\rm Myr}) &  & & \\
    \hline
    10 & no & $550$ &  1.7\% & $0.5$ & $0$   \\
    50 & no & $540$ & 0.7\%  & $0.5$ & $0$   \\
    100 & no & $575$ & 1\%  & $0.5$ & $0$   \\
    10 & yes & $542$ &  1.4\%  & $0.51 $ & 0.3\%  \\
    50 & yes & $537$ & 1.0\% & $0.53 $ & 0.3\%  \\
    100 & yes & $566$ & 0.3\% & $0.54$ & 0.5\%  \\
    \hline
    \end{tabular}
    \label{tab:reso}
\end{table}

\begin{figure}[t]
\includegraphics[width=0.5\textwidth]{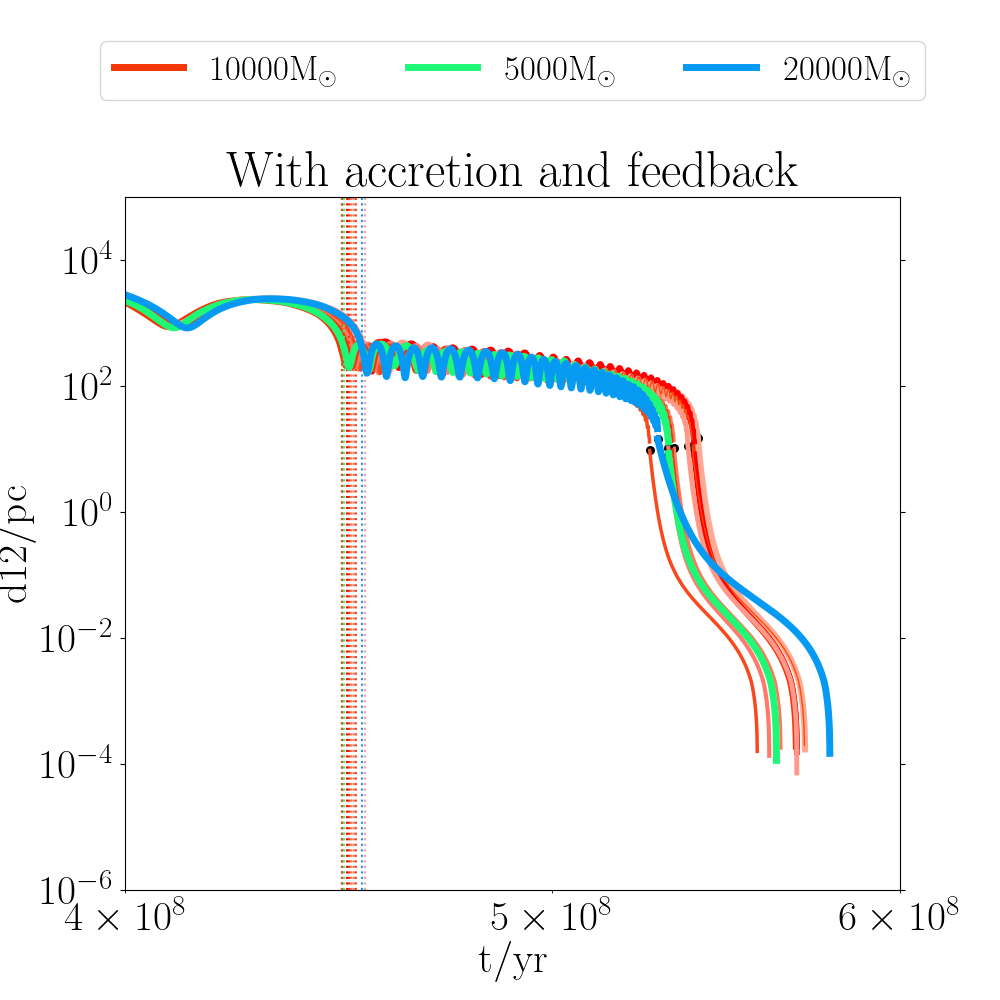}
\caption{The separation evolution for all $10$ runs of simulation M1q0.5fg0.1 (with accretion and feedback) using a stellar particle mass resolution of $10^{4} \rm M_\odot$, and two counterpart simulations in the same setup but with a stellar particle mass resolution of $2\times 10^{4}\, \rm M_\odot$ and $5\times10^{3}\, \rm M_\odot$. }
\label{plot:reso}
\end{figure}

In this section, we present the resolution tests for the merger galaxy M1q0.5fg0.1 (see Table~\ref{tab:params}). Figure~\ref{fig:100_50_d12} shows the evolution of separation at $100$, $50$, and $10$\,pc resolutions, with 10 runs for each resolution, using different initial random seeds for star formation to capture stochastically-driven statistical variations. As shown in both panels, the coalescence times at $50$ and $10$\,pc are similar, while the coalescence time at $100$\,pc is slightly longer than at the higher resolutions. Therefore, resolution effects in this merger experiment saturate at a resolution of $50$\,pc. 
It is interesting to compare these results to those for the isolated galaxies discussed in Sect.~\ref{sec:resolution}. In the isolated case, the trend with resolution is either non-monotonic, as seen in the absence of MBH accretion and feedback (see Fig.~\ref{fig:iso_noa}), or shows increasing coalescence times with better resolution, as in the case with MBH accretion and feedback (i.e., the opposite trend compared to the merger setup, see Fig.~\ref{fig:iso_acc}).
The main difference is that, in the merger setup, the interaction between galaxies generates gravitational torques that efficiently funnel gas through nuclear inflows (see Fig.~\ref{fig:movie}), leading to higher gas densities around the secondary MBH. In contrast, in the isolated case, AGN feedback efficiently clears the gas ($n_{\rm gas} \lesssim 0.1\,\rm H\,cm^{-3}$, see Fig.~\ref{fig:iso_acc}), whereas in the merger case, gas accumulates ($n_{\rm gas} \gtrsim 1\,\rm H\,cm^{-3}$, not shown here) regardless of AGN feedback. This results in a more robust evolution of the MBH pair driven by gas (viscous drag), leading to a more consistent trend with resolution.
Additionally, the stellar densities in the merger case show a similar trend with resolution (not shown here) as in the isolated case, though with slightly larger values due to enhanced nuclear star formation at a given resolution.
Finally, it is worth noting that despite an 8-fold change in spatial resolution (equivalent to a 512-fold change in mass resolution at a given density), the coalescence times vary by no more than $20\%$ in any of the tested cases.

The spread in coalescence time at $10$\,pc and $50$\,pc resolution without accretion and feedback (see the left panel of Fig.~\ref{fig:100_50_d12}) is due to the same factors that explain the time spread at $100$\,pc resolution, as mentioned previously. Without accretion and feedback, the MBH may encounter gas clumps or voids along its trajectory, which can increase or decrease the gaseous DF, thus speeding up or slowing down the orbital decay. There are three runs with relatively longer coalescence times at $10$\,pc resolution without accretion and feedback (left panel of Fig.~\ref{fig:100_50_d12}). This is because the sink particle happens to be in a void when the influence radius is reached, resulting in lower influence density, less efficient orbital decay, and longer coalescence times. In the absence of AGN feedback, we checked that the gas reservoir near the MBHs is unregulated, making it more likely to contain clumps and voids, leading to a wider spread in coalescence times among different galaxy realizations. 

Beside the spatial resolution test we also show a mass resolution test in Fig.~\ref{plot:reso} using a stellar particle mass resolution half and double the original value of $10^{4}\,\rm M_{\odot}$. As shown by the results, the level of convergence in coalescence time is high. The time difference between particle mass of $2\times 10^{4} \,\rm M_\odot$ and $5\times 10^{3}\,\rm M_\odot$  is of the same order of the stochastic noise in galaxy environment realizations represented by the variation between red curves.

As shown in Fig.~\ref{fig:100_50_e}, the eccentricity at coalescence is generally higher at $10$\,pc resolution compared to the other resolutions. Since $10$\,pc resolution is sufficient to resolve fine structures, there is little or no need for compensatory sub-grid density profiles (see second panels of Figs.~\ref{fig:iso_noa} and \ref{fig:iso_acc}). As a result, the stellar and gaseous densities within the resolution sphere at $10$\,pc (at stage 1 and 2) are lower compared to the other resolutions, where sub-grid density profiles are triggered in order to compensate unresolved densities in low-resolution runs. The efficiency of orbital circularization is proportional to these densities, so the lower densities at $10$\,pc lead to slightly higher eccentricities throughout the orbital decay process.

Lastly, Fig.~\ref{fig:100_50_q} illustrates the growth in mass ratio and individual MBH masses. There is a noticeable trend in both mass ratio and mass with resolution, when all runs across resolutions used the same AGN feedback efficiency ($\eta = 0.15$). This is consistent with the findings of \citet{Negri2017} and \citet{Lupi2019}, who suggest that lower AGN feedback efficiencies are required to achieve the same BH evolution at a better spatial resolution. \citet{Biernacki2017} discuss instead how if the same energy is 
injected in a smaller volume where the escape velocity remained the same, a smaller efficiency would be needed at better spatial resolution. Material accumulating in the resolution sphere due to better force resolution can push the resolution-efficiency trend in the other direction, as in our case. In practice, our simulation results align with decreasing feedback efficiency trend. At $10$\,pc resolution, MBH masses grow only minimally when using the same feedback efficiency as the lower resolution runs.

These considerations led us to consider a resolution-dependent AGN feedback efficiency.
In the right panels of Figs.~\ref{fig:100_50_d12} and \ref{fig:100_50_e}, and in both panels of Fig.~\ref{fig:100_50_q}, we also present cases where the AGN efficiency is reduced based on empirical trends observed in previous simulations. Specifically, we set the AGN efficiency to $\eta=0.09$, $0.055$, and $0.005$ for the $100$, $50$, and $10$\,pc resolutions, respectively. The results from these rescaled simulations are overlaid as dashed lines in the left panels of Figs.~\ref{fig:100_50_d12}, \ref{fig:100_50_e}, and \ref{fig:100_50_q}.
These rescaled runs show better convergence, not only in terms of coalescence time but also in the mass ratio for the $100$ and $50$\,pc resolutions. However, for the $10$\,pc resolution, setting the AGN feedback efficiency to $0.005$ appears to be too low, suggesting that further adjustment is necessary for optimal results at this high resolution. Determining the exact value of $\eta$ as a function of resolution is beyond the scope of this paper and will be addressed in future works.

In the right panel of Fig.~\ref{fig:100_50_d12}, when the AGN feedback efficiency $\eta$ is lowered from $0.15$ to $0.005$, the feedback effect weakens, and accretion becomes stronger due to a richer gas reservoir around the MBHs. Stronger accretion leads to a higher binary mass and a larger influence radius. Consequently, the binary crosses the influence radius at a point farther from the center, where the stellar density is lower. According to the simulations, the influence stellar density in the $\eta = 0.005$ run is nearly half of that in the $\eta = 0.15$ case, resulting in a longer coalescence time despite the larger mass and mass ratio. In future developments, we will consider incorporating a time-evolution model for the quantities at the influence radius that affect binary evolution during stage 2.

\begin{table*}
\centering
\caption{Merging galaxy simulation parameters\label{tab:params}}
\begin{tabular}{ccccccccc}
    \hline
    {Simulation} & 
    {$M_{\rm 200,1}$} & 
    {$M_{\rm BH,1}$} & 
    {$q$} & 
    {$f_{\rm g}$} & 
    {$t_{\rm coa}$} & 
    {$f_{\rm DA}$} & 
    {${\Delta q/q}$}\\
    {} &
    {($10^{\rm 11}\,{\rm M_\odot}$)} &
    {($10^{\rm 6}\,{\rm M_\odot}$)} &
    {} & 
    {} & 
    {(${\rm Gyr}$)} & 
    {} & 
    {} \\
    \hline
    \hline
M1q0.5fg0.1   & $1.1$ & $3.3$ & $0.5$ & $0.1$ & $0.57$& $0.81$ & $0.07$ \\
M1q0.1fg0.1   & $1.1$ & $3.3$ & $0.1$ & $0.1$ & $2.17$ & $1.0$ & $0.2$\\
M0.1q0.5fg0.1 & $0.11$ & $0.33$ & $0.5$ & $0.1$ & $1.16$ & $0.81$ & $0.04$ \\
M0.1q0.1fg0.1 & $0.11$ & $0.33$ & $0.1$ & $0.1$ & $2.5$ & $1.0$ & $0.01$ \\
M1q0.1fg0.5   & $1.1$ & $3.3$ & $0.1$ & $0.5$ & $2.7$ & $0.97$ & $0.8$\\
M0.1q0.1fg0.5 & $0.11$ & $0.33$ & $0.1$ & $0.5$ & $4.2$ & $0.94$  & $-0.30$\\
M0.1q0.5fg0.5 & $0.11$ & $0.33$ & $0.5$ & $0.5$ & $0.99$ & $0.8$ & $0.04$ \\
    \hline
\end{tabular}
\end{table*}

\begin{figure*}[t]
  \centering
  \begin{tabular}{@{}cc@{}}
    \includegraphics[width=0.9\textwidth]{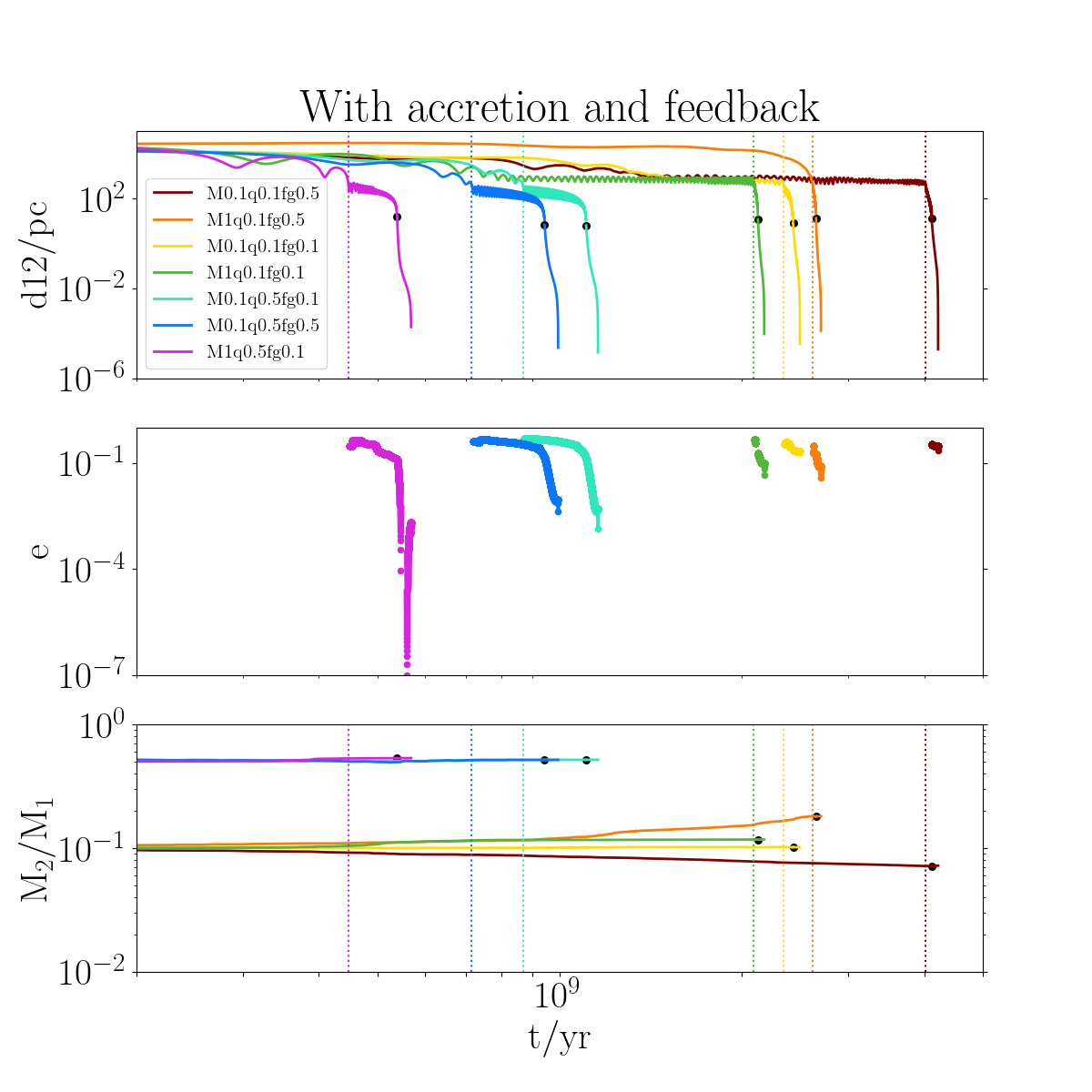}
  \end{tabular}
\caption{The orbital decay process of MBHs in $7$ different galaxy mergers listed in Table~\ref{tab:params}. The vertical dotted lines indicate when the MBHB evolution is handed over to RAMCOAL from RAMSES at the boundary of the resolution sphere. The panels from top to bottom show the time evolution of separation, eccentricity, and mass ratio. }
\label{fig:case1234}
\end{figure*}

Quantitative results from the coalescence simulations are presented in Table~\ref{tab:reso}. The mean coalescence time $\bar t_{\rm coa}$ and fractional standard deviation $\delta t_{\rm coa}=\sigma_{t_{\rm coa}}/\bar t_{\rm coa}$ for each resolution are listed in columns $3-4$. As expected, turning on accretion and feedback shortens the coalescence time (see column 3), as the increase in mass and mass ratio enhances DF. While a higher mass ratio should increase circularization efficiency, the accompanying decrease in gas density acts as a counterbalance, leading to slower overall circularization when MBH accretion and feedback are included.

Without MBH accretion and feedback, the mean coalescence time across all resolutions converges well, showing no significant trend with resolution. The largest fractional standard deviation in coalescence time is $1.7\%$ at $10$\,pc resolution, and the largest fractional discrepancy between resolutions is $6\%$ (between $50$ and $100$\,pc resolution). With MBH accretion and feedback, the largest fractional standard deviation in coalescence time is $1.4\%$ at $10$\,pc resolution, and the largest discrepancy between resolutions is $5\%$ (between $50$ and $100$\,pc resolution). Notably, the fractional discrepancy between $50$ and $10$\,pc runs is minor -- $1.8\%$ without MBH accretion and feedback, and only $0.09\%$ with MBH accretion and feedback.  If we focus on the RAMCOAL part the fractional deviation of the coalescence time is between 5\% and 35\%.  Based on these results, we can conclude that RAMCOAL is robust in modeling the orbital decay and evolution of MBHBs below the resolution limit, all the way to realistic coalescence. Overall, RAMCOAL is largely resolution-independent, with the remaining resolution dependence primarily due to the dependence of AGN feedback efficiency on resolution, setting which is beyond the scope of this paper.

Finally, the mean mass ratio $\bar q$ and fractional standard deviation $\delta q=\sigma_{q}/\bar q$ are  provided in columns $5-6$ of Table~\ref{tab:reso}. As previously mentioned, the feedback efficiency depends on resolution and has a significant impact on the growth of MBH mass and mass ratio. In this table, a constant feedback efficiency is used across all three resolutions, resulting in a trend in mass ratio with resolution. Implementing a reasonable resolution-dependent feedback efficiency would reduce this trend, as shown in Fig.~\ref{fig:100_50_q}. We note that the largest fractional standard deviation in coalescence mass ratio at fixed feedback efficiency is $0.5\%$, indicating a minor impact from the randomness in star formation location on the growth of the mass ratio.

\subsection{Coalescence of massive black holes for different merging galaxies}
\label{subsec:8pairs}

To test RAMCOAL in different galactic environments, we run simulations of all $7$ pairs of galaxies listed in Table~\ref{tab:params} with $100$\,pc resolution with accretion and feedback. The coalescence time is shown in column $6$ of Table~\ref{tab:params}. Figure~\ref{fig:case1234} illustrates the coalescence process of these $7$ pairs of MBHs in different galaxy mergers. 

We mainly focus on the effect of three parameters on the coalescence time: size of the system, mass ratio, and gas fraction. The total mass and size of the system is scaled up and down by changing the halo mass (the second column of Table~\ref{tab:params}), and the mass of MBH (and gas and stars) is also scaled to the halo mass (column $3$ of Table~\ref{tab:params}). According to equations~\ref{eq:bulgeforce}-\ref{eq:maxw}, the DF from stars and DM is proportional to the MBH mass $M_{\rm BH}^{\rm 2}$ and inversely proportional to the velocity dispersion $\sigma_\star^{\rm 3}$. Since $\sigma_\star\propto M_{\rm BH}^{\rm 1/2}$ (through the BH-to-bulge mass relation), the DF from stars and DM is proportional to $M_{\rm BH}^{\rm 1/2}$. Larger MBHs experience larger DF and coalesce faster. For instance, simulation ${\rm M1q0.1fg0.1}$ has a coalescence time twice shorter than that of ${\rm M0.1q0.5fg0.1}$, as the MBHs in the former are ten times more massive.

According to \citet{LBB20b}, gaseous DF can reverse into an acceleration, termed ``dynamical acceleration" (DA), due to AGN feedback under certain conditions (see equation~\ref{eq:DF} and the related discussion). Under these circumstances, the MBHs are accelerated along their orbits, slowing the orbital decay. We show the fraction of time DA is triggered during stage 1 (where DF dominates) in column $7$ of Table~\ref{tab:params}. All simulations with $q=0.5$ are less affected by DA ($f_{\rm DA} \sim 0.8$), meaning DA is triggered for $\sim 80\%$ of their stage 1 evolution time. On the other hand, the $q=0.1$ simulations have their entire stage 1 evolution dominated by DA. This is because low mass ratio systems more easily satisfy the criteria for DA. 
Indeed, in the simulations, the mean MBH velocity and Mach number in low mass ratio systems are nearly $10$ times smaller than their high mass ratio counterparts. 

In these systems, the secondary MBH spends most of its time at the apocenter, where velocity and Mach number are low, and the gas density is also reduced. Consequently, systems with eccentric orbits satisfy the DA conditions (${\cal M}< 4$, and $(1+{\cal M}^2)\,M_{\rm BH}\,n_{\rm \infty} < 10^9\, {\rm M_\odot}\, {\rm cm}^{-3}$) more easily, leading to longer coalescence times. Examples include: simulation ${\rm M1q0.1fg0.1}$ having a much longer coalescence time compared to ${\rm M1q0.5fg0.1}$, and simulation ${\rm M0.1q0.1fg0.1}$ having nearly double the coalescence time of ${\rm M0.1q0.5fg0.1}$. We recall that, however, the dynamical evolution is dominated by stellar dynamical friction, and this is why MBHs coalesce even when $f_{\rm DA}=1$. 

The effect of gas fraction is conditional, depending on whether DA is on or off. When DA is off, a larger gas fraction increases gas density, leading to larger gaseous DF, and therefore shorter coalescence times. However, when DA is on, a higher gas fraction actually prolongs the coalescence time because the gaseous DF reverses direction, accelerating the MBH. Therefore, the effect of gas fraction on coalescence time depends on the DA state. Since low mass ratio systems are more severely affected by DA, increasing the gas fraction in such systems leads to longer coalescence times. This is evident in comparisons like ${\rm M1q0.1fg0.1}$ versus ${\rm M1q0.1fg0.5}$ and ${\rm M0.1q0.1fg0.1}$ versus ${\rm M0.1q0.1fg0.5}$, where higher gas fraction results in longer coalescence times, illustrating the significant role DA plays in the orbital decay of MBH pairs.
In summary, high mass ratio systems are less affected by DA and generally have a coalescence time $1/3$ that of their low mass ratio counterparts.

${\rm M0.1q0.1fg0.5}$ has the longest coalescence time due to a combination of low mass ratio and high gas fraction. The mass ratio at coalescence is $30\%$ smaller than its initial value because the secondary MBH is too small to retain its gas reservoir and is stripped during interactions with the primary MBH. As a result, the secondary MBH does not grow, while the primary MBH nearly doubles in mass. The high gas fraction also amplifies the negative effect of DA, further slowing the orbital decay.

We note that although ${\rm M1q0.1fg0.1}$ has a total galaxy mass $10$ times larger than that of ${\rm M0.1q0.5fg0.1}$, the $5$ times lower mass ratio still leads to a longer coalescence time. This suggests that mass ratio has a greater influence on coalescence time than total galaxy mass. This is further illustrated by the comparison between ${\rm M1q0.1fg0.1}$ and ${\rm M1q0.5fg0.1}$, where the coalescence time of ${\rm M1q0.1fg0.1}$ is nearly $3$ times longer than that of ${\rm M1q0.5fg0.1}$, due to the $5$ times lower mass ratio.

\section{Discussion}
\label{sec:discussion}

Our results demonstrate that RAMCOAL, as implemented in RAMSES, is robust in modeling MBHB coalescence across a range of scales, from galactic environments down to sub-grid, realistic coalescence events. Currently, RAMCOAL stands out as the only code capable of tracking MBHB orbits to coalescence in both gas-poor and gas-rich environments. By enabling the resolution of sub-grid dynamics of MBHs in large-scale galactic mergers simulations, RAMCOAL fills a critical gap left by conventional codes, such as RAMSES, which lack the necessary treatment of MBH dynamics via direct N-body approaches.

The dominant phase of MBH dynamical evolution is generally the dynamical friction part, which has shown good convergence in both RAMSES \citep{Hugo2019} and RAMCOAL, as discussed. Regarding the hardening phase, it depends on the properties at the sphere of influence, which are currently kept fixed over time in Stage 2, but we will include time variation in future updates. The implementation of the MBHB evolution in circumbinary discs has the largest number of parameters, but the time scale at this stage is much shorter than that in the dynamical friction-dominated stage. We therefore consider that the results are not strongly dependent on specific parameter and model choices. We briefly discuss the possible effect on our results due to the model and parameter variation below.

Simulation of MBHBs and their evolution in circumbinary discs is still an area of on-going research. Different simulations have shown quantitative and qualitative difference in the MBHB orbital evolution depending on the parameters and assumptions made in modeling the circumbinary disc. Simulations found that gas within the cavity of the circumbinary discs, including streams and mini-discs around each MBH provide positive torque on the binary, while the gravitational torque from the circumbinary disc on the binary is negative. Thus the total torque can be positive or negative depending on the parameters of the binary and circumbinary disc \citep{C2009, Shi2012, Roedig2011, Roedig2012, DOrazio2013,Farris2014, Munoz2019, Tiede2020, Franchini2022, Dittmann2022, Dittmann2024, Siwek2023}. Some simulations have found the MBHB to out-spiral under some specific configurations \citep{Tang2017, C2009, Moody2019, Munoz2019, Munoz2020}. More recent simulations find the evolution of MBHB orbit depends on the mass ratio, circular binaries with mass ratio larger than $\sim 0.1-0.3$ expand, while those less than $\sim 0.05-0.1$ shrink \citep{Duffell2020, Munoz2020, Siwek2023, Lai2023}. Other simulations have also shown that the evolution of MBHBs depends strongly on the thickness and viscosity of the disc \citep{Tiede2020, Heath2020, Franchini2022,Dittmann2022, Dittmann2024}. Eccentricity is found to be another parameter influencing the expansion of the orbit in many works as well \citep{Munoz2019, Dora2021, Roedig2011, Roedig2012, Siwek2023}.

The assumptions made in modeling the gas also influence the evolution of MBHBs in the circumbinary disc. Simulations which neglect self-gravity in the disc and assume an isothermal equation of state of the gas, find typical binary orbital expansion under a net positive torque \citep{Miranda2017, Tang2017, Moody2019, Munoz2019, Munoz2020, Duffell2020, Tiede2020, Heath2020, Dora2021, Siwek2023, wang2023}. On the other hand, simulations assuming a higher level of $\beta$-cooling of the gas (in which the gas cools radiatively with a local cooling time-scale proportional to the local orbital time $t_{\rm cool}(r) = \beta(G M_{\rm bin} r^{-3})^{-1/2}$, here $\beta \simeq10$ for a high-level of cooling) find shrinking orbits instead \citep{Cuadra2009, Roedig2012, Roedig2014, Franchini2021}. 

In this work, we followed the models in \citet{Haiman2009, Roedig2012, Roedig2014} for the MBHB evolution in circumbinary discs. If the uncertainties in models and parameters above is considered in RAMCOAL, we expect a larger uncertainty in the time scale of stage 2. At this time the lack of consensus among different modeling approaches to circumbinary discs therefore limits the accuracy of RAMCOAL in stage 2. As the understanding of the physical evolution in circumbinary discs improves, we can simply update the equations in RAMCOAL, given its modular and flexible approach.

For the binary hardening in stellar environment, we follow the model in \citet{Sesana2006} in RAMCOAL. Some N-body simulations have found the tidal effects from the merging nuclear star clusters around the MBH accelerate the orbital hardening of MBHBs \citep{Fuji2006, Fell2007, VanOgiya2018, Ogiya2019}. The exchange of angular momentum during the nuclear star cluster merger leads to an expansion of the orbit of the stripped stars. This reduces the angular momentum of the MBHB, and leads to faster orbital decay. 

In \citet{Ogiya2020}, they find the formation of a hard binary occurs faster, accelerating the whole process of orbital decay into the GW regime in the presence of nuclear star clusters through the so-called `Ouroboros effect'. Similar results have also been found in \citet{Muk2024} that the tidal interaction triggered by nuclear star clusters around MBHs generates torque and accelerate the sinking. In this work, we assume both MBHs are in one big sub-grid star cluster, if taking the ``Ouroboros effect" into account, we expect the time scale of stage 2 to be even shorter. 

When the merger rate of galaxies is high, explicitly at high redshifts, it is possible for a third or fourth MBH to inspiral and join the orbital decay of a MBHB. In these situations, the formation of triplets or multiplets can arise, resulting in more complicated few-body dynamics \citep{Mikkola1990, Hoff2007, Amaro2010, Kul2012, Rantala2017, Ryu2018, Bonetti2018, Mannerkoski2021}. The system may undergo  Kozai-Lidov oscillations, which could boost the eccentricity of the central binary and in such way speed up its coalescence \citep{Kozai1962}. Besides the Kozai-Lidov oscillations, the chaotic three-body interactions can also boost the coalescence rate \citep{Bla2002,Hoffman2007, Amaro2010, Kul2012, Bonetti2016,Ryu2018}. More simulations have shown that three-body interactions can bring a sizable number of stalled MBHBs to coalescence when other mechanisms fail to work \citep{Bonetti2018, Bonetti2019}. In this work, we have not considered triple interactions, but they will be added to RAMCOAL in an upcoming work.

Recent works have shown the hardening time scale is in general $\sim 100-500\, \rm Myr$ \citep{Gualandris2022, Liao2024a}. The time scale for hardening and GW emission in our simulations are in general much shorter ($< 100\,\rm Myr$) because of the additional effect of viscous torque in circumbinary discs which is not included in the aforementioned simulations. As shown in the bottom panels of Fig.~\ref{fig:10seed_100pc_d}, the acceleration due to viscous torque dominates the binary hardening for a short period of time ($\sim 10\,\rm Myr$, $\sim 20\%$ of stage 2 evolution time) before GW emission takes over. By turning off the viscous torque completely, the coalescence time is increased by $15\%$, and the binary hardening time becomes $\sim 500\,\rm Myr$ without the effect of viscous torque.

Simulations in \citet{Liao2024a,Liao2024b} show that including the gas cooling and the resulting nuclear star formation at the galactic center  shorten the hardening time scale by a factor of $\sim 1.7$. This is due to the gas condensing at the center of galaxy, leading to nuclear star formation. These newly formed stars inherit low angular momenta from the gas, contribute to the loss cone and further assist in the SMBH hardening via three-body interactions. Our results agrees well with their simulations including gas (the ``CoolStarThmAGN" run with hardening time scales of $\sim 100\,\rm Myr$). However, our time scale is slightly shorter because on top of nuclear star cluster, we also include the the viscous torque from the circumbinary discs on the binary, which as illustrated in bottom panels of Fig.~\ref{fig:10seed_100pc_d}, can be as prominent as the stellar scattering in the final stages of the  binary hardening phase. 

Furthermore, the density profile of individual star cluster around each MBH ($\rho(r)\propto r^{-\gamma}$) can also affect the coalescence time scale. In \citet{Chen2024}, they find the sinking time almost doubled when using a shallower star cluster density profile (changing $\gamma =2$ to $1.5$). We run the M1q0.5fg0.1 simulation at $100$\,pc resolution with the same setup but with a power index of 1.5 instead of 2. Using $\gamma =1.5$ results in a coalescence time nearly twice longer than $\gamma =2$. This result agrees with that in \citet{Chen2024}. From the aspect of observational constraint, by analyzing NANOGrav 15-year data using MCMC and assuming a power-law profile of the galactic center density, \citet{Chenyifan2024} find a parsec-scale galactic center density of $\sim 10^{5}-10^{6} \,\rm M_\odot\,pc^{-3}$ and a power index of $-0.5$. In our work, the size of the star cluster core is not fixed to certain value. It is set to be the same as the molecular cloud core which is determined by the total enclosed mass in the resolution sphere. On average, the radius of the core is $\sim 3$\,pc in simulation M1q0.5fg0.1, and the corresponding  $\rho_{\rm c, star}$ is $\sim 10^{4} \,\rm M_\odot\,pc^{-3}$. If using a $1$\,pc core radius, the central density is $\ge 10^{5} \,\rm M_\odot\,pc^{-3}$, in the same range as predicted using NANOGrav data in \citet{Chenyifan2024}. 

From the aspect of observational constraints, analysis of the Horizon-AGN simulation assuming a central density profile with a power index of -2 is shown to give a good match with observations of nearby galaxy central stellar density (i.e. figure~B1 of \citealp{Marta2020}). In this work, we use $\gamma =2$ for simplicity and generalization. We defer a more detailed study of the effect of $\gamma$ to future works. In our work, the size of the star cluster core is not fixed to certain value. It is set to be the same as the molecular cloud core which is determined by the total enclosed mass in the resolution sphere. On average, the radius of the core is $\sim 3 $ pc in simulation M1q0.5fg0.1, and the corresponding  $\rho_{\rm c, star}$ is $\sim 10^{4} M_{\odot}\rm\,pc^{-3}$.

Complementing RAMCOAL, the KETJU code implemented in GADGET-4 provides a detailed and high-accuracy simulation of SMBH dynamics, particularly in stellar environments, incorporating post-Newtonian corrections \citep{Matias2023}. KETJU integrates the dynamics of SMBHs and surrounding stars using a regularized algorithm \citep{Rantala2017,Rantala2020}, while the broader particle interactions are computed using GADGET-4’s fast multipole method \citep{Springel2021}. 
The DF from collision-less particles, loss-cone scattering, and GW emission are taken into account in the MBHB orbital decay process in KETJU. Preferential accretion in circumbinary discs and AGN feedback are added into KETJU in RABBITS \citep{Liao2024a, Liao2024b}. Recently, KETJU has also been used in high resolution re-simulations of galaxies extracted from cosmological simulations \citep{2024OJAp....7E..28C}.  The dynamical evolution of MBHB in circumbinary discs is however not included in RABBITS and in the published simulations the gas softening length is $20\,{\rm pc}$, without an explicit gas DF added to account for unresolved DF, which typically contributes when the Bondi radius is not resolved \citep{Beckmann2018}. This emphasizes how KETJU and RAMCOAL are complementary in treating MBH dynamics in different environments.

Comparing the orbital decay and eccentricity evolution of MBHBs between RAMCOAL and KETJU, we find similar trends. Our results (Figs.~\ref{fig:100_50_d12} and \ref{fig:100_50_e}) align with the findings in \citet{Matias2023} and \citet{Liao2024b}, particularly regarding separation evolution. Notably, our simulations display an increasing eccentricity driven by viscous drag in circumbinary disks before GW emission begins to circularize the orbit -- an effect not seen in KETJU (figure 9 of \citealp{Liao2024a}), where viscous drag from circumbinary disks has yet to be fully incorporated. This highlights the importance of gas dynamics in MBHB evolution.

Eccentricity evolution is a key factor in understanding MBHB mergers, and our findings are consistent with previous studies. \citet{Rawling2023} showed that MBHB eccentricities in equal-mass galaxy mergers can vary significantly due to parsec-scale variations in the merger orbit, leading to eccentricities spanning nearly the full range from 0 to 1. We observe similar variations, with eccentricities strongly influenced by the specific characteristics of each galaxy, such as nuclear gas and substructures, which perturb the merger orbits. This variability underlines the sensitivity of MBHB dynamics to initial conditions, reinforcing the conclusions drawn by \citet{Rawling2023}.

In terms of computational performance, RAMCOAL's integration into RAMSES proves efficient. As a sub-grid model, RAMCOAL retrieves only the galactic properties around the center of mass from RAMSES, resulting in negligible additional computational overhead. 
For instance,  in our M1q0.5fg0.1 galaxy merger simulation (Sect.~\ref{subsec:8pairs}), the measured CPU time per coarse time step of the simulation at $100\,{\rm pc}$ resolution running with $128$ CPU cores is 2.94\,seconds in RAMSES before the two MBHs enter RAMCOAL, and is 3.06\,seconds during the RAMCOAL stage. According to our results, for $100$, $50$, and $10$\,pc resolutions, adding RAMCOAL to RAMSES does not significantly increase the CPU time of the simulation (only by 4\%). 
This efficiency would allow RAMCOAL to scale effectively in large cosmological simulations, ensuring that the simulation of MBHB coalescence does not hinder the broader computational demands of cosmological models.

Overall, RAMCOAL and KETJU are complementary tools that provide a comprehensive approach to modeling MBH dynamics. KETJU’s strength lies in its detailed treatment of stellar dynamics, albeit at a higher computational cost, while RAMCOAL focuses on gas-rich environments, circumbinary dynamics, and computational efficiency. Together, they offer a multi-faceted understanding of MBHB coalescence across different environments.

\section{Conclusions}
\label{sec:conclusion}

In this work, we introduced the RAMCOAL module for simulating the sub-grid dynamics of MBHBs within the RAMSES code. We demonstrated that RAMCOAL can track the entire coalescence process of MBHBs in real-time simulations without adding significant computational overhead. At present, RAMCOAL is the only code capable of realistically simulating the dynamics of all MBHBs down to coalescence in any environment, gas rich and gas poor. When introduced in cosmological simulations, it will allow for a significant advancement for studying MBH evolution in a realistic galactic context.

We tested RAMCOAL in both isolated galaxy and galaxy merger simulations at resolutions of $10$, $50$, and $100$\,pc, with and without MBH accretion and feedback. Our key findings include:
\begin{enumerate}
\item[-] The MBHB coalescence times at different resolutions are consistent, with only a fractional deviation of 6\% without MBH accretion and feedback, and 5\% when including these processes for the full merger. Notably, no strong trend was observed between resolution and coalescence time, with a difference less than 10\% variation with spatial resolutions varying by a factor of 8.
\item[-] The presence of MBH accretion and  feedback shortens the coalescence time, as these processes increase the mass of the MBHs and the mass ratio, accelerating the DF and thus the coalescence.
\item[-] Different realizations of galactic environments, such as varying initial star formation random seeds, have minimal impact on the MBHB coalescence time, with the largest deviation being only 1.7\%.
\item[-] Viscous drag in the circumbinary disk leads to a temporary increase in eccentricity before GW emission eventually circularizes the orbit. This eccentricity boost could enhance the detectability of MBHBs by LISA.
\item[-] Eccentricity is highly sensitive to fluctuations in the interstellar gas, showing a fractional deviation of about 30\%.
\item[-] The mass ratio of the MBHBs is largely unaffected by resolution or small stochastic effects, with deviations of less than 1\%.
\item[-] AGN feedback efficiency $\eta$ must be carefully calibrated depending on resolution. Based on our results, for $10$\,pc resolution, the efficiency parameter $\eta$ should be within the range of $[0.005, 0.034]$, while for $50$ and $100$\,pc resolution, $\eta$ values of $0.055$ and $0.09$ respectively lead to convergence in the MBH mass growth.
\end{enumerate}

These results highlight the robustness and efficiency of RAMCOAL in simulating the coalescence of MBHBs within realistic galactic environments. Unlike post-processing approaches, RAMCOAL predicts key orbital parameters such as mass ratio and eccentricity on-the-fly, thus reducing uncertainties and allowing for more accurate forecasts of gravitational waveforms and electromagnetic signatures. This capability is crucial for preparing future observational campaigns, particularly in the context of the upcoming multi-messenger era of astronomy.

RAMCOAL offers a significant computational advantage, making it suitable for large-scale cosmological simulations that require efficient MBHB coalescence tracking. It is capable of generating MBHB coalescence catalogs, providing essential data for predicting GW signals and electromagnetic counterparts~\citep[see][for a post-processing approach]{DongPaez2023}.

Future work will focus on expanding RAMCOAL's capabilities, including adding multiple MBH interactions and coupling it with the MBH spin evolution model of RAMSES~\citep{Dubois2014b,Dubois2021}. These enhancements will allow us to better predict the coalescence rate and GW background in cosmological simulations, and to further refine our predictions for the electromagnetic counterparts of MBH mergers that will be observable by facilities like LISA and PTAs, bridging the gap between electromagnetic and GW astronomy.

\begin{acknowledgements}
KL, MV, and YD acknowledge funding from the French National Research Agency (grant ANR-21-CE31-0026, project MBH\_waves). This work has received funding from the Centre National d’Etudes Spatiales.
RB acknowledges support from a UKRI Future Leaders Fellowship (grant code: MR/Y015517/1). This work has made use of the Infinity Cluster hosted by Institut d’Astrophysique de Paris; we thank Stéphane Rouberol for running smoothly this cluster for us. 

\end{acknowledgements}


\begin{thebibliography}{132}
\expandafter\ifx\csname natexlab\endcsname\relax\def\natexlab#1{#1}\fi

\bibitem[{{Abbott} {et~al.}(2016){Abbott}, {Abbott}, {Abbott}, {Abernathy},
  {Acernese}, {Ackley}, {Adams}, {Adams}, {Addesso}, {Adhikari}, {Adya},
  {Affeldt}, {Agathos}, {Agatsuma}, {Aggarwal}, {Aguiar}, {Aiello}, {Ain},
  {Ajith}, {Allen}, {Allocca}, {Altin}, {Anderson}, {Anderson}, {Arai},
  {Arain}, {Araya}, {Arceneaux}, {Areeda}, {Arnaud}, {Arun}, {Ascenzi},
  {Ashton}, {Ast}, {Aston}, {Astone}, {Aufmuth}, {Aulbert}, {Babak}, {Bacon},
  {Bader}, {Baker}, {Baldaccini}, {Ballardin}, {Ballmer}, {Barayoga},
  {Barclay}, {Barish}, {Barker}, {Barone}, {Barr}, {Barsotti}, {Barsuglia},
  {Barta}, {Bartlett}, {Barton}, {Bartos}, {Bassiri}, {Basti}, {Batch},
  {Baune}, {Bavigadda}, {Bazzan}, {Behnke}, {Bejger}, {Belczynski}, {Bell},
  {Bell}, {Berger}, {Bergman}, {Bergmann}, {Berry}, {Bersanetti}, {Bertolini},
  {Betzwieser}, {Bhagwat}, {Bhandare}, {Bilenko}, {Billingsley}, {Birch},
  {Birney}, {Birnholtz}, {Biscans}, {Bisht}, {Bitossi}, {Biwer}, {Bizouard},
  {Blackburn}, {Blair}, {Blair}, {Blair}, {Bloemen}, {Bock}, {Bodiya}, {Boer},
  {Bogaert}, {Bogan}, {Bohe}, {Bojtos}, {Bond}, {Bondu}, {Bonnand}, {Boom},
  {Bork}, {Boschi}, {Bose}, {Bouffanais}, {Bozzi}, {Bradaschia}, {Brady},
  {Braginsky}, {Branchesi}, {Brau}, {Briant}, {Brillet}, {Brinkmann},
  {Brisson}, {Brockill}, {Brooks}, {Brown}, {Brown}, {Brown}, {Buchanan},
  {Buikema}, {Bulik}, {Bulten}, {Buonanno}, {Buskulic}, {Buy}, {Byer},
  {Cabero}, {Cadonati}, {Cagnoli}, {Cahillane}, {Bustillo}, {Callister},
  {Calloni}, {Camp}, {Cannon}, {Cao}, {Capano}, {Capocasa}, {Carbognani},
  {Caride}, {Casanueva Diaz}, {Casentini}, {Caudill}, {Cavagli{\`a}},
  {Cavalier}, {Cavalieri}, {Cella}, {Cepeda}, {Baiardi}, {Cerretani},
  {Cesarini}, {Chakraborty}, {Chalermsongsak}, {Chamberlin}, {Chan}, {Chao},
  {Charlton}, {Chassand e-Mottin}, {Chen}, {Chen}, {Cheng}, {Chincarini},
  {Chiummo}, {Cho}, {Cho}, {Chow}, {Christensen}, {Chu}, {Chua}, {Chung},
  {Ciani}, {Clara}, {Clark}, {Cleva}, {Coccia}, {Cohadon}, {Colla}, {Collette},
  {Cominsky}, {Constancio}, {Conte}, {Conti}, {Cook}, {Corbitt}, {Cornish},
  {Corsi}, {Cortese}, {Costa}, {Coughlin}, {Coughlin}, {Coulon}, {Countryman},
  {Couvares}, {Cowan}, {Coward}, {Cowart}, {Coyne}, {Coyne}, {Craig},
  {Creighton}, {Creighton}, {Cripe}, {Crowder}, {Cruise}, {Cumming},
  {Cunningham}, {Cuoco}, {Dal Canton}, {Danilishin}, {D'Antonio}, {Danzmann},
  {Darman}, {Da Silva Costa}, {Dattilo}, {Dave}, {Daveloza}, {Davier},
  {Davies}, {Daw}, {Day}, {De}, {DeBra}, {Debreczeni}, {Degallaix}, {De
  Laurentis}, {Del{\'e}glise}, {Del Pozzo}, {Denker}, {Dent}, {Dereli},
  {Dergachev}, {DeRosa}, {De Rosa}, {DeSalvo}, {Dhurandhar}, {D{\'\i}az}, {Di
  Fiore}, {Di Giovanni}, {Di Lieto}, {Di Pace}, {Di Palma}, {Di Virgilio},
  {Dojcinoski}, {Dolique}, {Donovan}, {Dooley}, {Doravari}, {Douglas},
  {Downes}, {Drago}, {Drever}, {Driggers}, {Du}, {Ducrot}, {Dwyer}, {Edo},
  {Edwards}, {Effler}, {Eggenstein}, {Ehrens}, {Eichholz}, {Eikenberry},
  {Engels}, {Essick}, {Etzel}, {Evans}, {Evans}, {Everett}, {Factourovich},
  {Fafone}, {Fair}, {Fairhurst}, {Fan}, {Fang}, {Farinon}, {Farr}, {Farr},
  {Favata}, {Fays}, {Fehrmann}, {Fejer}, {Feldbaum}, {Ferrante}, {Ferreira},
  {Ferrini}, {Fidecaro}, {Finn}, {Fiori}, {Fiorucci}, {Fisher}, {Flaminio},
  {Fletcher}, {Fong}, {Fournier}, {Franco}, {Frasca}, {Frasconi}, {Frede},
  {Frei}, {Freise}, {Frey}, {Frey}, {Fricke}, {Fritschel}, {Frolov}, {Fulda},
  {Fyffe}, {Gabbard}, {Gair}, {Gammaitoni}, {Gaonkar}, {Garufi}, {Gatto},
  {Gaur}, {Gehrels}, {Gemme}, {Gendre}, {Genin}, {Gennai}, {George}, {Gergely},
  {Germain}, {Ghosh}, {Ghosh}, {Ghosh}, {Giaime}, {Giardina}, {Giazotto},
  {Gill}, {Glaefke}, {Gleason}, {Goetz}, {Goetz}, {Gondan}, {Gonz{\'a}lez},
  {Castro}, {Gopakumar}, {Gordon}, {Gorodetsky}, {Gossan}, {Gosselin},
  {Gouaty}, {Graef}, {Graff}, {Granata}, {Grant}, {Gras}, {Gray}, {Greco},
  {Green}, {Greenhalgh}, {Groot}, {Grote}, {Grunewald}, {Guidi}, {Guo},
  {Gupta}, {Gupta}, {Gushwa}, {Gustafson}, {Gustafson}, {Hacker}, {Hall},
  {Hall}, {Hammond}, {Haney}, {Hanke}, {Hanks}, {Hanna}, {Hannam}, {Hanson},
  {Hardwick}, {Harms}, {Harry}, {Harry}, {Hart}, {Hartman}, {Haster},
  {Haughian}, {Healy}, {Heefner}, {Heidmann}, {Heintze}, {Heinzel}, {Heitmann},
  {Hello}, {Hemming}, {Hendry}, {Heng}, {Hennig}, {Heptonstall}, {Heurs},
  {Hild}, {Hoak}, {Hodge}, {Hofman}, {Hollitt}, {Holt}, {Holz}, {Hopkins},
  {Hosken}, {Hough}, {Houston}, {Howell}, {Hu}, {Huang}, {Huerta}, {Huet},
  {Hughey}, {Husa}, {Huttner}, {Huynh-Dinh}, {Idrisy}, {Indik}, {Ingram},
  {Inta}, {Isa}, {Isac}, {Isi}, {Islas}, {Isogai}, {Iyer}, {Izumi}, {Jacobson},
  {Jacqmin}, {Jang}, {Jani}, {Jaranowski}, {Jawahar}, {Jim{\'e}nez-Forteza},
  {Johnson}, {Johnson-McDaniel}, {Jones}, {Jones}, {Jonker}, {Ju}, {Haris},
  {Kalaghatgi}, {Kalogera}, {Kandhasamy}, {Kang}, {Kanner}, {Karki},
  {Kasprzack}, {Katsavounidis}, {Katzman}, {Kaufer}, {Kaur}, {Kawabe},
  {Kawazoe}, {K{\'e}f{\'e}lian}, {Kehl}, {Keitel}, {Kelley}, {Kells},
  {Kennedy}, {Keppel}, {Key}, {Khalaidovski}, {Khalili}, {Khan}, {Khan},
  {Khan}, {Khazanov}, {Kijbunchoo}, {Kim}, {Kim}, {Kim}, {Kim}, {Kim}, {Kim},
  {King}, {King}, {Kinzel}, {Kissel}, {Kleybolte}, {Klimenko}, {Koehlenbeck},
  {Kokeyama}, {Koley}, {Kondrashov}, {Kontos}, {Koranda}, {Korobko}, {Korth},
  {Kowalska}, {Kozak}, {Kringel}, {Krishnan}, {Kr{\'o}lak}, {Krueger}, {Kuehn},
  {Kumar}, {Kumar}, {Kuo}, {Kutynia}, {Kwee}, {Lackey}, {Landry}, {Lange},
  {Lantz}, {Lasky}, {Lazzarini}, {Lazzaro}, {Leaci}, {Leavey}, {Lebigot},
  {Lee}, {Lee}, {Lee}, {Lee}, {Lenon}, {Leonardi}, {Leong}, {Leroy},
  {Letendre}, {Levin}, {Levine}, {Li}, {Libson}, {Littenberg}, {Lockerbie},
  {Logue}, {Lombardi}, {London}, {Lord}, {Lorenzini}, {Loriette}, {Lormand},
  {Losurdo}, {Lough}, {Lousto}, {Lovelace}, {L{\"u}ck}, {Lundgren}, {Luo},
  {Lynch}, {Ma}, {MacDonald}, {Machenschalk}, {MacInnis}, {Macleod},
  {Maga{\~n}a-Sandoval}, {Magee}, {Mageswaran}, {Majorana}, {Maksimovic},
  {Malvezzi}, {Man}, {Mandel}, {Mandic}, {Mangano}, {Mansell}, {Manske},
  {Mantovani}, {Marchesoni}, {Marion}, {M{\'a}rka}, {M{\'a}rka}, {Markosyan},
  {Maros}, {Martelli}, {Martellini}, {Martin}, {Martin}, {Martynov}, {Marx},
  {Mason}, {Masserot}, {Massinger}, {Masso-Reid}, {Matichard}, {Matone},
  {Mavalvala}, {Mazumder}, {Mazzolo}, {McCarthy}, {McClelland}, {McCormick},
  {McGuire}, {McIntyre}, {McIver}, {McManus}, {McWilliams}, {Meacher},
  {Meadors}, {Meidam}, {Melatos}, {Mendell}, {Mendoza-Gandara}, {Mercer},
  {Merilh}, {Merzougui}, {Meshkov}, {Messenger}, {Messick}, {Meyers},
  {Mezzani}, {Miao}, {Michel}, {Middleton}, {Mikhailov}, {Milano}, {Miller},
  {Millhouse}, {Minenkov}, {Ming}, {Mirshekari}, {Mishra}, {Mitra},
  {Mitrofanov}, {Mitselmakher}, {Mittleman}, {Moggi}, {Mohan}, {Mohapatra},
  {Montani}, {Moore}, {Moore}, {Moraru}, {Moreno}, {Morriss}, {Mossavi},
  {Mours}, {Mow-Lowry}, {Mueller}, {Mueller}, {Muir}, {Mukherjee}, {Mukherjee},
  {Mukherjee}, {Mukund}, {Mullavey}, {Munch}, {Murphy}, {Murray}, {Mytidis},
  {Nardecchia}, {Naticchioni}, {Nayak}, {Necula}, {Nedkova}, {Nelemans},
  {Neri}, {Neunzert}, {Newton}, {Nguyen}, {Nielsen}, {Nissanke}, {Nitz},
  {Nocera}, {Nolting}, {Normandin}, {Nuttall}, {Oberling}, {Ochsner}, {O'Dell},
  {Oelker}, {Ogin}, {Oh}, {Oh}, {Ohme}, {Oliver}, {Oppermann}, {Oram},
  {O'Reilly}, {O'Shaughnessy}, {Ott}, {Ottaway}, {Ottens}, {Overmier}, {Owen},
  {Pai}, {Pai}, {Palamos}, {Palashov}, {Palomba}, {Pal-Singh}, {Pan}, {Pan},
  {Pankow}, {Pannarale}, {Pant}, {Paoletti}, {Paoli}, {Papa}, {Paris},
  {Parker}, {Pascucci}, {Pasqualetti}, {Passaquieti}, {Passuello},
  {Patricelli}, {Patrick}, {Pearlstone}, {Pedraza}, {Pedurand }, {Pekowsky},
  {Pele}, {Penn}, {Perreca}, {Pfeiffer}, {Phelps}, {Piccinni}, {Pichot},
  {Pickenpack}, {Piergiovanni}, {Pierro}, {Pillant}, {Pinard}, {Pinto},
  {Pitkin}, {Poeld}, {Poggiani}, {Popolizio}, {Post}, {Powell}, {Prasad},
  {Predoi}, {Premachandra}, {Prestegard}, {Price}, {Prijatelj}, {Principe},
  {Privitera}, {Prix}, {Prodi}, {Prokhorov}, {Puncken}, {Punturo}, {Puppo},
  {P{\"u}rrer}, {Qi}, {Qin}, {Quetschke}, {Quintero}, {Quitzow-James}, {Raab},
  {Rabeling}, {Radkins}, {Raffai}, {Raja}, {Rakhmanov}, {Ramet}, {Rapagnani},
  {Raymond}, {Razzano}, {Re}, {Read}, {Reed}, {Regimbau}, {Rei}, {Reid},
  {Reitze}, {Rew}, {Reyes}, {Ricci}, {Riles}, {Robertson}, {Robie}, {Robinet},
  {Rocchi}, {Rolland}, {Rollins}, {Roma}, {Romano}, {Romano}, {Romanov},
  {Romie}, {Rosi{\'n}ska}, {Rowan}, {R{\"u}diger}, {Ruggi}, {Ryan}, {Sachdev},
  {Sadecki}, {Sadeghian}, {Salconi}, {Saleem}, {Salemi}, {Samajdar}, {Sammut},
  {Sampson}, {Sanchez}, {Sandberg}, {Sandeen}, {Sand ers}, {Sanders},
  {Sassolas}, {Sathyaprakash}, {Saulson}, {Sauter}, {Savage}, {Sawadsky},
  {Schale}, {Schilling}, {Schmidt}, {Schmidt}, {Schnabel}, {Schofield},
  {Sch{\"o}nbeck}, {Schreiber}, {Schuette}, {Schutz}, {Scott}, {Scott},
  {Sellers}, {Sengupta}, {Sentenac}, {Sequino}, {Sergeev}, {Serna},
  {Setyawati}, {Sevigny}, {Shaddock}, {Shaffer}, {Shah}, {Shahriar}, {Shaltev},
  {Shao}, {Shapiro}, {Shawhan}, {Sheperd}, {Shoemaker}, {Shoemaker}, {Siellez},
  {Siemens}, {Sigg}, {Silva}, {Simakov}, {Singer}, {Singer}, {Singh}, {Singh},
  {Singhal}, {Sintes}, {Slagmolen}, {Smith}, {Smith}, {Smith}, {Smith}, {Son},
  {Sorazu}, {Sorrentino}, {Souradeep}, {Srivastava}, {Staley}, {Steinke},
  {Steinlechner}, {Steinlechner}, {Steinmeyer}, {Stephens}, {Stevenson},
  {Stone}, {Strain}, {Straniero}, {Stratta}, {Strauss}, {Strigin}, {Sturani},
  {Stuver}, {Summerscales}, {Sun}, {Sutton}, {Swinkels}, {Szczepa{\'n}czyk},
  {Tacca}, {Talukder}, {Tanner}, {T{\'a}pai}, {Tarabrin}, {Taracchini},
  {Taylor}, {Theeg}, {Thirugnanasambandam}, {Thomas}, {Thomas}, {Thomas},
  {Thorne}, {Thorne}, {Thrane}, {Tiwari}, {Tiwari}, {Tokmakov}, {Tomlinson},
  {Tonelli}, {Torres}, {Torrie}, {T{\"o}yr{\"a}}, {Travasso}, {Traylor},
  {Trifir{\`o}}, {Tringali}, {Trozzo}, {Tse}, {Turconi}, {Tuyenbayev},
  {Ugolini}, {Unnikrishnan}, {Urban}, {Usman}, {Vahlbruch}, {Vajente},
  {Valdes}, {Vallisneri}, {van Bakel}, {van Beuzekom}, {van den Brand}, {Van
  Den Broeck}, {Vand er-Hyde}, {van der Schaaf}, {van Heijningen}, {van
  Veggel}, {Vardaro}, {Vass}, {Vas{\'u}th}, {Vaulin}, {Vecchio}, {Vedovato},
  {Veitch}, {Veitch}, {Venkateswara}, {Verkindt}, {Vetrano}, {Vicer{\'e}},
  {Vinciguerra}, {Vine}, {Vinet}, {Vitale}, {Vo}, {Vocca}, {Vorvick}, {Voss},
  {Vousden}, {Vyatchanin}, {Wade}, {Wade}, {Wade}, {Waldman}, {Walker},
  {Wallace}, {Walsh}, {Wang}, {Wang}, {Wang}, {Wang}, {Wang}, {Ward}, {Ward},
  {Warner}, {Was}, {Weaver}, {Wei}, {Weinert}, {Weinstein}, {Weiss}, {Welborn},
  {Wen}, {We{\ss}els}, {Westphal}, {Wette}, {Whelan}, {Whitcomb}, {White},
  {Whiting}, {Wiesner}, {Wilkinson}, {Willems}, {Williams}, {Williams},
  {Williamson}, {Willis}, {Willke}, {Wimmer}, {Winkelmann}, {Winkler}, {Wipf},
  {Wiseman}, {Wittel}, {Woan}, {Worden}, {Wright}, {Wu}, {Yablon}, {Yakushin},
  {Yam}, {Yamamoto}, {Yancey}, {Yap}, {Yu}, {Yvert}, {Zadro{\.Z}ny},
  {Zangrando}, {Zanolin}, {Zendri}, {Zevin}, {Zhang}, {Zhang}, {Zhang},
  {Zhang}, {Zhao}, {Zhou}, {Zhou}, {Zhu}, {Zucker}, {Zuraw}, {Zweizig}, {LIGO
  Scientific Collaboration}, \& {Virgo Collaboration}}]{LIGO2016}
{Abbott}, B.~P., {Abbott}, R., {Abbott}, T.~D., {et~al.} 2016, \prl, 116,
  061102

\bibitem[{Agazie {et~al.}(2023)Agazie, Anumarlapudi, Archibald, Arzoumanian,
  Baker, Bécsy, Blecha, Brazier, Brook, Burke-Spolaor, Burnette, Case,
  Charisi, Chatterjee, Chatziioannou, Cheeseboro, Chen, Cohen, Cordes, Cornish,
  Crawford, Cromartie, Crowter, Cutler, DeCesar, DeGan, Demorest, Deng, Dolch,
  Drachler, Ellis, Ferrara, Fiore, Fonseca, Freedman, Garver-Daniels, Gentile,
  Gersbach, Glaser, Good, Gültekin, Hazboun, Hourihane, Islo, Jennings,
  Johnson, Jones, Kaiser, Kaplan, Kelley, Kerr, Key, Klein, Laal, Lam, Lamb,
  Lazio, Lewandowska, Littenberg, Liu, Lommen, Lorimer, Luo, Lynch, Ma,
  Madison, Mattson, McEwen, McKee, McLaughlin, McMann, Meyers, Meyers,
  Mingarelli, Mitridate, Natarajan, Ng, Nice, Ocker, Olum, Pennucci, Perera,
  Petrov, Pol, Radovan, Ransom, Ray, Romano, Sardesai, Schmiedekamp,
  Schmiedekamp, Schmitz, Schult, Shapiro-Albert, Siemens, Simon, Siwek, Stairs,
  Stinebring, Stovall, Sun, Susobhanan, Swiggum, Taylor, Taylor, Turner, Unal,
  Vallisneri, van Haasteren, Vigeland, Wahl, Wang, Witt, Young, \&
  Collaboration}]{Agazie_2023}
Agazie, G., Anumarlapudi, A., Archibald, A.~M., {et~al.} 2023, \apjl, 951, L8

\bibitem[{{Amaro-Seoane} {et~al.}(2023){Amaro-Seoane}, {Andrews}, {Arca Sedda},
  {Askar}, {Baghi}, {Balasov}, {Bartos}, {Bavera}, {Bellovary}, {Berry},
  {Berti}, {Bianchi}, {Blecha}, {Blondin}, {Bogdanovi{\'c}}, {Boissier},
  {Bonetti}, {Bonoli}, {Bortolas}, {Breivik}, {Capelo}, {Caramete},
  {Cattorini}, {Charisi}, {Chaty}, {Chen}, {Chru{\'s}li{\'n}ska}, {Chua},
  {Church}, {Colpi}, {D'Orazio}, {Danielski}, {Davies}, {Dayal}, {De Rosa},
  {Derdzinski}, {Destounis}, {Dotti}, {Dutan}, {Dvorkin}, {Fabj}, {Foglizzo},
  {Ford}, {Fouvry}, {Franchini}, {Fragos}, {Fryer}, {Gaspari}, {Gerosa},
  {Graziani}, {Groot}, {Habouzit}, {Haggard}, {Haiman}, {Han}, {Istrate},
  {Johansson}, {Khan}, {Kimpson}, {Kokkotas}, {Kong}, {Korol}, {Kremer},
  {Kupfer}, {Lamberts}, {Larson}, {Lau}, {Liu}, {Lloyd-Ronning}, {Lodato},
  {Lupi}, {Ma}, {Maccarone}, {Mandel}, {Mangiagli}, {Mapelli}, {Mathis},
  {Mayer}, {McGee}, {McKernan}, {Miller}, {Mota}, {Mumpower}, {Nasim},
  {Nelemans}, {Noble}, {Pacucci}, {Panessa}, {Paschalidis}, {Pfister},
  {Porquet}, {Quenby}, {Ricarte}, {R{\"o}pke}, {Regan}, {Rosswog}, {Ruiter},
  {Ruiz}, {Runnoe}, {Schneider}, {Schnittman}, {Secunda}, {Sesana}, {Seto},
  {Shao}, {Shapiro}, {Sopuerta}, {Stone}, {Suvorov}, {Tamanini}, {Tamfal},
  {Tauris}, {Temmink}, {Tomsick}, {Toonen}, {Torres-Orjuela}, {Toscani},
  {Tsokaros}, {Unal}, {V{\'a}zquez-Aceves}, {Valiante}, {van Putten}, {van
  Roestel}, {Vignali}, {Volonteri}, {Wu}, {Younsi}, {Yu}, {Zane}, {Zwick},
  {Antonini}, {Baibhav}, {Barausse}, {Bonilla Rivera}, {Branchesi},
  {Branduardi-Raymont}, {Burdge}, {Chakraborty}, {Cuadra}, {Dage}, {Davis}, {de
  Mink}, {Decarli}, {Doneva}, {Escoffier}, {Gandhi}, {Haardt}, {Lousto},
  {Nissanke}, {Nordhaus}, {O'Shaughnessy}, {Portegies Zwart}, {Pound},
  {Schussler}, {Sergijenko}, {Spallicci}, {Vernieri}, \&
  {Vigna-G{\'o}mez}}]{Amaro2023}
{Amaro-Seoane}, P., {Andrews}, J., {Arca Sedda}, M., {et~al.} 2023, Living
  Reviews in Relativity, 26, 2

\bibitem[{{Amaro-Seoane} {et~al.}(2017){Amaro-Seoane}, {Audley}, {Babak},
  {Baker}, {Barausse}, {Bender}, {Berti}, {Binetruy}, {Born}, {Bortoluzzi},
  {Camp}, {Caprini}, {Cardoso}, {Colpi}, {Conklin}, {Cornish}, {Cutler},
  {Danzmann}, {Dolesi}, {Ferraioli}, {Ferroni}, {Fitzsimons}, {Gair}, {Gesa
  Bote}, {Giardini}, {Gibert}, {Grimani}, {Halloin}, {Heinzel}, {Hertog},
  {Hewitson}, {Holley-Bockelmann}, {Hollington}, {Hueller}, {Inchauspe},
  {Jetzer}, {Karnesis}, {Killow}, {Klein}, {Klipstein}, {Korsakova}, {Larson},
  {Livas}, {Lloro}, {Man}, {Mance}, {Martino}, {Mateos}, {McKenzie},
  {McWilliams}, {Miller}, {Mueller}, {Nardini}, {Nelemans}, {Nofrarias},
  {Petiteau}, {Pivato}, {Plagnol}, {Porter}, {Reiche}, {Robertson},
  {Robertson}, {Rossi}, {Russano}, {Schutz}, {Sesana}, {Shoemaker}, {Slutsky},
  {Sopuerta}, {Sumner}, {Tamanini}, {Thorpe}, {Troebs}, {Vallisneri},
  {Vecchio}, {Vetrugno}, {Vitale}, {Volonteri}, {Wanner}, {Ward}, {Wass},
  {Weber}, {Ziemer}, \& {Zweifel}}]{LISA2017}
{Amaro-Seoane}, P., {Audley}, H., {Babak}, S., {et~al.} 2017, arXiv e-prints,
  arXiv:1702.00786

\bibitem[{{Amaro-Seoane} {et~al.}(2010){Amaro-Seoane}, {Sesana}, {Hoffman},
  {Benacquista}, {Eichhorn}, {Makino}, \& {Spurzem}}]{Amaro2010}
{Amaro-Seoane}, P., {Sesana}, A., {Hoffman}, L., {et~al.} 2010, \mnras, 402,
  2308

\bibitem[{{Antonini} \& {Merritt}(2012)}]{AM2012}
{Antonini}, F. \& {Merritt}, D. 2012, \apj, 745, 83

\bibitem[{{Armitage} \& {Natarajan}(2005)}]{A2005}
{Armitage}, P.~J. \& {Natarajan}, P. 2005, \apj, 634, 921

\bibitem[{{Bah{\'e}} {et~al.}(2022){Bah{\'e}}, {Schaye}, {Schaller}, {Bower},
  {Borrow}, {Chaikin}, {Kugel}, {Nobels}, \&
  {Ploeckinger}}]{2022MNRAS.516..167B}
{Bah{\'e}}, Y.~M., {Schaye}, J., {Schaller}, M., {et~al.} 2022, \mnras, 516,
  167

\bibitem[{{Barausse}(2012)}]{B2012}
{Barausse}, E. 2012, \mnras, 423, 2533

\bibitem[{{Beckmann} {et~al.}(2018){Beckmann}, {Slyz}, \&
  {Devriendt}}]{Beckmann2018}
{Beckmann}, R.~S., {Slyz}, A., \& {Devriendt}, J. 2018, \mnras, 478, 995

\bibitem[{{Begelman} {et~al.}(1980){Begelman}, {Blandford}, \&
  {Rees}}]{BBR1980}
{Begelman}, M.~C., {Blandford}, R.~D., \& {Rees}, M.~J. 1980, \nat, 287, 307

\bibitem[{{Berczik} {et~al.}(2006){Berczik}, {Merritt}, {Spurzem}, \&
  {Bischof}}]{B2006}
{Berczik}, P., {Merritt}, D., {Spurzem}, R., \& {Bischof}, H.-P. 2006, \apjl,
  642, L21

\bibitem[{{Biernacki} {et~al.}(2017){Biernacki}, {Teyssier}, \&
  {Bleuler}}]{Biernacki2017}
{Biernacki}, P., {Teyssier}, R., \& {Bleuler}, A. 2017, \mnras, 469, 295

\bibitem[{{Binney} \& {Tremaine}(2008)}]{BT1987}
{Binney}, J. \& {Tremaine}, S. 2008, {Galactic Dynamics: Second Edition}

\bibitem[{{Blaes} {et~al.}(2002){Blaes}, {Lee}, \& {Socrates}}]{Bla2002}
{Blaes}, O., {Lee}, M.~H., \& {Socrates}, A. 2002, \apj, 578, 775

\bibitem[{{Bonetti} {et~al.}(2016){Bonetti}, {Haardt}, {Sesana}, \&
  {Barausse}}]{Bonetti2016}
{Bonetti}, M., {Haardt}, F., {Sesana}, A., \& {Barausse}, E. 2016, \mnras, 461,
  4419

\bibitem[{{Bonetti} {et~al.}(2018){Bonetti}, {Sesana}, {Barausse}, \&
  {Haardt}}]{Bonetti2018}
{Bonetti}, M., {Sesana}, A., {Barausse}, E., \& {Haardt}, F. 2018, \mnras, 477,
  2599

\bibitem[{{Bonetti} {et~al.}(2019){Bonetti}, {Sesana}, {Haardt}, {Barausse}, \&
  {Colpi}}]{Bonetti2019}
{Bonetti}, M., {Sesana}, A., {Haardt}, F., {Barausse}, E., \& {Colpi}, M. 2019,
  \mnras, 486, 4044

\bibitem[{{Bortolas} {et~al.}(2021){Bortolas}, {Franchini}, {Bonetti}, \&
  {Sesana}}]{Bortolas2021}
{Bortolas}, E., {Franchini}, A., {Bonetti}, M., \& {Sesana}, A. 2021, \apjl,
  918, L15

\bibitem[{{Capelo} {et~al.}(2015){Capelo}, {Volonteri}, {Dotti}, {Bellovary},
  {Mayer}, \& {Governato}}]{C2015}
{Capelo}, P.~R., {Volonteri}, M., {Dotti}, M., {et~al.} 2015, \mnras, 447, 2123

\bibitem[{{Chandrasekhar}(1943)}]{C1943}
{Chandrasekhar}, S. 1943, \apj, 97, 255

\bibitem[{{Chapon} {et~al.}(2013){Chapon}, {Mayer}, \&
  {Teyssier}}]{chaponetal13}
{Chapon}, D., {Mayer}, L., \& {Teyssier}, R. 2013, \mnras, 429, 3114

\bibitem[{{Chen} {et~al.}(2024{\natexlab{a}}){Chen}, {Mukherjee}, {Di Matteo},
  {Ni}, {Bird}, \& {Croft}}]{Chen2024}
{Chen}, N., {Mukherjee}, D., {Di Matteo}, T., {et~al.} 2024{\natexlab{a}}, The
  Open Journal of Astrophysics, 7, 28

\bibitem[{{Chen} {et~al.}(2024{\natexlab{b}}){Chen}, {Mukherjee}, {Matteo},
  {Ni}, {Bird}, \& {Croft}}]{2024OJAp....7E..28C}
{Chen}, N., {Mukherjee}, D., {Matteo}, T.~D., {et~al.} 2024{\natexlab{b}}, The
  Open Journal of Astrophysics, 7, 28

\bibitem[{{Chen} {et~al.}(2021){Chen}, {Ni}, {Holgado}, {Di Matteo}, {Tremmel},
  {DeGraf}, {Bird}, {Croft}, \& {Feng}}]{Chen2021}
{Chen}, N., {Ni}, Y., {Holgado}, A.~M., {et~al.} 2021, arXiv e-prints,
  arXiv:2112.08555

\bibitem[{{Chen} {et~al.}(2024{\natexlab{c}}){Chen}, {Daniel}, {D'Orazio},
  {Mitridate}, {Sagunski}, {Xue}, {Agazie}, {Baier}, {Baker}, {B{\'e}csy},
  {Blecha}, {Brazier}, {Brook}, {Burke-Spolaor}, {Burnette}, {Casey-Clyde},
  {Charisi}, {Chatterjee}, {Cohen}, {Cordes}, {Cornish}, {Crawford},
  {Cromartie}, {DeCesar}, {Demorest}, {Deng}, {Dey}, {Dolch}, {Ferrara},
  {Fiore}, {Fonseca}, {Freedman}, {Gardiner}, {Gersbach}, {Glaser}, {Good},
  {G{\"u}ltekin}, {Hazboun}, {Jennings}, {Johnson}, {Kaplan}, {Kelley}, {Key},
  {Laal}, {Lam}, {Lamb}, {Larsen}, {Lazio}, {Lewandowska}, {Liu}, {Luo},
  {Lynch}, {Ma}, {Madison}, {McEwen}, {McKee}, {McLaughlin}, {Meyers},
  {Mingarelli}, {Nice}, {Ocker}, {Olum}, {Pennucci}, {Petrov}, {Pol},
  {Radovan}, {Ransom}, {Ray}, {Romano}, {Runnoe}, {Saffer}, {Sardesai},
  {Schmitz}, {Siemens}, {Simon}, {Siwek}, {Sosa Fiscella}, {Stairs},
  {Stinebring}, {Susobhanan}, {Swiggum}, {Taylor}, {Taylor}, {Turner}, {Unal},
  {Vallisneri}, {van Haasteren}, {Verbiest}, {Vigeland}, {Witt}, {Wright}, \&
  {Young}}]{Chenyifan2024}
{Chen}, Y., {Daniel}, M., {D'Orazio}, D.~J., {et~al.} 2024{\natexlab{c}}, arXiv
  e-prints, arXiv:2411.05906

\bibitem[{{Cuadra} {et~al.}(2009{\natexlab{a}}){Cuadra}, {Armitage},
  {Alexander}, \& {Begelman}}]{C2009}
{Cuadra}, J., {Armitage}, P.~J., {Alexander}, R.~D., \& {Begelman}, M.~C.
  2009{\natexlab{a}}, \mnras, 393, 1423

\bibitem[{{Cuadra} {et~al.}(2009{\natexlab{b}}){Cuadra}, {Armitage},
  {Alexander}, \& {Begelman}}]{Cuadra2009}
{Cuadra}, J., {Armitage}, P.~J., {Alexander}, R.~D., \& {Begelman}, M.~C.
  2009{\natexlab{b}}, \mnras, 393, 1423

\bibitem[{{De Rosa} {et~al.}(2020){De Rosa}, {Vignali}, {Bogdanovi{\'c}},
  {Capelo}, {Charisi}, {Dotti}, {Husemann}, {Lusso}, {Mayer}, {Paragi},
  {Runnoe}, {Sesana}, {Steinborn}, {Bianchi}, {Colpi}, {Del Valle}, {Frey},
  {Gab{\'a}nyi}, {Giustini}, {Guainazzi}, {Haiman}, {Herrera Ruiz},
  {Herrero-Illana}, {Iwasawa}, {Komossa}, {Lena}, {Loiseau}, {Perez-Torres},
  {Piconcelli}, \& {Volonteri}}]{review2020}
{De Rosa}, A., {Vignali}, C., {Bogdanovi{\'c}}, T., {et~al.} 2020, arXiv
  e-prints, arXiv:2001.06293

\bibitem[{{del Valle} \& {Volonteri}(2018)}]{volonteri&delvalle2018}
{del Valle}, L. \& {Volonteri}, M. 2018, \mnras, 480, 439

\bibitem[{{Dittmann} \& {Ryan}(2022)}]{Dittmann2022}
{Dittmann}, A.~J. \& {Ryan}, G. 2022, \mnras, 513, 6158

\bibitem[{{Dittmann} \& {Ryan}(2024)}]{Dittmann2024}
{Dittmann}, A.~J. \& {Ryan}, G. 2024, \apj, 967, 12

\bibitem[{{Dong-P{\'a}ez} {et~al.}(2023){Dong-P{\'a}ez}, {Volonteri},
  {Beckmann}, {Dubois}, {Mangiagli}, {Trebitsch}, {Vergani}, \&
  {Webb}}]{DongPaez2023}
{Dong-P{\'a}ez}, C.~A., {Volonteri}, M., {Beckmann}, R.~S., {et~al.} 2023,
  \aap, 676, A2

\bibitem[{{D'Orazio} \& {Duffell}(2021)}]{Dora2021}
{D'Orazio}, D.~J. \& {Duffell}, P.~C. 2021, \apjl, 914, L21

\bibitem[{{D'Orazio} {et~al.}(2013){D'Orazio}, {Haiman}, \&
  {MacFadyen}}]{DOrazio2013}
{D'Orazio}, D.~J., {Haiman}, Z., \& {MacFadyen}, A. 2013, \mnras, 436, 2997

\bibitem[{{Dosopoulou} \& {Antonini}(2017)}]{DA2017}
{Dosopoulou}, F. \& {Antonini}, F. 2017, \apj, 840, 31

\bibitem[{{Dotti} {et~al.}(2007){Dotti}, {Colpi}, {Haardt}, \& {Mayer}}]{D2007}
{Dotti}, M., {Colpi}, M., {Haardt}, F., \& {Mayer}, L. 2007, \mnras, 379, 956

\bibitem[{{Dotti} {et~al.}(2013){Dotti}, {Colpi}, {Pallini}, {Perego}, \&
  {Volonteri}}]{Dotti2013}
{Dotti}, M., {Colpi}, M., {Pallini}, S., {Perego}, A., \& {Volonteri}, M. 2013,
  \apj, 762, 68

\bibitem[{{Dubois} {et~al.}(2021){Dubois}, {Beckmann}, {Bournaud}, {Choi},
  {Devriendt}, {Jackson}, {Kaviraj}, {Kimm}, {Kraljic}, {Laigle}, {Martin},
  {Park}, {Peirani}, {Pichon}, {Volonteri}, \& {Yi}}]{Dubois2021}
{Dubois}, Y., {Beckmann}, R., {Bournaud}, F., {et~al.} 2021, \aap, 651, A109

\bibitem[{{Dubois} {et~al.}(2012){Dubois}, {Devriendt}, {Slyz}, \&
  {Teyssier}}]{Dubois2012}
{Dubois}, Y., {Devriendt}, J., {Slyz}, A., \& {Teyssier}, R. 2012, \mnras, 420,
  2662

\bibitem[{{Dubois} {et~al.}(2013){Dubois}, {Pichon}, {Devriendt}, {Silk},
  {Haehnelt}, {Kimm}, \& {Slyz}}]{Dubois2013}
{Dubois}, Y., {Pichon}, C., {Devriendt}, J., {et~al.} 2013, \mnras, 428, 2885

\bibitem[{{Dubois} \& {Teyssier}(2008)}]{dubois&teyssier08winds}
{Dubois}, Y. \& {Teyssier}, R. 2008, \aap, 477, 79

\bibitem[{Dubois {et~al.}(2014)Dubois, Volonteri, \& Silk}]{Dubois2014a}
Dubois, Y., Volonteri, M., \& Silk, J. 2014, \mnras, 440, 1590

\bibitem[{{Dubois} {et~al.}(2014){Dubois}, {Volonteri}, {Silk}, {Devriendt}, \&
  {Slyz}}]{Dubois2014b}
{Dubois}, Y., {Volonteri}, M., {Silk}, J., {Devriendt}, J., \& {Slyz}, A. 2014,
  \mnras, 440, 2333

\bibitem[{{Duffell} {et~al.}(2020){Duffell}, {D'Orazio}, {Derdzinski},
  {Haiman}, {MacFadyen}, {Rosen}, \& {Zrake}}]{Duffell2020}
{Duffell}, P.~C., {D'Orazio}, D., {Derdzinski}, A., {et~al.} 2020, \apj, 901,
  25

\bibitem[{{EPTA Collaboration} {et~al.}(2024){EPTA Collaboration}, {InPTA
  Collaboration}, {Antoniadis}, {Arumugam}, {Arumugam}, {Babak}, {Bagchi}, {Bak
  Nielsen}, {Bassa}, {Bathula}, {Berthereau}, {Bonetti}, {Bortolas}, {Brook},
  {Burgay}, {Caballero}, {Chalumeau}, {Champion}, {Chanlaridis}, {Chen},
  {Cognard}, {Dandapat}, {Deb}, {Desai}, {Desvignes}, {Dhanda-Batra},
  {Dwivedi}, {Falxa}, {Ferdman}, {Franchini}, {Gair}, {Goncharov}, {Gopakumar},
  {Graikou}, {Grie{\ss}meier}, {Gualandris}, {Guillemot}, {Guo}, {Gupta},
  {Hisano}, {Hu}, {Iraci}, {Izquierdo-Villalba}, {Jang}, {Jawor}, {Janssen},
  {Jessner}, {Joshi}, {Kareem}, {Karuppusamy}, {Keane}, {Keith}, {Kharbanda},
  {Kikunaga}, {Kolhe}, {Kramer}, {Krishnakumar}, {Lackeos}, {Lee}, {Liu},
  {Liu}, {Lyne}, {McKee}, {Maan}, {Main}, {Mickaliger}, {Ni{\c{t}}u},
  {Nobleson}, {Paladi}, {Parthasarathy}, {Perera}, {Perrodin}, {Petiteau},
  {Porayko}, {Possenti}, {Prabu}, {Quelquejay Leclere}, {Rana}, {Samajdar},
  {Sanidas}, {Sesana}, {Shaifullah}, {Singha}, {Speri}, {Spiewak},
  {Srivastava}, {Stappers}, {Surnis}, {Susarla}, {Susobhanan}, {Takahashi},
  {Tarafdar}, {Theureau}, {Tiburzi}, {van der Wateren}, {Vecchio}, {Venkatraman
  Krishnan}, {Verbiest}, {Wang}, {Wang}, {Wu}, {Auclair}, {Barausse},
  {Caprini}, {Crisostomi}, {Fastidio}, {Khizriev}, {Middleton}, {Neronov},
  {Postnov}, {Roper Pol}, {Semikoz}, {Smarra}, {Steer}, {Truant}, \&
  {Valtolina}}]{EPTA2024}
{EPTA Collaboration}, {InPTA Collaboration}, {Antoniadis}, J., {et~al.} 2024,
  \aap, 685, A94

\bibitem[{{Escala} {et~al.}(2005){Escala}, {Larson}, {Coppi}, \&
  {Mardones}}]{E2005}
{Escala}, A., {Larson}, R.~B., {Coppi}, P.~S., \& {Mardones}, D. 2005, \apj,
  630, 152

\bibitem[{{Farris} {et~al.}(2014){Farris}, {Duffell}, {MacFadyen}, \&
  {Haiman}}]{Farris2014}
{Farris}, B.~D., {Duffell}, P., {MacFadyen}, A.~I., \& {Haiman}, Z. 2014, \apj,
  783, 134

\bibitem[{{Fellhauer} \& {Lin}(2007)}]{Fell2007}
{Fellhauer}, M. \& {Lin}, D.~N.~C. 2007, \mnras, 375, 604

\bibitem[{{Franchini} {et~al.}(2022){Franchini}, {Lupi}, \&
  {Sesana}}]{Franchini2022}
{Franchini}, A., {Lupi}, A., \& {Sesana}, A. 2022, \apjl, 929, L13

\bibitem[{{Franchini} {et~al.}(2021){Franchini}, {Sesana}, \&
  {Dotti}}]{Franchini2021}
{Franchini}, A., {Sesana}, A., \& {Dotti}, M. 2021, \mnras, 507, 1458

\bibitem[{{Fujii} {et~al.}(2006){Fujii}, {Funato}, \& {Makino}}]{Fuji2006}
{Fujii}, M., {Funato}, Y., \& {Makino}, J. 2006, \pasj, 58, 743

\bibitem[{{Gabor} {et~al.}(2016){Gabor}, {Capelo}, {Volonteri}, {Bournaud},
  {Bellovary}, {Governato}, \& {Quinn}}]{Gabor2016}
{Gabor}, J.~M., {Capelo}, P.~R., {Volonteri}, M., {et~al.} 2016, \aap, 592, A62

\bibitem[{Graham {et~al.}(2003)Graham, Erwin, Trujillo, \& Ramos}]{Graham2003}
Graham, A.~W., Erwin, P., Trujillo, I., \& Ramos, A.~A. 2003, The Astronomical
  Journal, 125, 2951

\bibitem[{{Gualandris} {et~al.}(2022){Gualandris}, {Khan}, {Bortolas},
  {Bonetti}, {Sesana}, {Berczik}, \& {Holley-Bockelmann}}]{Gualandris2022}
{Gualandris}, A., {Khan}, F.~M., {Bortolas}, E., {et~al.} 2022, \mnras, 511,
  4753

\bibitem[{{Haardt} \& {Madau}(1996)}]{haardt&madau96}
{Haardt}, F. \& {Madau}, P. 1996, \apj, 461, 20

\bibitem[{{Haiman} {et~al.}(2009){Haiman}, {Kocsis}, \& {Menou}}]{Haiman2009}
{Haiman}, Z., {Kocsis}, B., \& {Menou}, K. 2009, \apj, 700, 1952

\bibitem[{{Heath} \& {Nixon}(2020)}]{Heath2020}
{Heath}, R.~M. \& {Nixon}, C.~J. 2020, \aap, 641, A64

\bibitem[{{Hernquist}(1990)}]{Hernquist1990}
{Hernquist}, L. 1990, \apj, 356, 359

\bibitem[{{Hoffman} \& {Loeb}(2007{\natexlab{a}})}]{Hoff2007}
{Hoffman}, L. \& {Loeb}, A. 2007{\natexlab{a}}, \mnras, 377, 957

\bibitem[{{Hoffman} \& {Loeb}(2007{\natexlab{b}})}]{Hoffman2007}
{Hoffman}, L. \& {Loeb}, A. 2007{\natexlab{b}}, \mnras, 377, 957

\bibitem[{{Izquierdo-Villalba} {et~al.}(2020){Izquierdo-Villalba}, {Bonoli},
  {Dotti}, {Sesana}, {Rosas-Guevara}, \& {Spinoso}}]{2020MNRAS.495.4681I}
{Izquierdo-Villalba}, D., {Bonoli}, S., {Dotti}, M., {et~al.} 2020, \mnras,
  495, 4681

\bibitem[{{Katz} \& {Larson}(2019)}]{KL2018}
{Katz}, M.~L. \& {Larson}, S.~L. 2019, \mnras, 483, 3108

\bibitem[{{Khan} {et~al.}(2013){Khan}, {Holley-Bockelmann}, {Berczik}, \&
  {Just}}]{K2013}
{Khan}, F.~M., {Holley-Bockelmann}, K., {Berczik}, P., \& {Just}, A. 2013,
  \apj, 773, 100

\bibitem[{{Khan} {et~al.}(2011){Khan}, {Just}, \& {Merritt}}]{KJM2011}
{Khan}, F.~M., {Just}, A., \& {Merritt}, D. 2011, \apj, 732, 89

\bibitem[{{Kim} \& {Kim}(2007)}]{KK2007}
{Kim}, H. \& {Kim}, W.-T. 2007, \apj, 665, 432

\bibitem[{Kimm \& Cen(2014)}]{Kimm&cen2014}
Kimm, T. \& Cen, R. 2014, \apj, 788, 121

\bibitem[{{Klein} {et~al.}(2016){Klein}, {Barausse}, {Sesana}, {Petiteau},
  {Berti}, {Babak}, {Gair}, {Aoudia}, {Hinder}, {Ohme}, \& {Wardell}}]{K2016}
{Klein}, A., {Barausse}, E., {Sesana}, A., {et~al.} 2016, \prd, 93, 024003

\bibitem[{{Kormendy} \& {Richstone}(1995)}]{KR1995}
{Kormendy}, J. \& {Richstone}, D. 1995, \araa, 33, 581

\bibitem[{{Kozai}(1962)}]{Kozai1962}
{Kozai}, Y. 1962, \aj, 67, 591

\bibitem[{{Kulkarni} \& {Loeb}(2012)}]{Kul2012}
{Kulkarni}, G. \& {Loeb}, A. 2012, \mnras, 422, 1306

\bibitem[{{Lai} \& {Mu{\~n}oz}(2023)}]{Lai2023}
{Lai}, D. \& {Mu{\~n}oz}, D.~J. 2023, \araa, 61, 517

\bibitem[{{Lescaudron} {et~al.}(2023){Lescaudron}, {Dubois}, {Beckmann}, \&
  {Volonteri}}]{Lescaudron23}
{Lescaudron}, S., {Dubois}, Y., {Beckmann}, R.~S., \& {Volonteri}, M. 2023,
  \aap, 674, A217

\bibitem[{{Li} {et~al.}(2020{\natexlab{a}}){Li}, {Bogdanovi{\'c}}, \&
  {Ballantyne}}]{LBB20a}
{Li}, K., {Bogdanovi{\'c}}, T., \& {Ballantyne}, D.~R. 2020{\natexlab{a}},
  \apj, 896, 113 (LBB20)

\bibitem[{{Li} {et~al.}(2020{\natexlab{b}}){Li}, {Bogdanovi{\'c}}, \&
  {Ballantyne}}]{LBB20b}
{Li}, K., {Bogdanovi{\'c}}, T., \& {Ballantyne}, D.~R. 2020{\natexlab{b}},
  \apj, 905, 123

\bibitem[{{Li} {et~al.}(2022){Li}, {Bogdanovi{\'c}}, {Ballantyne}, \&
  {Bonetti}}]{LBB2022}
{Li}, K., {Bogdanovi{\'c}}, T., {Ballantyne}, D.~R., \& {Bonetti}, M. 2022,
  \apj, 933, 104

\bibitem[{{Liao} {et~al.}(2024{\natexlab{a}}){Liao}, {Irodotou}, {Johansson},
  {Naab}, {Rizzuto}, {Hislop}, {Rawlings}, \& {Wright}}]{Liao2024a}
{Liao}, S., {Irodotou}, D., {Johansson}, P.~H., {et~al.} 2024{\natexlab{a}},
  \mnras, 528, 5080

\bibitem[{{Liao} {et~al.}(2024{\natexlab{b}}){Liao}, {Irodotou}, {Johansson},
  {Naab}, {Rizzuto}, {Hislop}, {Wright}, \& {Rawlings}}]{Liao2024b}
{Liao}, S., {Irodotou}, D., {Johansson}, P.~H., {et~al.} 2024{\natexlab{b}},
  \mnras, 530, 4058

\bibitem[{{Lupi} {et~al.}(2019){Lupi}, {Volonteri}, {Decarli}, {Bovino},
  {Silk}, \& {Bergeron}}]{Lupi2019}
{Lupi}, A., {Volonteri}, M., {Decarli}, R., {et~al.} 2019, \mnras, 488, 4004

\bibitem[{{Magorrian} {et~al.}(1998){Magorrian}, {Tremaine}, {Richstone},
  {Bender}, {Bower}, {Dressler}, {Faber}, {Gebhardt}, {Green}, {Grillmair},
  {Kormendy}, \& {Lauer}}]{M1998}
{Magorrian}, J., {Tremaine}, S., {Richstone}, D., {et~al.} 1998, \aj, 115, 2285

\bibitem[{{Mannerkoski} {et~al.}(2021){Mannerkoski}, {Johansson}, {Rantala},
  {Naab}, \& {Liao}}]{Mannerkoski2021}
{Mannerkoski}, M., {Johansson}, P.~H., {Rantala}, A., {Naab}, T., \& {Liao}, S.
  2021, \apjl, 912, L20

\bibitem[{{Mannerkoski} {et~al.}(2023){Mannerkoski}, {Rawlings}, {Johansson},
  {Naab}, {Rantala}, {Springel}, {Irodotou}, \& {Liao}}]{Matias2023}
{Mannerkoski}, M., {Rawlings}, A., {Johansson}, P.~H., {et~al.} 2023, \mnras,
  524, 4062

\bibitem[{{Mikkola} \& {Valtonen}(1990)}]{Mikkola1990}
{Mikkola}, S. \& {Valtonen}, M.~J. 1990, \apj, 348, 412

\bibitem[{Miranda {et~al.}(2016)Miranda, Muñoz, \& Lai}]{Miranda2017}
Miranda, R., Muñoz, D.~J., \& Lai, D. 2016, \mnras, 466, 1170

\bibitem[{{Moody} {et~al.}(2019){Moody}, {Shi}, \& {Stone}}]{Moody2019}
{Moody}, M. S.~L., {Shi}, J.-M., \& {Stone}, J.~M. 2019, \apj, 875, 66

\bibitem[{{Mu{\~n}oz} {et~al.}(2020){Mu{\~n}oz}, {Lai}, {Kratter}, \&
  {Miranda}}]{Munoz2020}
{Mu{\~n}oz}, D.~J., {Lai}, D., {Kratter}, K., \& {Miranda}, R. 2020, \apj, 889,
  114

\bibitem[{{Mu{\~n}oz} {et~al.}(2019){Mu{\~n}oz}, {Miranda}, \&
  {Lai}}]{Munoz2019}
{Mu{\~n}oz}, D.~J., {Miranda}, R., \& {Lai}, D. 2019, \apj, 871, 84

\bibitem[{{Mukherjee} {et~al.}(2025){Mukherjee}, {Zhou}, {Chen}, {Di Carlo}, \&
  {Di Matteo}}]{Muk2024}
{Mukherjee}, D., {Zhou}, Y., {Chen}, N., {Di Carlo}, U.~N., \& {Di Matteo}, T.
  2025, \apj, 981, 203

\bibitem[{Navarro {et~al.}(1997)Navarro, Frenk, \& White}]{NFW1997}
Navarro, J.~F., Frenk, C.~S., \& White, S. D.~M. 1997, \apj, 490, 493

\bibitem[{{Negri} \& {Volonteri}(2017)}]{Negri2017}
{Negri}, A. \& {Volonteri}, M. 2017, \mnras, 467, 3475

\bibitem[{{Ogiya} {et~al.}(2020){Ogiya}, {Hahn}, {Mingarelli}, \&
  {Volonteri}}]{Ogiya2020}
{Ogiya}, G., {Hahn}, O., {Mingarelli}, C. M.~F., \& {Volonteri}, M. 2020,
  \mnras, 493, 3676

\bibitem[{{Ogiya} {et~al.}(2019){Ogiya}, {van den Bosch}, {Hahn}, {Green},
  {Miller}, \& {Burkert}}]{Ogiya2019}
{Ogiya}, G., {van den Bosch}, F.~C., {Hahn}, O., {et~al.} 2019, \mnras, 485,
  189

\bibitem[{{Ostriker}(1999)}]{O1999}
{Ostriker}, E.~C. 1999, \apj, 513, 252

\bibitem[{{Park} \& {Bogdanovi{\'c}}(2017)}]{PB2017}
{Park}, K. \& {Bogdanovi{\'c}}, T. 2017, \apj, 838, 103

\bibitem[{Peters(1964)}]{Peters1964}
Peters, P.~C. 1964, Phys. Rev., 136, B1224

\bibitem[{{Pfister} {et~al.}(2019){Pfister}, {Volonteri}, {Dubois}, {Dotti}, \&
  {Colpi}}]{Hugo2019}
{Pfister}, H., {Volonteri}, M., {Dubois}, Y., {Dotti}, M., \& {Colpi}, M. 2019,
  \mnras, 486, 101

\bibitem[{{Quinlan}(1996)}]{Q1996}
{Quinlan}, G.~D. 1996, \na, 1, 35

\bibitem[{{Quinlan} \& {Hernquist}(1997)}]{QH1997}
{Quinlan}, G.~D. \& {Hernquist}, L. 1997, \na, 2, 533

\bibitem[{{Rantala} {et~al.}(2017){Rantala}, {Pihajoki}, {Johansson}, {Naab},
  {Lah{\'e}n}, \& {Sawala}}]{Rantala2017}
{Rantala}, A., {Pihajoki}, P., {Johansson}, P.~H., {et~al.} 2017, \apj, 840, 53

\bibitem[{{Rantala} {et~al.}(2020){Rantala}, {Pihajoki}, {Mannerkoski},
  {Johansson}, \& {Naab}}]{Rantala2020}
{Rantala}, A., {Pihajoki}, P., {Mannerkoski}, M., {Johansson}, P.~H., \&
  {Naab}, T. 2020, \mnras, 492, 4131

\bibitem[{{Rasera} \& {Teyssier}(2006)}]{rasera&teyssier06}
{Rasera}, Y. \& {Teyssier}, R. 2006, \aap, 445, 1

\bibitem[{{Rawlings} {et~al.}(2023){Rawlings}, {Mannerkoski}, {Johansson}, \&
  {Naab}}]{Rawling2023}
{Rawlings}, A., {Mannerkoski}, M., {Johansson}, P.~H., \& {Naab}, T. 2023,
  \mnras, 526, 2688

\bibitem[{{Reardon} {et~al.}(2023){Reardon}, {Zic}, {Shannon}, {Hobbs},
  {Bailes}, {Di Marco}, {Kapur}, {Rogers}, {Thrane}, {Askew}, {Bhat},
  {Cameron}, {Cury{\l}o}, {Coles}, {Dai}, {Goncharov}, {Kerr}, {Kulkarni},
  {Levin}, {Lower}, {Manchester}, {Mandow}, {Miles}, {Nathan}, {Os{\l}owski},
  {Russell}, {Spiewak}, {Zhang}, \& {Zhu}}]{Reardon2023}
{Reardon}, D.~J., {Zic}, A., {Shannon}, R.~M., {et~al.} 2023, \apjl, 951, L6

\bibitem[{{Roedig} {et~al.}(2011){Roedig}, {Dotti}, {Sesana}, {Cuadra}, \&
  {Colpi}}]{Roedig2011}
{Roedig}, C., {Dotti}, M., {Sesana}, A., {Cuadra}, J., \& {Colpi}, M. 2011,
  \mnras, 415, 3033

\bibitem[{{Roedig} \& {Sesana}(2014)}]{Roedig2014}
{Roedig}, C. \& {Sesana}, A. 2014, \mnras, 439, 3476

\bibitem[{{Roedig} {et~al.}(2012){Roedig}, {Sesana}, {Dotti}, {Cuadra},
  {Amaro-Seoane}, \& {Haardt}}]{Roedig2012}
{Roedig}, C., {Sesana}, A., {Dotti}, M., {et~al.} 2012, \aap, 545, A127

\bibitem[{{Rosdahl} {et~al.}(2017){Rosdahl}, {Schaye}, {Dubois}, {Kimm}, \&
  {Teyssier}}]{Rosdahl17}
{Rosdahl}, J., {Schaye}, J., {Dubois}, Y., {Kimm}, T., \& {Teyssier}, R. 2017,
  \mnras, 466, 11

\bibitem[{{Rosen} \& {Bregman}(1995)}]{Rosen1995}
{Rosen}, A. \& {Bregman}, J.~N. 1995, \apj, 440, 634

\bibitem[{{Ryu} {et~al.}(2018){Ryu}, {Perna}, {Haiman}, {Ostriker}, \&
  {Stone}}]{Ryu2018}
{Ryu}, T., {Perna}, R., {Haiman}, Z., {Ostriker}, J.~P., \& {Stone}, N.~C.
  2018, \mnras, 473, 3410

\bibitem[{{Sesana} {et~al.}(2006){Sesana}, {Haardt}, \& {Madau}}]{Sesana2006}
{Sesana}, A., {Haardt}, F., \& {Madau}, P. 2006, \apj, 651, 392

\bibitem[{{Shakura} \& {Sunyaev}(1973)}]{SS1973}
{Shakura}, N.~I. \& {Sunyaev}, R.~A. 1973, \aap, 500, 33

\bibitem[{{Shapiro} \& {Teukolsky}(1983)}]{Shapiro1983}
{Shapiro}, S.~L. \& {Teukolsky}, S.~A. 1983, {Black holes, white dwarfs, and
  neutron stars : the physics of compact objects}

\bibitem[{{Shi} {et~al.}(2012){Shi}, {Krolik}, {Lubow}, \& {Hawley}}]{Shi2012}
{Shi}, J.-M., {Krolik}, J.~H., {Lubow}, S.~H., \& {Hawley}, J.~F. 2012, \apj,
  749, 118

\bibitem[{{Siwek} {et~al.}(2023){Siwek}, {Weinberger}, \&
  {Hernquist}}]{Siwek2023}
{Siwek}, M., {Weinberger}, R., \& {Hernquist}, L. 2023, \mnras, 522, 2707

\bibitem[{{Soltan}(1982)}]{S1982}
{Soltan}, A. 1982, \mnras, 200, 115

\bibitem[{{Springel} {et~al.}(2005){Springel}, {Di Matteo}, \&
  {Hernquist}}]{Springel05}
{Springel}, V., {Di Matteo}, T., \& {Hernquist}, L. 2005, \mnras, 361, 776

\bibitem[{{Springel} {et~al.}(2021){Springel}, {Pakmor}, {Zier}, \&
  {Reinecke}}]{Springel2021}
{Springel}, V., {Pakmor}, R., {Zier}, O., \& {Reinecke}, M. 2021, \mnras, 506,
  2871

\bibitem[{{Sutherland} \& {Dopita}(1993)}]{Sutherland1993}
{Sutherland}, R.~S. \& {Dopita}, M.~A. 1993, \apjs, 88, 253

\bibitem[{{Tang} {et~al.}(2017){Tang}, {MacFadyen}, \& {Haiman}}]{Tang2017}
{Tang}, Y., {MacFadyen}, A., \& {Haiman}, Z. 2017, \mnras, 469, 4258

\bibitem[{{Teyssier}(2002)}]{Teyssier2002}
{Teyssier}, R. 2002, \aap, 385, 337

\bibitem[{{Thorne} \& {Braginskii}(1976)}]{KT1976}
{Thorne}, K.~S. \& {Braginskii}, V.~B. 1976, \apjl, 204, L1

\bibitem[{{Tiede} {et~al.}(2020){Tiede}, {Zrake}, {MacFadyen}, \&
  {Haiman}}]{Tiede2020}
{Tiede}, C., {Zrake}, J., {MacFadyen}, A., \& {Haiman}, Z. 2020, \apj, 900, 43

\bibitem[{{Toomre}(1964)}]{T1964}
{Toomre}, A. 1964, \apj, 139, 1217

\bibitem[{{Toyouchi} {et~al.}(2020){Toyouchi}, {Hosokawa}, {Sugimura}, \&
  {Kuiper}}]{T2020}
{Toyouchi}, D., {Hosokawa}, T., {Sugimura}, K., \& {Kuiper}, R. 2020, \mnras
  [\eprint[arXiv]{2002.08017}]

\bibitem[{{Tremmel} {et~al.}(2015){Tremmel}, {Governato}, {Volonteri}, \&
  {Quinn}}]{Tre2015}
{Tremmel}, M., {Governato}, F., {Volonteri}, M., \& {Quinn}, T.~R. 2015,
  \mnras, 451, 1868

\bibitem[{Trujillo {et~al.}(2004)Trujillo, Erwin, Ramos, \&
  Graham}]{Trujillo2004}
Trujillo, I., Erwin, P., Ramos, A.~A., \& Graham, A.~W. 2004, The Astronomical
  Journal, 127, 1917

\bibitem[{{van den Bosch} \& {Ogiya}(2018)}]{VanOgiya2018}
{van den Bosch}, F.~C. \& {Ogiya}, G. 2018, \mnras, 475, 4066

\bibitem[{{Volonteri} {et~al.}(2003){Volonteri}, {Haardt}, \&
  {Madau}}]{Volonteri2003}
{Volonteri}, M., {Haardt}, F., \& {Madau}, P. 2003, \apj, 582, 559

\bibitem[{{Volonteri} {et~al.}(2020){Volonteri}, {Pfister}, {Beckmann},
  {Dubois}, {Colpi}, {Conselice}, {Dotti}, {Martin}, {Jackson}, {Kraljic},
  {Pichon}, {Trebitsch}, {Yi}, {Devriendt}, \& {Peirani}}]{Marta2020}
{Volonteri}, M., {Pfister}, H., {Beckmann}, R.~S., {et~al.} 2020, \mnras, 498,
  2219

\bibitem[{{Wang} {et~al.}(2023){Wang}, {Bai}, \& {Lai}}]{wang2023}
{Wang}, H.-Y., {Bai}, X.-N., \& {Lai}, D. 2023, \apj, 943, 175

\bibitem[{{Yu}(2002)}]{Y2002}
{Yu}, Q. 2002, \mnras, 331, 935

\bibitem[{{Zrake} {et~al.}(2021){Zrake}, {Tiede}, {MacFadyen}, \&
  {Haiman}}]{Zrake2021}
{Zrake}, J., {Tiede}, C., {MacFadyen}, A., \& {Haiman}, Z. 2021, \apjl, 909,
  L13

\end{thebibliography}
\end{document}